\documentclass[a4paper,12pt]{article}

\usepackage[english]{babel}
\usepackage[utf8x]{inputenc}
\usepackage[T1]{fontenc}

\usepackage[a4paper,top=2.5cm,bottom=2cm,left=2.5cm,right=2.5cm,marginparwidth=1.75cm]{geometry}

\usepackage{etoolbox}
\makeatletter
\patchcmd{\maketitle}{\@makefntext}{\fakecommand}{}{}
\patchcmd{\maketitle}{\rlap}{\hbox}{}{}
\patchcmd{\@maketitle}{\@author}{\hspace*{5pt}\@author}{}{}
\makeatother

\usepackage{amsmath}
\usepackage{graphicx}
\usepackage[colorlinks=true, allcolors=blue]{hyperref}
\usepackage{setspace}
\usepackage{apacite}
\usepackage{dsfont}
\usepackage{booktabs}
\usepackage[titletoc]{appendix}
\usepackage{tabularx} 
\usepackage{colortbl}
\usepackage{longtable}
\usepackage{threeparttablex} 
\usepackage{threeparttable}
\usepackage{chngcntr}		
\usepackage[section]{placeins}
\usepackage{microtype}     
\usepackage{caption}        
\usepackage{rotating}
\usepackage{pdflscape}    
\usepackage{afterpage}
\usepackage{geometry}
\usepackage{color} 
\usepackage{float} 
\usepackage{lmodern}
\usepackage{hyperref}
\usepackage{subcaption}
\usepackage{bm}

\usepackage[hang,flushmargin]{footmisc}

\DeclareMathOperator*{\argmax}{\arg\max}
\DeclareMathOperator*{\argmin}{\arg\min}
\newcommand{\bigCI}{\mathrel{\text{\scalebox{1.07}{$\perp\mkern-10mu\perp$}}}} 

\newcommand\blfootnote[1]{%
  \begingroup
  \renewcommand\thefootnote{}\footnote{#1}%
  \endgroup
}

\newtheorem{assump}{Assumption}
\newtheorem{algorithm}{Algorithm}

\title{Double Machine Learning based Program Evaluation under Unconfoundedness\blfootnote{Financial support from the Swiss National Science Foundation (SNSF) is gratefully acknowledged (grant number SNSF 407540\_166999). I thank Petyo Bonev, Martin Huber, Edward Kennedy, Michael Lechner, Vira Semenova, Anthony Strittmatter, Stefan Wager, and Michael Zimmert for helpful comments and suggestions. The usual disclaimer applies.}}

\author{Michael C. Knaus\thanks{University of St. Gallen. Michael C. Knaus is also affiliated with IZA, Bonn, \href{mailto:michael.knaus@unisg.ch}{michael.knaus@unisg.ch}. }}

 \date{First version: March 9, 2020 \\ \smallskip This version: \today \\
}

\begin{document}
\maketitle

\doublespacing

\begin{abstract}

This paper reviews, applies and extends recently proposed methods based on Double Machine Learning (DML) with a focus on program evaluation under unconfoundedness. DML based methods leverage flexible prediction models to adjust for confounding variables in the estimation of (i) standard average effects, (ii) different forms of heterogeneous effects, and (iii) optimal treatment assignment rules. An evaluation of multiple programs of the Swiss Active Labour Market Policy illustrates how DML based methods enable a comprehensive program evaluation. Motivated by extreme individualised treatment effect estimates of the DR-learner, we propose the normalised DR-learner (NDR-learner) to address this issue. The NDR-learner acknowledges that individualised effect estimates can be stabilised by an individualised normalisation of inverse probability weights.


\end{abstract}

\newpage

\section{Introduction}

The adaptation of so-called machine learning to causal inference has been a productive area of methodological research in recent years. The resulting new methods complement the existing econometric toolbox for program evaluation along at least two dimensions \cite<see for recent overviews>{Athey2017,Athey2019MachineAbout,Abadie2018EconometricEvaluation}. On the one hand, they provide flexible methods to estimate standard average effects. In particular, they provide a data-driven approach to variable and model selection in studies that rely on an unconfoundedness assumption\footnote{Also known as exogeneity, selection on observables, ignorability, or conditional independence assumption.} for identification. On the other hand, they enable a more comprehensive evaluation by providing new methods for the flexible estimation of heterogeneous effects and of treatment assignment rules.

This paper considers Double Machine Learning (DML) \cite{Chernozhukov2018} as a framework for flexible and comprehensive program evaluation. The DML framework seems attractive because (i) it can be combined with a variety of standard supervised machine learning methods, (ii) it covers average effects for binary \cite<e.g.>{Belloni2014InferenceControls,Belloni2017,Chernozhukov2018}, multiple \cite<e.g.>{Farrell2015} as well as continuous treatments \cite<e.g.>{Kennedy2017Non-parametricEffects,Colangelo2019DoubleTreatments,Semenova2021DebiasedFunctions}, (iii) it naturally extends to the estimation of heterogeneous treatment effects of different forms like canonical subgroup effects, the best linear prediction of effect heterogeneity, or nonparametric effect heterogeneity \cite<e.g>{fan2020EstimationData,Zimmert2019NonparametricConfounding,Foster2019OrthogonalLearning,Oprescu2019OrthogonalInference,Semenova2021DebiasedFunctions,Kennedy2020OptimalEffects,Curth2021NonparametricAlgorithms}, and (iv) it can be used to estimate optimal treatment assignment rules \cite<e.g.>{Dudik2011DoublyLearning,Athey2021PolicyData,Zhou2018OfflineOptimization}. 
All these DML based methods have favourable statistical properties and allow the use of standard tools like \textit{t}-tests, OLS, kernel regression, series regression, or supervised machine learning for estimating causal parameters of interest after flexibly adjusting for confounding.

\begin{figure}[t] 
\centering
\caption{Stylised workflow of Double Machine Learning based program evaluation} \label{fig:arrows}
\includegraphics[width=0.8\textwidth]{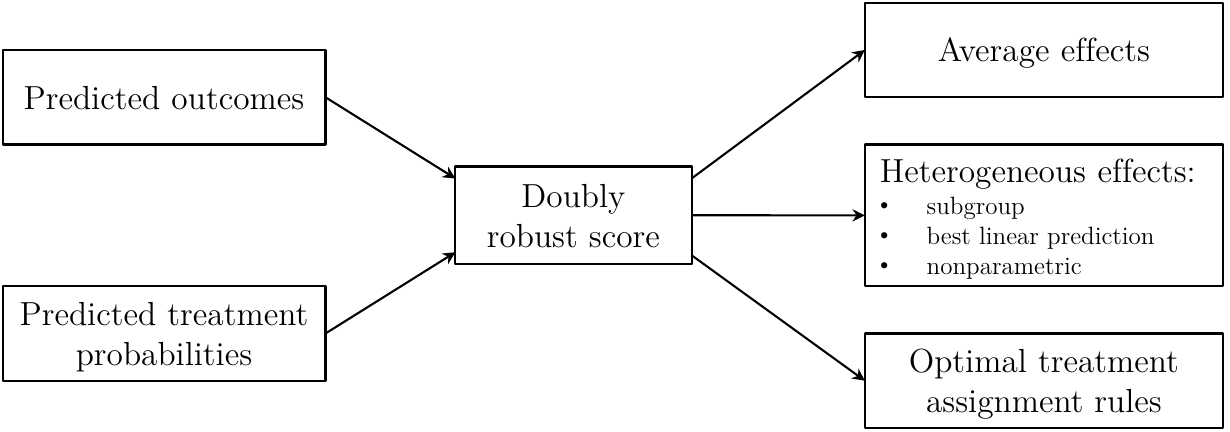}
\end{figure}

This paper starts with a review of DML based methods, then applies these methods in a standard labour economic setting, and comes back to the methods by proposing the normalised DR-learner as a potential fix to a finite sample problem encountered in the application. Thus, it contributes to the steadily growing literature of causal machine learning for program evaluation in three ways. First, the review highlights that methods for different parameters build on the same doubly robust score. The construction of this score might be computationally expensive because it requires the estimation of outcomes and treatment probabilities via machine learning methods. However, once constructed the score can be reused to estimate a variety of interesting parameters. This paper focuses on methods that build on the doubly robust score because they allow to leverage conceptual and computational synergies. The result is a comprehensive pipeline for program evaluation within the same framework as Figure \ref{fig:arrows} illustrates. This is currently not possible with the variety of more specialised alternatives that integrate machine learning in the estimation of average treatment effects \cite<e.g.>{vanderLaan2006TargetedLearning,Athey2018ApproximateDimensions,Avagyan2017HonestEstimation,Tan2020Model-assistedData,Ning2020RobustScore}, heterogeneous treatment effects \cite<e.g.>{Tian2014,Athey2016,Chernozhukov2017GenericExperiments,Wager2017,Athey2017a,Kunzel2017,Nie2021} and optimal treatment assignment \cite<e.g.>{Bansak2018ImprovingAssignment,Kallus2018BalancedLearning}.

Second, we use DML based methods to provide a comprehensive and computationally convenient evaluation of four programs of the Swiss Active Labour Market Policy (ALMP) in a standard dataset \cite{Lechner2020SwissDataset}. The evaluation in this paper illustrates the potential of DML based methods for program evaluations under unconfoundedness and provides a potential blueprint for similar analyses. This adds to a small but steadily growing literature that applies causal machine learning to program evaluation in general \cite<e.g.>{Bertrand2017,Strittmatter2018WhatEvaluation,Gulyas2019UnderstandingApproach,Knittel2019UsingUse,Davis2020RethinkingJobs,Baiardi2021TheStudies,Farbmacher2021HeterogeneousCognition} and to evaluations based on unconfoundedness in particular \cite<e.g>{Kreif2019MachineInference,Cockx2020PriorityBelgium,Knaus2020HeterogeneousApproach,Knaus2021ASkills}. 

Third, we contribute to the methodological literature on the flexible estimation of individualised treatment effects \cite<see for a recent overview>{Knaus2021} by proposing the normalised DR-learner (NDR-learner), which builds on the recent DR-learner of \citeA{Kennedy2020OptimalEffects}. The application reveals that the plain DR-learner produces few extreme effect estimates. It turns out that individualised effect estimates can be stabilised by an individualised normalisation of inverse probability weights. Thus, the NDR-learner can be considered as a generalisation of the popular \citeA{Hajek1971CommentOne} normalisation for inverse probability weighting estimators for average effects. The increased stability comes at the price that the NDR-learner limits the class of permissible machine learning methods for effect heterogeneity estimation to methods that form predictions as convex combination of outcomes (e.g. Random Forests).

Overall, we find that DML based methods provide a promising set of methods for program evaluation. The estimated average program effects are in line with the previous literature. We find that computer, vocational and language courses increase employment in the 31 months after programs start, while the effects of job search trainings are mostly negative. The heterogeneity analysis additionally reveals substantial heterogeneities by gender, nationality, previous labour market success and qualification. These are picked up by the estimated optimal assignment rules.

The paper proceeds as follows. Section \ref{sec:estimands} defines the estimands of interest and their identification under unconfoundedness. Section \ref{sec:dml} reviews DML based methods for estimation and introduces the NDR-learner. Section \ref{sec:appl} presents the application. Section \ref{sec:imp} describes the implementation of the methods. Section \ref{sec:res} reports the results. Section \ref{sec:conc} concludes. The Appendix provides additional explanations and results. The R-package  \href{https://github.com/MCKnaus/causalDML}{\color{blue}\texttt{causalDML}} implements the applied estimators. An \href{https://mcknaus.github.io/assets/code/Notebook_DML_ALMP_MCK2020.html}{\color{blue}R notebook} replicating the analysis is provided.

\section{Estimands of interest} \label{sec:estimands}

\subsection{Definition} \label{sec:def}

We define the estimands of interest in the multiple treatment version of the potential outcomes framework \cite{Rubin1974,Imbens2000TheFunctions,Lechner2001}. Let $\mathcal{W} = \{0,...,T\}$ denote a set of multiple programs and $D_i(w) = \mathds{1}(W_i = w)$ a binary variable indicating in which program individual $i$ ($i=1, ..., N$) is actually observed.\footnote{For DML based estimation with continuous treatments see, e.g \citeA{Kennedy2017Non-parametricEffects}, \citeA{Colangelo2019DoubleTreatments}, and \citeA{Semenova2021DebiasedFunctions}.} We assume that each individual has a potential outcome $Y_i(w)$ for all $w \in \mathcal{W}$. Without loss of generality, the discussion below assumes that higher outcome values are desirable.

The first estimand of interest is the average potential outcome (APO), $\gamma_w = E[Y_i(w)]$. It answers the question about the average outcome if the whole population was assigned to program $w$. However, the more interesting question is usually to compare different programs $w$ and $w'$. To this end, we take the difference of the according individual potential outcomes, $Y_i(w) - Y_i(w')$,\footnote{This would be $Y_i(1) - Y_i(0)$ in the canonical binary treatment setting.} and aggregate them to different estimands: First, the average treatment effect (ATE), $\delta_{w,w'} = E[Y_i(w) - Y_i(w')]$. Second, the average treatment effect on the treated (ATET), $\theta_{w,w'} = E[Y_i(w) - Y_i(w') \mid W_i = w]$. Third, the conditional average treatment effect (CATE), $\tau_{w,w'}(z) = E[Y_i(w) - Y_i(w') \mid Z_i=z]$, where $Z_i \in \mathcal{Z}$ is a vector of observed pre-treatment variables.\footnote{We focus in this study on expectations of the individual treatment effects. DML based methods for quantile treatment effects can be found, e.g. in \citeA{Belloni2017} and \citeA{Kallus2019LocalizedBeyond}.}

The different aggregations accommodate the notion that treatment effects might be heterogeneous. ATE represents the average effect in the population, while ATET shows it for the subpopulation that is actually observed in program $w$. Thus, the comparison of ATE and ATET can be informative about the quality of the program assignment mechanism. For example, ATET being larger than ATE indicates that the observed program assignment is better than random. 

The ATET is defined by the observed program assignment and thus not subject to the choice of the researcher. In contrast, the conditioning variables $Z_i$ of the CATE are specified by the researcher to investigate potentially heterogeneous effects across the groups of individuals that are defined by different values of $Z_i$. Such heterogeneous effects can be indicative for underlying mechanisms. Further, CATEs characterise which groups win and which lose by how much by receiving program $w$ instead of $w'$.

The different average effects above provide a comprehensive evaluation of programs under the current program assignment policy. In many applications, however, we want to conclude the analysis with a recommendation how the assignment policy could be improved. This can either be done using the evidence on the different average effects defined above or by formally defining the objective of an optimal assignment rule. The latter is pursued by the literature on statistical treatment rules \cite<e.g.>[and references therein]{Manski2004StatisticalPopulations,Hirano2009AsymptoticsRules,Stoye2009MinimaxSamples,Stoye2012MinimaxExperiments,Kitagawa2018WhoChoice,Athey2021PolicyData}. Here we focus on the case with multiple treatment options as considered by \citeA{Zhou2018OfflineOptimization}. 

Let $\pi(Z_i)$ be a policy that assigns individuals to programs according to their characteristics $Z_i$ or, put more formally, the function $\pi(Z_i)$ maps observable characteristics to a program: $\pi: \mathcal{Z} \rightarrow \mathcal{W}$. In principle, the policy rule can be completely flexible and in the ideal world we would assign each individual to the program with the highest conditional APO, $E[Y_i(w) \mid Z_i =z]$. However, in many cases we want to restrict the set of candidate policy rules denoted by $\Pi$ to be interpretable for the communication with decision makers or to incorporate costs or fairness constraints. Each of these candidate policy rules has a policy value function denoted by $Q(\pi) = E[Y_i(\pi(Z_i))] = E \left[ \sum_w \mathds{1}(\pi(Z_i) = w) Y_i(w) \right]$. 
$Q(\pi)$ quantifies the average population outcome if policy rule $\pi$ would be used to assign programs. The estimand of interest is then the optimal policy rule $\pi^*$ with the highest value function for the set of candidate policy rules, or formally $\pi^* = \argmax_{\pi \in \Pi} Q(\pi)$. 

\subsection{Identification}

The previous section defined the estimands of interest in terms of potential outcomes. However, each individual is only observed in one program. Thus, only one potential outcome per individual is observable and the other potential outcomes remain latent. This is the fundamental problem of causal inference \cite{Holland1986StatisticsInference} and we need further assumptions to identify the estimands of interest. In this paper, we consider the unconfoundedness assumption that assumes access to a vector of pre-treatment variables $X_i \in \mathcal{X}$ containing $Z_i$ such that the following standard assumptions hold \cite<e.g.>{Imbens2015CausalSciences}: 

\begin{assump}\label{ass:cia} $~$ 

(a) Unconfoundedness: $Y_i(w) \bigCI W_i \mid X_i=x$, $\forall$ $w \in \mathcal{W}$, and $x \in \mathcal{X}$.

(b) Common support: $0 < P[W_i = w \mid X_i =x] \equiv e_w(x)$, $\forall$ $w \in \mathcal{W}$ and $x \in \mathcal{X}$.

(c) Stable Unit Treatment Value Assumption (SUTVA): $Y_i=Y_i(W_i)$.
\end{assump}

The unconfoundedness assumption requires that $X_i$ contains all confounding variables that jointly affect program assignment and the outcome. Common support states that it must be possible to observe each individual in all programs. SUTVA rules out interference. These assumptions allow the identification of the average potential outcome (APO) conditional on confounders in three common ways:
\begin{align}
    E[Y_i(w) \mid X_i=x] & = E\left[Y_i \mid W_i = w, X_i = x \right] \equiv \mu(w,x) \label{eq:id1} \\
                         & = E \left[ \dfrac{D_i(w) Y_i }{e_w(x)} \middle| X_i = x \right]  \label{eq:id2} \\
                         & = E \biggl[ \underbrace{\mu(w,x) + \dfrac{D_i(w) (Y_i - \mu(w,x))}{e_w(x)}}_{%
    \let\scriptstyle\textstyle
    \substack{\equiv \Gamma(w,x)}} \biggm|  X_i = x \biggr]   \label{eq:id3}
\end{align}

Equation \ref{eq:id1} shows that the conditional APO is identified as a conditional expectation of the observed outcome. Equation \ref{eq:id2} shows that it is identified by reweighting the observed outcome with the inverse treatment probability. Finally, Equation \ref{eq:id3} adds the reweighted outcome residual to the conditional outcome representation of Equation \ref{eq:id1}. This seems redundant because we can check that the reweighted residual has expectation zero under unconfoundedness. However, this identification result is doubly robust in the sense that it still holds if we replace either $\mu(w,x)$ or $e_w(x)$ in Equation \ref{eq:id3} by arbitrary functions of $x$.\footnote{Appendix \ref{sec:app-a} reviews identification and identification double robustness of Equation \ref{eq:id3} for completeness.} This doubly robust structure plays a crucial role for the estimation procedures that we discuss in the next section.

From an identification perspective, $\Gamma(w,x)$ defined in Equation \ref{eq:id3} suffices to identify all estimands of interest stated in the previous subsection:
\begin{itemize}
    \item APO: $\gamma_w = E[Y_i(w)] = E[\Gamma(w,X_i)]$
    \item ATE: $\delta_{w,w'} = E[Y_i(w) - Y_i(w')] = E[\Gamma(w,X_i) - \Gamma(w',X_i)]$
    \item ATET: $\theta_{w,w'} = E[Y_i(w) - Y_i(w') \mid W_i = w] = E[\Gamma(w,X_i) - \Gamma(w',X_i) \mid W_i = w] $
    \item CATE: $\tau_{w,w'}(z) = E[Y_i(w) - Y_i(w') \mid Z_i=z] = E[\Gamma(w,X_i) - \Gamma(w',X_i) \mid Z_i = z]$
    \item Policy value: $Q(\pi) = E[Y_i(\pi(Z_i))] = E [ \sum_w \mathds{1}(\pi(Z_i) = w) \Gamma(w,X_i) ]$
    \item Optimal policy: $\pi^* = \argmax_{\pi \in \Pi} Q(\pi) = \argmax_{\pi \in \Pi} E [ \sum_w \mathds{1}(\pi(Z_i) = w) \Gamma(w,X_i) ] $
\end{itemize}

\section{Estimation based on Double Machine Learning}  \label{sec:dml}

\subsection{The doubly robust scores} \label{sec:dr-score}

All Double Machine Learning (DML) based estimators for the estimands of interest build on the doubly robust scores of \citeA{Robins1994,Robins1995AnalysisData} and their Augmented Inverse Probability Weighting (AIPW) estimator in particular. In the following, large Greek letters denote the scores corresponding to the small Greek letters used to define the estimands in Section \ref{sec:def}.

The construction of the doubly robust scores requires the input of so-called nuisance parameters that are usually of secondary interest and considered as tool to eventually obtain the parameters of interest. In our case, the two nuisance parameters are ${\mu(w,x) = E[Y_i \mid W_i = w, X_i = x]}$ and $e_w(x) = P[W_i = w \mid X_i = x]$ for all $w$. $\mu(w,x)$ is the conditional outcome mean for the subgroup observed in program $w$. $e_w(x)$ is the conditional probability to be observed in program $w$, also known as the propensity score. Usually these functions are unknown and need to be estimated. Following \citeA{Chernozhukov2018} they are estimated based on $K$-fold cross-fitting: (i) randomly divide the sample in $K$ folds of similar size, (ii) leave out fold $k$ and estimate models for the nuisance parameters in the remaining $K-1$ folds, (iii) use these models to predict $\hat{\mu}^{-k}(w,x)$ and $\hat{e}^{-k}_w(x)$ in the left out fold $k$, and (iv) repeat (i) to (iii) such that each fold is left out once. This procedure avoids overfitting in the sense that no observation is used to predict its own nuisance parameters. To avoid notational clutter, we ignore the dependence on the specific fold in the following notation and refer to the cross-fitted nuisance parameters as $\hat{\mu}(w,x)$ and $\hat{e}_w(x)$.  

The main building block of the following estimators is the doubly robust score of the \textit{APO}, which replaces the true nuisance parameters in Equation \ref{eq:id3} by their cross-fitted predictions:
\begin{equation} \label{eq:dr}
    \hat{\Gamma}_{i,w} = \hat{\mu}(w,X_i) + \dfrac{D_i(w) (Y_i - \hat{\mu}(w,X_i))}{\hat{e}_w(X_i)}.
\end{equation}




The \textit{ATE} score for the comparison of treatment $w$ and $w'$ is then constructed as the difference of the respective APO scores:
\begin{equation} \label{eq:delta}
\hat{\Delta}_{i,w,w'} = \hat{\Gamma}_{i,w} - \hat{\Gamma}_{i,w'}
\end{equation}

The only estimator we consider that uses the same nuisance parameter but plugs them into a different score is the \textit{ATET} estimator. Although the identification result with the doubly robust APO score in the previous section holds, it is not doubly robust. However, the doubly robust score for the ATET exists and is defined as
\begin{equation} \label{eq:dr-atet}
    \hat{\Theta}_{i,w,w'} = \dfrac{D_i(w) (Y_i - \hat{\mu}(w',X_i))}{\hat{e}_w} - \dfrac{ D_i(w') \hat{e}_{w}(X_i) (Y_i - \hat{\mu}(w',X_i))}{\hat{e}_w \hat{e}_{w'}(X_i)},
\end{equation}

where $\hat{e}_w = N_w / N$ is the unconditional treatment probability with $N_w$ counting the number of individuals observed in program $w$ \cite<see also, e.g.>{Farrell2015}.

\subsection{Average potential outcomes and treatment effects} \label{sec:estimator}

The estimation of the APOs, ATEs and ATETs boils down to taking the means of the previously defined doubly robust scores. For statistical inference, we can rely on standard one-sample \textit{t}-tests. Thus, the score's mean and the variance of this mean are the point and the variance estimate of the respective estimand of interest:
\begin{itemize}
    \item APO: $\hat{\mu}_w = N^{-1} \sum_i \hat{\Gamma}_{i,w}$ and $\hat{\sigma}^2_{\mu_w} = N^{-1} \sum_i (\hat{\Gamma}_{i,w} - \hat{\mu}_w)^2$
    \item ATE:  $\hat{\delta}_{w,w'} = N^{-1} \sum_i \hat{\Delta}_{i,w,w'}$ and $\hat{\sigma}^2_{\delta_{w,w'}} = N^{-1} \sum_i (\hat{\Delta}_{i,w,w'} - \hat{\delta}_{w,w'})^2$
    \item ATET:  $\hat{\theta}_{w,w'} = N^{-1} \sum_i \hat{\Theta}_{i,w,w'}$ and $\hat{\sigma}^2_{\theta_{w,w'}} = N^{-1} \sum_i (\hat{\Theta}_{i,w,w'} - \hat{\theta}_{w,w'})^2$
\end{itemize}

Note that the estimated variances require no adjustment for the fact that we have estimated the nuisance parameters in a first step. The resulting estimators are consistent, asymptotically normal and semiparametrically efficient under the main assumption that the estimators of the cross-fitted nuisance parameters are consistent and converge sufficiently fast \cite{Belloni2014InferenceControls,Farrell2015,Belloni2017,Chernozhukov2018}. In particular, the product of the convergence rates of the outcome and propensity score estimators must be faster than $n^{1/2}$. This allows to apply machine learning to estimate the nuisance parameters.\footnote{Further results, regularity conditions and discussions can be found in section 5.1 of \citeA{Chernozhukov2018}.} Flexible machine learning estimators converge usually slower than the parametric rate $n^{1/2}$ but several are known to be able to achieve $n^{1/4}$ and faster, which would be sufficient if both nuisance parameter estimators achieve it.\footnote{For example, versions of Lasso \cite{Belloni2013LeastModels}, Boosting \cite{Luo2016High-DimensionalConvergence}, Random Forests \cite{Wager2015AdaptiveForests,Syrgkanis2020EstimationDimensions}, Neural Nets \cite{Farrell2021DeepInference}, forward model selection \cite{Kozbur2020AnalysisSelection} or ensembles of those can be shown to achieve the required rates under conditions stated in the original papers.} 

It is well known that estimators using doubly robust scores and parametric models for the nuisance parameters are doubly robust in the sense that they remain consistent if one of the parametric models is misspecified \cite<see, e.g.>{Glynn2009AnEstimator}. The difference of the DML version is that it exploits what \citeA{Smucler2019AContrasts} call 'rate double robustness'. This robustness allows to estimate the parameters of interest at the parametric rate $n^{1/2}$ even if the nuisance parameters are estimated at slower rates using machine learning methods that do not require the specification of an actual parametric model.

The rate double robustness is the consequence of the so-called Neyman orthogonality of the doubly robust score. Neyman orthogonality is at the heart of the general DML framework of \cite{Chernozhukov2018}. Scores with this orthogonality are immune against small errors in the estimation of nuisance parameters and thus allow them to be estimated via machine learning. Appendix \ref{sec:app-neyman} revisits what this means in formal terms.

\subsection{Conditional average treatment effects} \label{sec:est-cate}

\subsubsection{DR-learner} \label{sec:drl}

We can reuse the ATE score of Equation \ref{eq:delta} to estimate conditional effects. The so-called DR-learner was introduced for binary treatments but directly translates also to multiple treatments settings. It exploits that the conditional expectation of the score with known nuisance parameters equals CATE: $\tau_{w,w'}(z) = E[\Delta_{i,w,w'} \mid Z_i = z]$.\footnote{Note that this does not work for the ATET score in Equation \ref{eq:dr-atet} and suitable adaptations are beyond the scope of this paper.} Thus, a natural way to estimate CATEs is to use the score with estimated nuisance parameters, $\hat{\Delta}_{i,w,w'}$, as pseudo-outcome in a general regression framework:
\begin{equation}\label{eq:drl}
    \hat{\tau}^{dr}(Z_i)  = \argmin_{\tau} \sum_{i=1}^N \left(\hat{\Delta}_{i,w,w'} -  \tau(Z_i) \right)^2
\end{equation}

From a conceptual and estimation perspective it is instructive to distinguish two special cases of CATEs at this point  \cite<see also>{Knaus2021}: (i) Group average treatment effects (GATE) provide the average effects for pre-specified, usually low-dimensional, groups.\footnote{Note that the GATE is different to the Sorted Group Average Treatment Effect (GATE\textbf{S}) of \citeA{Chernozhukov2017GenericExperiments}.} This covers standard subgroup analysis comparing, e.g., effects of men and women, or heterogeneity along pre-specified continuous variables like age. (ii) Individualised average treatment effects (IATEs) aim for the most detailed effect heterogeneity considering all confounders as heterogeneity variables, i.e. $Z_i = X_i$ and thus $IATE(x) = \tau_{w,w'}(x) = E[Y_i(w) - Y_i(w') \mid X_i=x]$. 

OLS, series or kernel regressions of the pseudo-outcome on low-dimensional heterogeneity variables estimate \textit{GATEs}. The outputs of such regressions can be interpreted in the standard way. The only difference is that instead of modelling the level of an outcome, they now model the level of a causal effect. Most importantly standard statistical inference applies as is shown for OLS and series regression by \citeA{Semenova2021DebiasedFunctions} as well as for kernel regression by \citeA{fan2020EstimationData} and \citeA{Zimmert2019NonparametricConfounding}. Similar to the discussion in the previous section, the Neyman orthogonality of $\Delta_{i,w,w'}$ allows to ignore that nuisance parameters are estimated with flexible methods potentially converging slower than $n^{1/2}$ when calculating standard errors. The details about the required convergence rates are discussed in the referenced papers.

IATEs may be estimated using the pseudo-outcome in supervised machine learning regressions with the full set of confounders as predictors. As discussed by \citeA{Chernozhukov2017GenericExperiments} statistical inference is not yet well understood for low-dimensional $Z_i$ and even harder for high-dimensional $Z_i$ when machine learning is used to solve Equation \ref{eq:drl}. However, \citeA{Kennedy2020OptimalEffects} shows that the doubly robust structure of the ATE score results in favourable bounds on the mean squared error for the estimated IATEs that would not be attainable by outcome regression or IPW based methods alone.\footnote{The Orthogonal Random Forest of \citeA{Oprescu2019OrthogonalInference} is another estimator that is based on the pseudo-outcome idea and can be asymptotically normal under the assumption of parameteric nuisance parameters. We focus in this paper on the more general DR-learner. See also \citeA{Curth2021NonparametricAlgorithms} for a more nuanced analysis of the DR-learner in comparison to other alternatives.} 

We consider two variants of the DR-learner for \textit{IATEs}. First, we reuse the pseudo-outcome in one supervised machine learning regression to estimate IATEs in-sample. This \textit{full sample procedure} is computationally convenient but prone to overfitting. Thus, the second variant produces out-of-sample IATE predictions for each individual in the sample. Following Algorithm 1 of \citeA{Kennedy2020OptimalEffects}, this requires a \textit{four-fold cross-fitting} scheme that is detailed in Algorithm 2 of Appendix \ref{sec:app-drl}.
The computational downside of this procedure is that we cannot reuse the same nuisance parameter predictions as for the average estimator and need to estimate them for the IATE only. However, the results below suggest that this computational effort is important to avoid severe overfitting.

\subsubsection{Normalised DR-learner}  \label{sec:ndr}

Note that the point estimates of the plain DR-learner can be expressed as $\hat{\tau}^{dr}(z) = \sum_{i=1}^N \alpha_i \hat{\Delta}_{i,w,w'}$ if the weight $\alpha_i$ that each observation receives can be calculated. For example, the ATE estimator as special case of the DR-learner with $Z_i$ being a constant uses $\alpha_i = 1/N$, the least squares regression uses $\alpha_i = z (\bm{Z'}\bm{Z})^{-1} \bm{Z'}$ with $\bm{Z}$ being the stacked covariate matrix, and the kernel regression uses $\alpha_i = \frac{\mathcal{K}_h \left(Z_i - z \right) }{ \sum_{i=1}^N \mathcal{K}_h \left( Z_i - z \right) }$ with $\mathcal{K}_h$ representing a proper kernel function. The class of estimators with a known weighted representation is called linear smoothers \cite<see e.g.>{Buja1989LinearModels}. 
Popular machine learners like tree-based methods (regression trees, Random Forests or boosted trees), Ridge or any method that runs OLS after variable selection like Post-Lasso \cite{Belloni2013LeastModels} have this structure.\footnote{In practice most of these methods are applied with data-driven selection of tuning parameters, which makes them strictly speaking non-linear smoothers \cite{Buja1989LinearModels}. However, this does not affect our results.} Also for these methods we know the weight $\alpha_i(x)$ that each observation receives in predicting the (pseudo-)outcome at $x$. These weights usually sum up to one, i.e. $\sum_{i=1}^N \alpha_i(x) = 1$. Using such outcome weighting predictors in the final step allows to express the DR-learner estimated IATE as
\begin{align}
    \hat{\tau}_{w,w'}^{drl}(x) & =\sum_{i=1}^N \alpha_i(x) \hat{\Delta}_{i,w,w'} \nonumber\\
                         & = \sum_{i=1}^N \alpha_i(x) [\hat{\mu}(w,X_i) - \hat{\mu}(w',X_i)] \nonumber\\
                         &\quad+ \sum_{i=1}^N \underbrace{\dfrac{\alpha_i(x) D_i(w) }{\hat{e}_w(X_i)}}_{\lambda^w_i(x)}  \Tilde{Y_i}(w,X_i)  - \sum_{i=1}^N \underbrace{\dfrac{\alpha_i(x) D_i(w') }{\hat{e}_{w'}(X_i)}}_{\lambda^{w'}_i(x)} \Tilde{Y_i}(w',X_i), \label{eq:dr-lambda}
\end{align}

where $\Tilde{Y_i}(w,X_i) = Y_i - \hat{\mu}(w,X_i)$ denotes the outcome residual. 

The DR-learner shares the problem of all estimators that involve reweighting by the inverse of the propensity score. In finite samples, ${\lambda^w_i}(x)$ and ${\lambda^{w'}_i}(x)$ usually do not sum to one, i.e. $\sum_{i=1}^N {\lambda^w_i}(x) \neq 1$ and $\sum_{i=1}^N {\lambda^{w'}_i}(x) \neq 1$. This is especially problematic if it sums to something much greater than one. In this case the weighted residuals receive much more weight than the outcome regressions. This might result in implausibly large effect estimates that even could fall outside of the possible bounds of a given outcome variable \cite{Kang2007,Robins2007Comment:Variable}.\footnote{For bounded outcomes, the effects must lie in the interval $[Y_{min} - Y_{max},Y_{max} - Y_{min}]$, with $Y_{min}$ and $Y_{max}$ denoting the minimum and maximum values of the outcome, respectively.} 

For average effects the \citeA{Hajek1971CommentOne} normalisation is recommended to stabilise estimators using inverse probability weights \cite<e.g.>{Imbens2004NonparametricReview,Lunceford2004StratificationStudy,Robins2007Comment:Variable,busso2014new}. However, Equation \ref{eq:dr-lambda} highlights that the inverse probability weights become $x$-specific and such a one time normalisation that targets the average effect does not solve the problem for the individualised effect. This can be problematic as finite sample imbalances are more likely to occur on the individualised level. Thus, we propose the normalised DR-learner (NDR-learner) as a stabilised complement to the DR-learner.

The \textit{NDR-learner} normalises the weighted residuals by the sum of weights:
\begin{align}
    \hat{\tau}_{w,w'}^{ndrl}(x) = & \sum_{i=1}^N \alpha_i(x) [\hat{\mu}(w,X_i) - \hat{\mu}(w',X_i)] \nonumber\\
    &+ \left(\sum_{i=1}^N \lambda^w_i(x) \right)^{-1} \sum_{i=1}^N \lambda^w_i(x)  \Tilde{Y_i}(w,X_i)  - \left(\sum_{i=1}^N \lambda^{w'}_i(x) \right)^{-1} \sum_{i=1}^N \lambda^{w'}_i(x) \Tilde{Y_i}(w',X_i) \label{eq:ndr}
\end{align}

This ensures that the weights of the residuals sum up to one under the condition that weights $\alpha_i(x)$ are non-negative. Thus, methods like Ridge or Post-Lasso with potentially negative weights might not be applicable. 

The NDR-learner is more demanding from a computational point of view because it requires to calculate the weights $\alpha_i(x)$ and the normalisation for each $x$ of interest (Algorithm 2 in Appendix \ref{sec:app-drl} provides the details of the implementation). However, the application below shows that the normalisation deals well with the cases where outcome residuals receive high weights leading to implausibly large effect estimates. Thus, the NDR-learner is an interesting alternative to the DR-learner if effect sizes become suspicious.

\subsection{Optimal treatment assignment} \label{sec:opt-pol}

The APO score of Section \ref{sec:dr-score} can also be reused to estimate optimal treatment assignment. To this end, note that the value function of any policy rule $\pi(Z_i)$ can be estimated as
\[ \hat{Q}(\pi) = N^{-1} \sum_{i=1}^N \sum_{w=0}^{T} \mathds{1}(\pi(Z_i) = w) \hat{\Gamma}_{i,w}. \]

This means each individual contributes the score of the treatment that she is assigned to under this policy rule. However, we are not necessarily interested in the value function of some policy rule, but want to estimate the optimal policy rule that maximises this value function, $\hat{\pi}^* = \argmax_{\pi \in \Pi} \hat{Q}(\pi)$. This requires to search over all candidate policy rules to find the optimum as there exists no closed form solution. 

\textit{Example:} Consider the case where $Z_i$ is a binary covariate and $W_i$ is a binary treatment. We have four different policy rules: treat nobody ($\pi^1$), treat only those with  $Z_i=1$ ($\pi^2$), treat only those with $Z_i=0$ ($\pi^3$), or treat everybody ($\pi^4$). We illustrate this using two representative observations, $i=1$ with $Z_1=0$, and $i=2$ with $Z_2=1$ in Table \ref{tab:example}. The columns three to six show the assignments under the four potential assignment rules. For example, the first observation receives no treatment under policy rules $\pi^1$ and $\pi^2$, but is treated under policy rules $\pi^3$ and $\pi^4$. To find the optimal rule, we compare the means of the APO scores in the last four columns and pick the policy rule that corresponds to the largest mean. The number of policy values to compare increases dramatically in settings with multiple treatments and $Z_i$ being a vector of potentially non-binary variables. 

\begin{table}[h]
\begin{center}
\caption{Example of DML based optimal treatment assignment}
\begin{tabular}{lccccccccc}
 $i$ & $Z_i$ & $\pi^1$ & $\pi^2$ & $\pi^3$ & $\pi^4$ & $\hat{Q}(\pi^1)$ & $\hat{Q}(\pi^2)$ & $\hat{Q}(\pi^3)$ & $\hat{Q}(\pi^4)$ \\  
 \midrule
 1 & 0 & 0 & 0 & 1 & 1 & $\hat{\Gamma}_{1,0}$ & $\hat{\Gamma}_{1,0}$ & $\hat{\Gamma}_{1,1}$ & $\hat{\Gamma}_{1,1}$ \\  
 2 & 1 & 0 & 1 & 0 & 1 & $\hat{\Gamma}_{2,0}$ & $\hat{\Gamma}_{2,1}$ & $\hat{\Gamma}_{2,0}$ & $\hat{\Gamma}_{2,1}$ \\  
 \vdots & \vdots & \vdots & \vdots & \vdots & \vdots & \vdots & \vdots & \vdots & \vdots \\  
\end{tabular}
\label{tab:example}
\end{center}
\end{table}

We expect that the estimated policy in finite samples and with estimated nuisance parameters does not coincide with the true optimal policy rule. This is conceptualised as the 'regret' defined as the difference between the true and the estimated optimal value function, $R(\hat{\pi}^*) = Q(\pi^*) - Q(\hat{\pi}^*)$. 

\citeA{Zhou2018OfflineOptimization} show that the DML based procedure minimises the maximum regret asymptotically under two main conditions: First, the same convergence conditions for the nuisance parameters that are required for ATE estimation (the product of the nuisance parameter convergence rates achieves $n^{1/2}$). Second, the set of candidate policy rules $\Pi$ is not too complex. In particular, \citeA{Zhou2018OfflineOptimization} show that decision trees with fixed depth are a suitable class of policy rules. Again the double robustness of the used scores results in statistical guarantees that are not achievable for methods based on outcome regressions or IPW alone.

\section{Application: Swiss Active Labour Market Policy}  \label{sec:appl}

We use a standard observational dataset of Swiss Active Labour Market Policy (ALMP) that is already basis of previous studies \cite{Huber2017,Lechner2018,Knaus2020HeterogeneousApproach} to estimate the effect of different programs on employment.\footnote{\citeA{gerfin2002microeconometric}, \citeA{Lalive2008TheUnemployment} and \citeA{Knaus2020HeterogeneousApproach} among others provide a more detailed description of the surrounding institutional setting.} In particular, we start with the sample of 100,120 unemployed individuals of \citeA{Huber2017} that consists of 24 to 55 year old individuals registered unemployed in 2003.\footnote{The dataset is available as restricted use file via the platform \href{https://forsbase.unil.ch/project/study-public-overview/17035/1/}{FORSbase} \cite{Lechner2020SwissDataset}.} We consider non-participants and participants of four different \textit{program types}: job search, vocational training, computer programs and language courses.\footnote{The dataset contains also participants of an employment program and personality training. However, we leave them out to keep the number of obtained results manageable.} As the assignment policies differ substantially across the three language regions, we focus only on individuals living in the German speaking part and remove those in the French and Italian speaking part to avoid common support problems.

We evaluate the first program participation within the first six months after the begin of the unemployment spell. One problem of this definition is that non-participants comprise people that quickly come back into employment before they would be assigned to a training program. This could result in an overly optimistic evaluation of non-participation. We follow \citeA{lechner1999earnings} and \citeA{lechner2007value} and assign pseudo program starting points to the non-participants and keep only those who are still unemployed at this point.\footnote{The assignment of the pseudo starting point is based on estimated probabilities to start a program at a specific time. The probability depends also on covariates and is estimated using the same random forest specification that is discussed later in Section \ref{sec:imp}.} This results in a final sample size of 62,497 observations.

\begin{table}[t]
  \centering  \footnotesize
\begin{threeparttable}[t]
  \caption{Descriptive statistics of selected variables by program type} 
  \label{tab:desc} 
\begin{tabular}{lccccccc} 
\toprule
\\[-1.8ex]& \multicolumn{1}{c}{No program} & \multicolumn{1}{c}{Job search} & \multicolumn{1}{c}{Vocational} & \multicolumn{1}{c}{Computer} & \multicolumn{1}{c}{Language}\\ 
\\[-1.8ex] & \multicolumn{1}{c}{(1)} & \multicolumn{1}{c}{(2)} & \multicolumn{1}{c}{(3)} & \multicolumn{1}{c}{(4)} & \multicolumn{1}{c}{(5)}\\ 
\midrule
No. of observations & 47,620 & 11,610 & 858 & 905 & 1504 \\ 
 [0.4em]
  Outcome: months employed of 31 & 14.7 & 14.4 & 18.4 & 19.2 & 13.5 \\ 
   [0.4em]
  Female (binary) & 0.44 & 0.44 & 0.33 & 0.60 & 0.55 \\ 
   [0.4em]
  Age & 36.6 & 37.3 & 37.5 & 39.1 & 35.3 \\ 
   [0.4em]
  Foreigner (binary) & 0.36 & 0.33 & 0.30 & 0.21 & 0.66 \\
   [0.4em]
  Employability & 1.93 & 1.98 & 1.93 & 1.97 & 1.85 \\ 
   [0.4em]
  Past income in CHF 10,000 & 4.25 & 4.67 & 4.87 & 4.32 & 3.73 \\
    \bottomrule
    \end{tabular}%
    \begin{tablenotes} \item \textit{Note:} Employability is an ordered variable with one
indicating low employability, two medium employability and three high employability. The exchange rate USD/CHF was roughly 1.3 at that time. The full set of variables is reported in Table \ref{tab:desc-app}. \end{tablenotes}  
\end{threeparttable}
\end{table}

The \textit{outcome} of interest is the cumulated number of months in employment in the 31 months after program start, which is the maximum available time span in the dataset. Row one of Table \ref{tab:desc} provides the number of observations in each group. Roughly 75\% participate in no program. By far the largest program is the job search program, which is also called basic program. The more specific programs are much smaller with roughly 1000 observations each. Row two shows that the average outcomes substantially differ by different groups. However, it is not clear whether this is only due to selection effects because the observable characteristics are not comparable across groups, as the remaining rows show. Especially the share of females, the share of foreigners and past income differ quite substantially across programs. The \textit{confounders} comprise 45 variables and are reported in Table \ref{tab:desc-app} of Appendix \ref{sec:app-res}. They consist of socio-economic characteristics of the unemployed individuals, caseworker characteristics, information about the assignment process, information about the previous job, and regional economic indicators.

\section{Implementation} \label{sec:imp}

The nuisance parameters are estimated via Random Forest \cite{Breiman2001} using the implementation with honest splitting in the \texttt{grf} R-package \cite{Athey2017a} and 5-fold cross-fitting. The tuning parameters in each regression are selected by out-of-bag validation. All regressions apply the full set of confounders. We run the outcome regressions for each treatment group separately to obtain $\hat{\mu}(w,x)$. Also the propensity scores are separately estimated for each treatment using a treatment indicator as outcome in the random forest. The propensity scores are then normalised to sum to one within an individual.

We estimate CATEs at different granularity. First, we investigate GATEs for subgroups by gender, foreigners and three categories of employability. These are regularly used in the program evaluation literature and usually investigated by re-estimating everything in the subgroups. However, it can be performed at very low computational costs after DML for average effects using only a standard OLS regression with the pseudo-outcome as described in Section \ref{sec:drl} and using dummy variables for all groups but the reference group as covariates. Second, we estimate kernel regression and spline regression GATEs for the continuous variables age and past income based on the R-packages \texttt{np} \cite{Hayfield2008NonparametricPackage} and \texttt{crs} \cite{Racine2021Crs:Splines}, respectively. The kernel regressions apply a second-order Gaussian kernel function and use 0.9 of the cross-validated bandwidth for undersmoothing as suggested by \citeA{Zimmert2019NonparametricConfounding}. The spline regressions use B-splines with cross-validated degree and number of knots. Third, we specify an OLS regression in which all the five previously used variables enter linearly. Finally, we go beyond the handpicked variables and estimate the IATEs using all 45 confounders in the DR-learner and the NDR-learner. Both are implemented with the honest Random Forest because the \texttt{grf} package allows to extract the prediction weights $\alpha_i(x)$ required for the NDR-learner. We apply both variants described in Section \ref{sec:drl}. Once we estimate the IATE for each observation using DR- and NDR-learner in the full sample and once we predict them out-of-sample. For the latter, Appendix \ref{sec:app-drl} provides a detailed description of the underlying DR- and NDR-learner algorithms.

\begin{table}[t]
  \centering \footnotesize
  \begin{threeparttable}
  \caption{Steps of implementation}     \label{tab:cookbook}
    \begin{tabularx}{\textwidth}{llXl}
        \toprule
         Step & Input & Operation & Output \\
         \midrule
		1. & $W_i$, $X_i$ & Predict treatment probabilities & $\hat{e}_w(x)$ \\
		[0.2em]
		2. & $Y_i$, $W_i$, $X_i$ & Predict treatment specific outcomes & $\hat{\mu}(w,x)$ \\
		[0.2em]
		3. & $Y_i$, $W_i$, $\hat{e}_w(x)$, $\hat{\mu}(w,x)$ & Plug into Equation \ref{eq:dr} & $\hat{\Gamma}_{i,w}$ \\
		[0.2em]
		4. & $\hat{\Gamma}_{i,w}$ & Mean, one-sample \textit{t}-test & APOs \\
		[0.2em]
		5. & $\hat{\Gamma}_{i,w}$ & Take difference & $\hat{\Delta}_{i,w,w'}$ \\
		[0.2em]
		6. & $\hat{\Delta}_{i,w,w'}$ & Mean, one-sample \textit{t}-test & ATEs \\
		[0.2em]
		7. & $\hat{\Delta}_{i,w,w'}$, $Z_i$ & OLS/Kernel/Series regression & GATEs \\
		[0.2em]
		8. & $\hat{\Delta}_{i,w,w'}$, $X_i$ & Supervised Machine Learning & IATEs \\
		[0.2em]
		9. & $\hat{\Gamma}_{i,w}$, $Z_i$ & Optimal decision tree & Optimal treament rule \\
    	\bottomrule
    \end{tabularx}%
      \end{threeparttable}
\end{table}%

The optimal treatment assignment rule is estimated as decision trees of depth one, two and three. We follow Algorithm 2 for exact tree-search of \citeA{Zhou2018OfflineOptimization} that is implemented in the \texttt{policytree} R-package \cite{Sverdrup2020Policytree:Trees}. We estimate the trees first with the five handpicked variables. However, these variables include gender and foreigner status that might be too sensitive to include in practice. Thus, we investigate another set of 16 variables that includes only the objective measures of education and labour market history of the unemployed persons that would be available for recommendations from the administrative records. 

Table \ref{tab:cookbook} summarises all required implementation steps. It highlights that a comprehensive DML based program evaluation can be run with few lines of code in any statistical software program that is capable of the operations in the third column. Thus, researchers can build their customised analyses in a modular fashion based on established code. Alternatively, the R-package \href{https://github.com/MCKnaus/causalDML}{\texttt{causalDML}} already implements the required steps as showcased in the \href{https://mcknaus.github.io/assets/code/Notebook_DML_ALMP_MCK2020.html}{replication notebook} accompanying this paper. Most importantly the package provides a fast implementation of the individualised normalisation required for the NDR-learner in C++.

\section{Results} \label{sec:res}

\subsection{Average effects}

\begin{figure}
    \centering
    \caption{Average treatment effects of participation vs. non-participation}
    \includegraphics[width=0.6\textwidth]{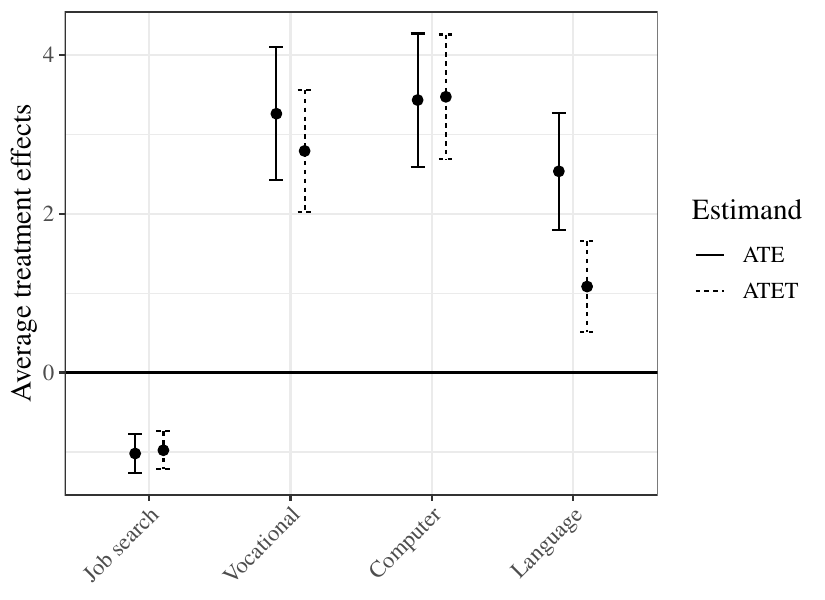}
    \subcaption*{\textit{Note:} The figure shows the point estimates of the average treatment effects of participating in the program labeled on the x-axis vs. non-participation and their 95\% confidence intervals. Numeric results in Panels B and C of Table \ref{tab:avg-eff} of  Appendix \ref{sec:app-res}.}
    \label{fig:ate}
\end{figure}

We focus here on the effect estimates and discuss the nuisance parameters in Appendix \ref{sec:app-res-np}. Throughout this section, we compare the four programs to non-participation.\footnote{The underlying APOs are shown in Figure \ref{tab:avg-eff} of Appendix \ref{sec:app-res}.} Recall that the outcome of interest is the cumulated number of months employed in the 31 months after program start. Figure \ref{fig:ate} depicts ATE and ATET estimates and shows substantial differences in the effectiveness of programs. The job search program decreases the months in employment on average by about one month. In contrast, other programs that teach hard skills show substantial improvements with roughly three additional months in employment on average.\footnote{For a better understanding of the underlying dynamics, Figure \ref{fig:ate31} of Appendix \ref{sec:app-res} reports and discusses the effects of program participation on the employment probabilities over time.}

Comparing ATE and ATET shows no big differences for most programs. This suggests that there is either no effect heterogeneity correlated with observables or that the assignment does not take advantage of this heterogeneity. We would expect to see ATETs being higher than ATEs if program assignment is well targeted. However, we find only evidence for the opposite as the actual participants of a language course show a 1.5 months lower treatment effect compared to the population. This difference suggests that there is substantial effect heterogeneity to uncover and the potential to improve treatment assignment.  

\subsection{Heterogeneous effects}

\subsubsection{Subgroup GATEs}

This subsection studies effect heterogeneity at different granularity. We start by estimating group average treatment effects (GATEs) for discrete subgroups. Panel A of Table \ref{tab:cate-sg} shows the result of an OLS regression with a female dummy as covariate, $\hat{\Delta}_{i,w,w'} = \beta_0 + \beta_1 female_i + error_i$. The constant ($\beta_0$) provides the GATE for the reference group men and the female coefficient ($\beta_1$) describes how much the GATE differs for women. The results show substantial gender differences in the effectiveness of programs. Women significantly suffer less or profit more from job search and computer program participation. This gender gap in the effectiveness of ALMPs is also well-documented in the literature \cite{crepon2016active,Card2018WhatEvaluations}. In contrast to this, we find that women profit on average significantly less from language courses than men.

\begin{table}[t]
  \centering  \footnotesize

\begin{threeparttable}
  \caption{Group average treatment effects} 
  \label{tab:cate-sg} 
\begin{tabular}{lcccc} 
\toprule
\\[-1.8ex] & \multicolumn{1}{c}{Job search} & \multicolumn{1}{c}{Vocational} & \multicolumn{1}{c}{Computer} & \multicolumn{1}{c}{Language}\\ 
\\[-1.8ex] & \multicolumn{1}{c}{(1)} & \multicolumn{1}{c}{(2)} & \multicolumn{1}{c}{(3)} & \multicolumn{1}{c}{(4)}\\ 
\midrule
\textit{Panel A:} & & & &  \\
 Constant & -1.29$^{***}$ &  3.82$^{***}$ & 2.33$^{***}$ &  3.40$^{***}$ \\ 
    & (0.17) & (0.55) & (0.60) & (0.46) \\
  [0.4em]
  Female &  0.60$^{**}$ & -1.27 & 2.49$^{***}$ & -1.97$^{**}$ \\ 
    & (0.25) & (0.87) & (0.85) & (0.77) \\ 

\textit{Panel B:} & & & &  \\
Constant & -1.27$^{***}$  & 2.48$^{***}$  &  3.75$^{***}$  &  3.56$^{***}$  \\ 
   & (0.16) & (0.53) & (0.50) & (0.52) \\ 
   [0.4em]
  Foreigner &  0.70$^{***}$  & 2.17$^{**}$  & -0.88 & -2.84$^{***}$  \\ 
   & (0.26) & (0.90) & (0.93) & (0.72) \\

\textit{Panel C:} & & & &  \\
Constant & -0.18 &  5.48$^{***}$ &  5.76$^{***}$ &  2.61$^{***}$ \\ 
  & (0.33) & (1.04) & (1.10) & (0.85) \\ 
  [0.4em]
  Medium employability & -0.93$^{***}$ & -2.41$^{**}$ & -2.67$^{**}$ & -0.18 \\ 
   & (0.36) & (1.16) & (1.20) & (0.96) \\ 
  [0.4em]
  High employability & -1.50$^{***}$ & -4.42$^{***}$ & -3.59$^{**}$ &  0.59 \\ 
   & (0.50) & (1.52) & (1.69) & (1.48) \\ 
   [0.4em]
  F-statistic & 5.04$^{***}$ & 4.31$^{**}$ & 2.96$^{*}$ & 0.18 \\ 

    \bottomrule
    \end{tabular}%
         \begin{tablenotes} \item \textit{Note:} This table shows OLS coefficients and their heteroscedasticity robust standard errors (in parentheses) of regressions run with the pseudo-outcome defined as described in Section \ref{sec:est-cate}. $^{*}$p$<$0.1; $^{**}$p$<$0.05; $^{***}$p$<$0.01 \end{tablenotes}  
\end{threeparttable}
\end{table}

Panel B replaces the female dummy in the regression by a foreigner dummy. Strikingly, Swiss citizens as reference group show a big positive effect for participating in language courses but the effect disappears for foreigners. After adding the coefficient for foreigners to the constant, the foreigners' GATE is only 0.72 ($3.56 - 2.84$, standard error: 0.69). A crucial information to better understand this finding would be to know which languages they learn.\footnote{See \citeA{Heiler2021EffectTreatments} for a discussion about how treatment heterogeneity could drive effect heterogeneity.} However, this information is unfortunately not available in the dataset.

Panel C shows the results of a similar regression but now with two dummies indicating medium and high employability such that low employability becomes the reference group. The F-statistic in the last line tests the joint significance of the two dummies. It is statistically significant at least at the 10\%-level for the programs in the first three columns. They all show a common gradient that individuals with low employability benefit substantially more or at least suffer less from program participation.

\subsubsection{Nonparametric GATEs for continuous heterogeneity variables}

\begin{figure}[h!] 
\caption{Effect heterogeneity regarding past income} \label{fig:cate-vverds}
\begin{subfigure}{0.399\textwidth}
\includegraphics[width=\linewidth]{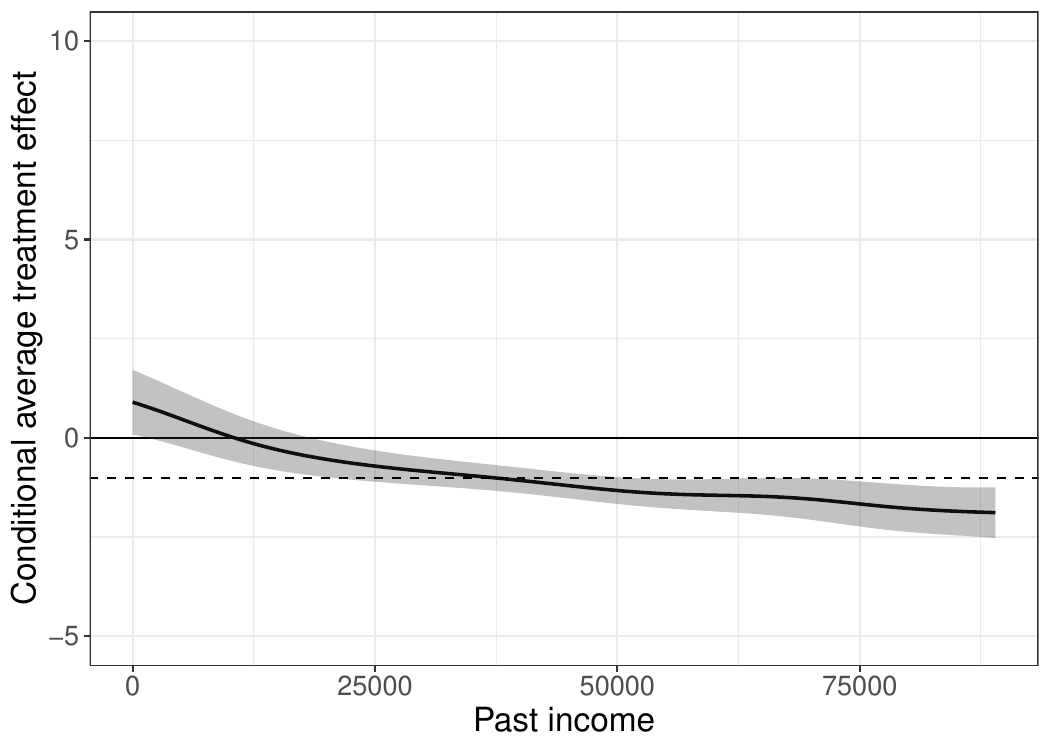}
\caption{Job search (Kernel)} 
\end{subfigure}\hspace*{\fill}
\begin{subfigure}{0.399\textwidth}
\includegraphics[width=\linewidth]{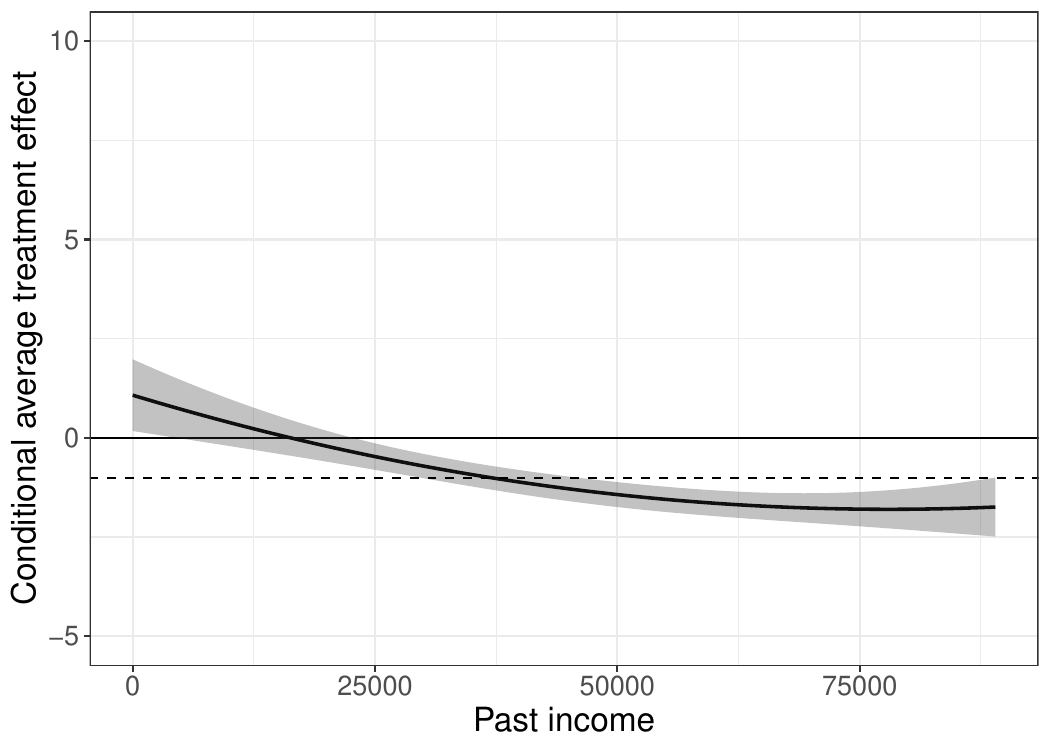}
\caption{Job search (Spline)} \label{fig:cate-vv-f}

\end{subfigure}

\medskip
\begin{subfigure}{0.399\textwidth}
\includegraphics[width=\linewidth]{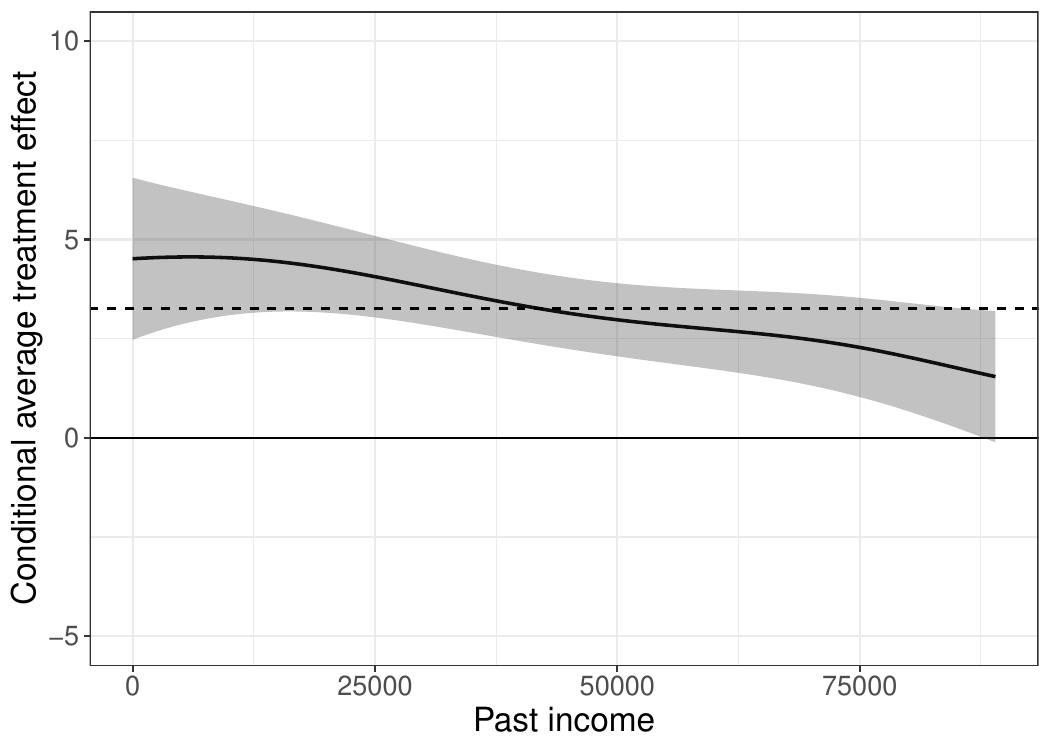}
\caption{Vocational (Kernel)}

\end{subfigure}\hspace*{\fill}
\begin{subfigure}{0.399\textwidth}
\includegraphics[width=\linewidth]{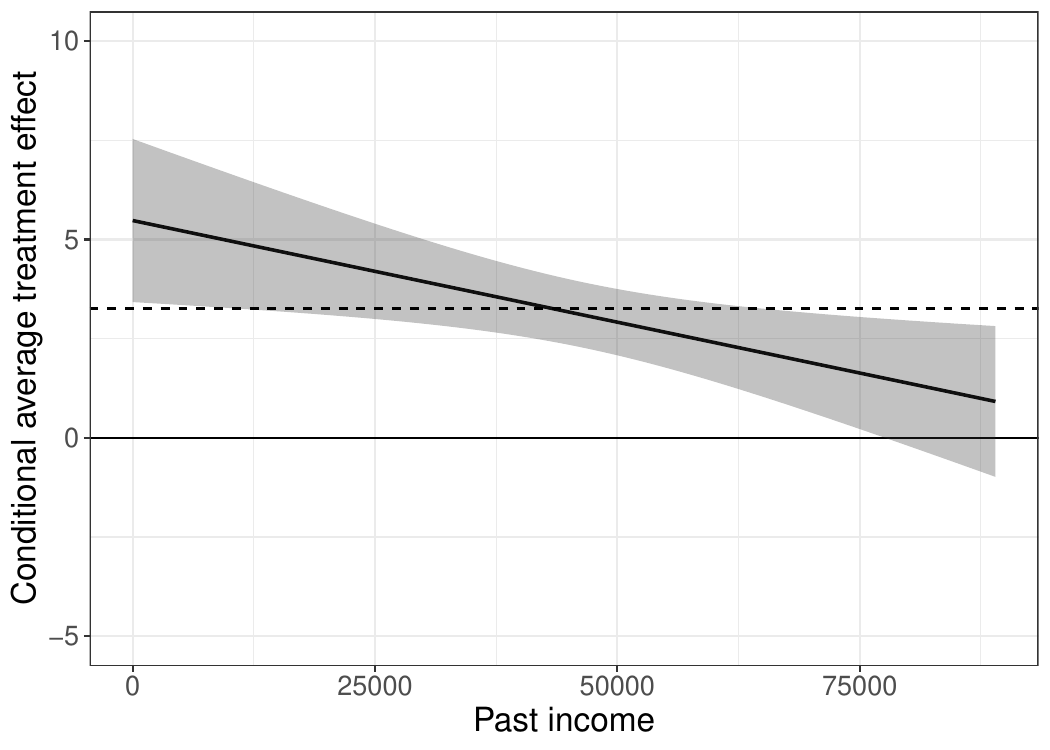}
\caption{Vocational (Spline)} \label{fig:cate-vv-g}

\end{subfigure}

\medskip
\begin{subfigure}{0.399\textwidth}
\includegraphics[width=\linewidth]{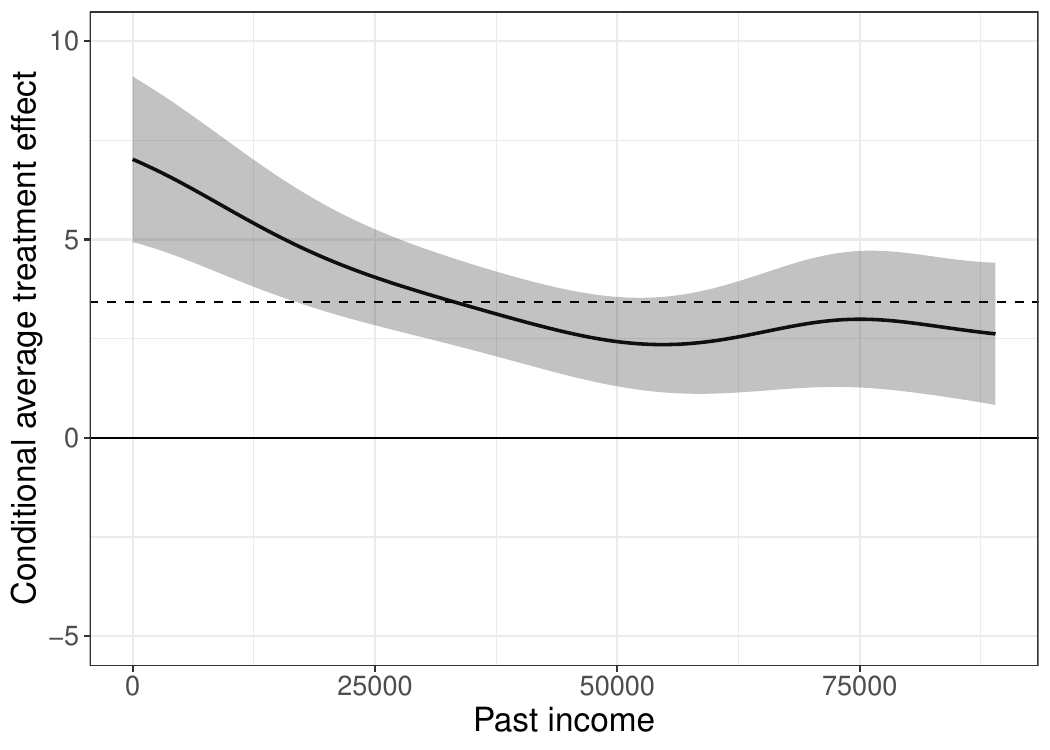}
\caption{Computer (Kernel)} \label{fig:cate-vv-c}
\end{subfigure}\hspace*{\fill}
\begin{subfigure}{0.399\textwidth}
\includegraphics[width=\linewidth]{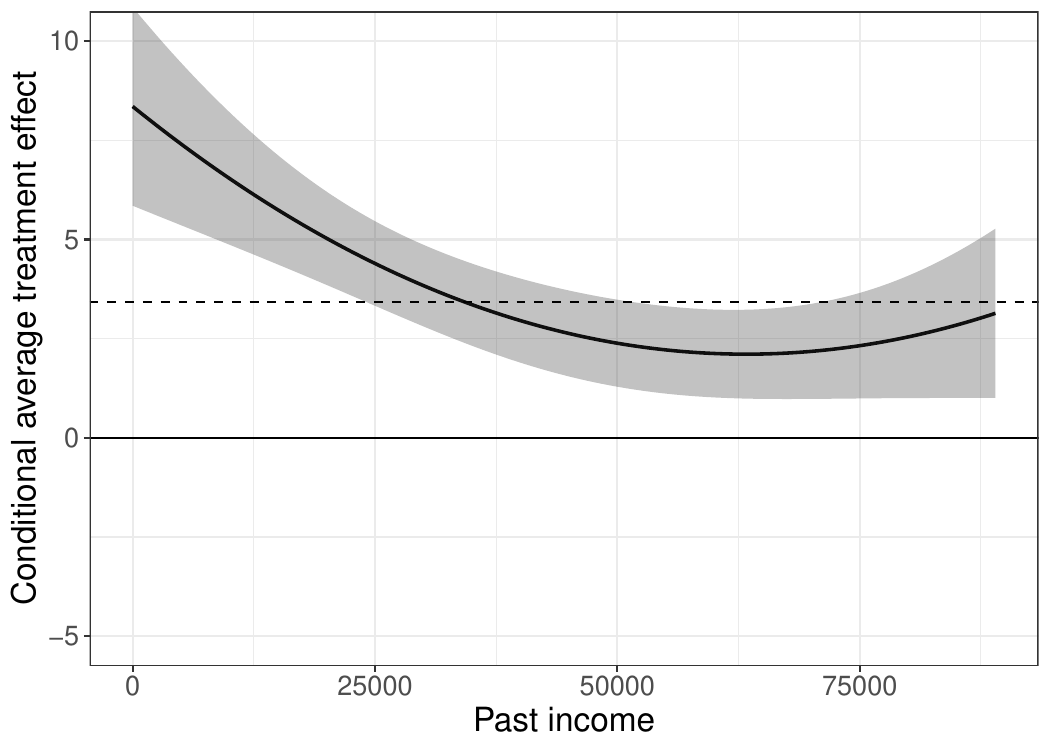}
\caption{Computer (Spline)} \label{fig:cate-vv-h}
\end{subfigure}

\medskip
\begin{subfigure}{0.399\textwidth}
\includegraphics[width=\linewidth]{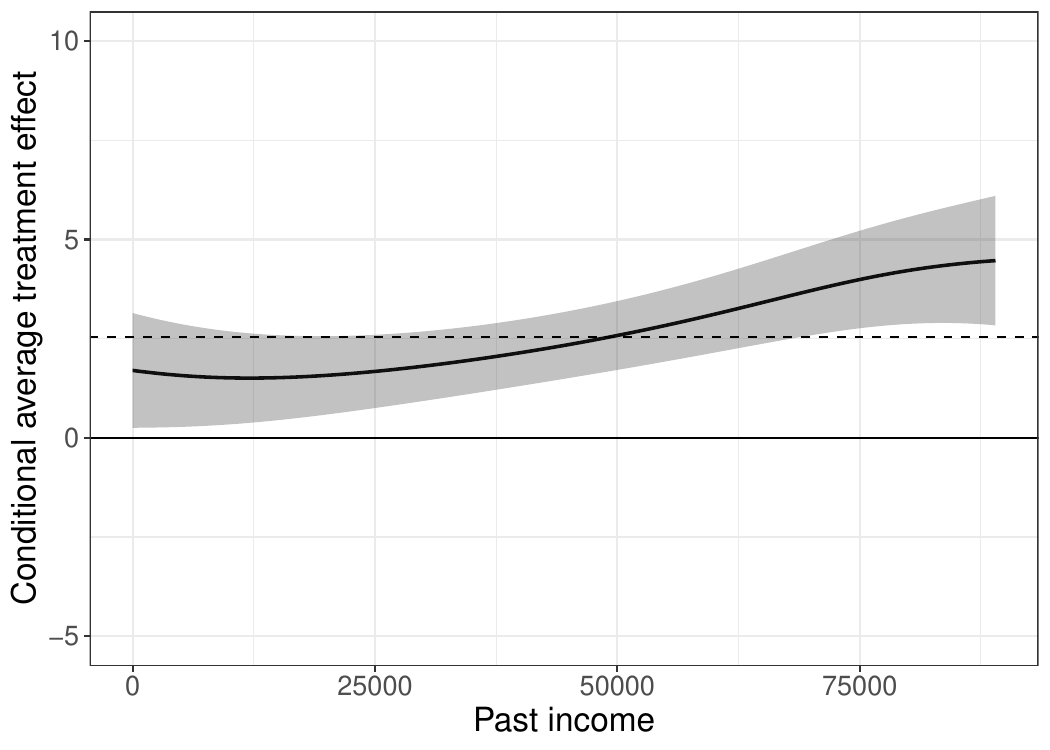}
\caption{Language (Kernel)} \label{fig:cate-vv-e}
\end{subfigure}\hspace*{\fill}
\begin{subfigure}{0.399\textwidth}
\includegraphics[width=\linewidth]{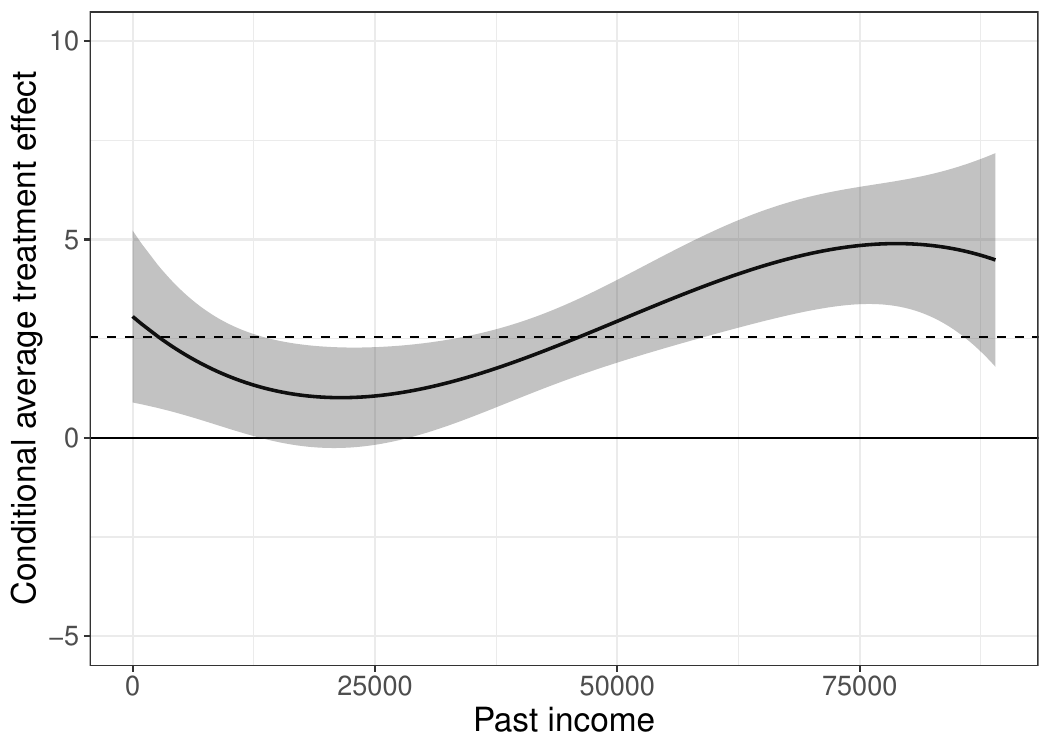}
\caption{Language (Spline)} \label{fig:cate-vv-i}
\end{subfigure}

\subcaption*{\textit{Note:} Dotted line indicates point estimate of the respective average treatment effect. Grey area shows 95\%-confidence interval.}
\end{figure}

While subgroup analyses are standard in program evaluations, the estimation of nonparametric GATEs using kernel or series regression is rarely pursued. We estimate such GATEs along two continuous variables past income and age. We find no notable heterogeneity for the latter.\footnote{Figure \ref{fig:cate-age} shows the according results.} However, effect sizes are clearly associated with past income. Figure \ref{fig:cate-vverds} shows on the left the results of the kernel regression and on the right of the series regression. Both estimators agree by and large. They document that effects decrease with higher past income for all but for language programs. The latter have only a small positive effect for individuals with low past income but it increases with higher income. One potential explanation for these findings is that the value of language skills is larger for high-skilled workers in multilingual countries like Switzerland because they reduce information costs across language borders \cite<see, e.g.>{Isphording2014LanguageSuccess}. 

\begin{table}[!t]
  \centering  \footnotesize
\begin{threeparttable}
  \caption{Best linear prediction of CATEs} 
  \label{tab:cate-ols} 
\begin{tabular}{lcccc} 
\toprule
\\[-1.8ex] & \multicolumn{1}{c}{Job search} & \multicolumn{1}{c}{Vocational} & \multicolumn{1}{c}{Computer} & \multicolumn{1}{c}{Language}\\ 
\\[-1.8ex] & \multicolumn{1}{c}{(1)} & \multicolumn{1}{c}{(2)} & \multicolumn{1}{c}{(3)} & \multicolumn{1}{c}{(4)} \\ 
\midrule
Constant & -0.48 &  4.46$^{*}$ &  4.93$^{**}$ &  6.07$^{***}$ \\ 
   & (0.70) & (2.40) & (2.37) & (2.09) \\ 
   [0.4em]
  Female &  0.25 & -2.12$^{**}$ &  1.89$^{**}$ & -1.61$^{**}$ \\ 
  & (0.27) & (0.92) & (0.90) & (0.80) \\ 
   [0.4em]
  Age &  0.02 &  0.09$^{*}$ &  0.05 & -0.07 \\ 
   & (0.01) & (0.05) & (0.05) & (0.05) \\ 
   [0.4em]
  Foreigner &  0.50$^{*}$ &  1.60$^{*}$ & -1.20 & -2.77$^{***}$ \\ 
  & (0.27) & (0.90) & (0.96) & (0.74) \\ 
   [0.4em]
  Medium employability & -0.65$^{*}$ & -1.64 & -2.36$^{*}$ & -0.78 \\ 
   & (0.37) & (1.18) & (1.22) & (0.99) \\ 
   [0.4em]
  High employability & -1.03$^{**}$ & -3.13$^{**}$ & -3.15$^{*}$ & -0.47 \\
   & (0.51) & (1.55) & (1.71) & (1.53) \\ 
   [0.4em]
  Past income in CHF 10,000  & -0.26$^{***}$ & -0.62$^{***}$ & -0.39$^{**}$ &  0.31$^{*}$ \\ 
   & (0.06) & (0.23) & (0.19) & (0.18) \\
   [0.6em]
  F-statistic & 6.95$^{***}$ & 4.12$^{***}$ & 3.35$^{***}$ & 5.22$^{***}$ \\ 
    \bottomrule
    \end{tabular}%
         \begin{tablenotes} \item \textit{Note:} This table shows OLS coefficients and their heteroscedasticity robust standard errors (in parentheses) of regressions run with the pseudo-outcome. $^{*}$p$<$0.1; $^{**}$p$<$0.05; $^{***}$p$<$0.01 \end{tablenotes}  
\end{threeparttable}
\end{table}

\subsubsection{Best linear prediction of GATEs}

The GATEs considered so far were nonparametric but only univariate. Now we model the GATE by specifying a multivariate OLS regression with the previously used covariates entering linearly. It is most likely misspecified and thus estimates the best linear predictor (BLP) of GATEs with respect to these variables. However, it provides a compact and accessible summary of the effect heterogeneities. Additionally, it holds the other included variables constant. Consider for example the coefficients for being female in Table \ref{tab:cate-ols}. Compared to Table \ref{tab:cate-sg}, the coefficients in the first three columns are smaller and the one for language courses is larger (for example for job search it is 0.25 instead of 0.60). The reason is that it represents a partial effect that holds other variables like past income fixed. The subgroup female coefficient in Table \ref{tab:cate-sg} partly picks up that women have lower past income and that lower income is associated with higher treatment effects for all but language courses. This example illustrates that the same strategies that are usually applied to interpret an outcome OLS regression can now be used to interpret the effect OLS regression.

\subsubsection{Individualised average treatment effects}

\begin{figure}[t]
    \centering
    \caption{Boxplot of out-of-sample predicted IATEs by DR- and NDR-learner}
    \includegraphics[width=0.7\textwidth]{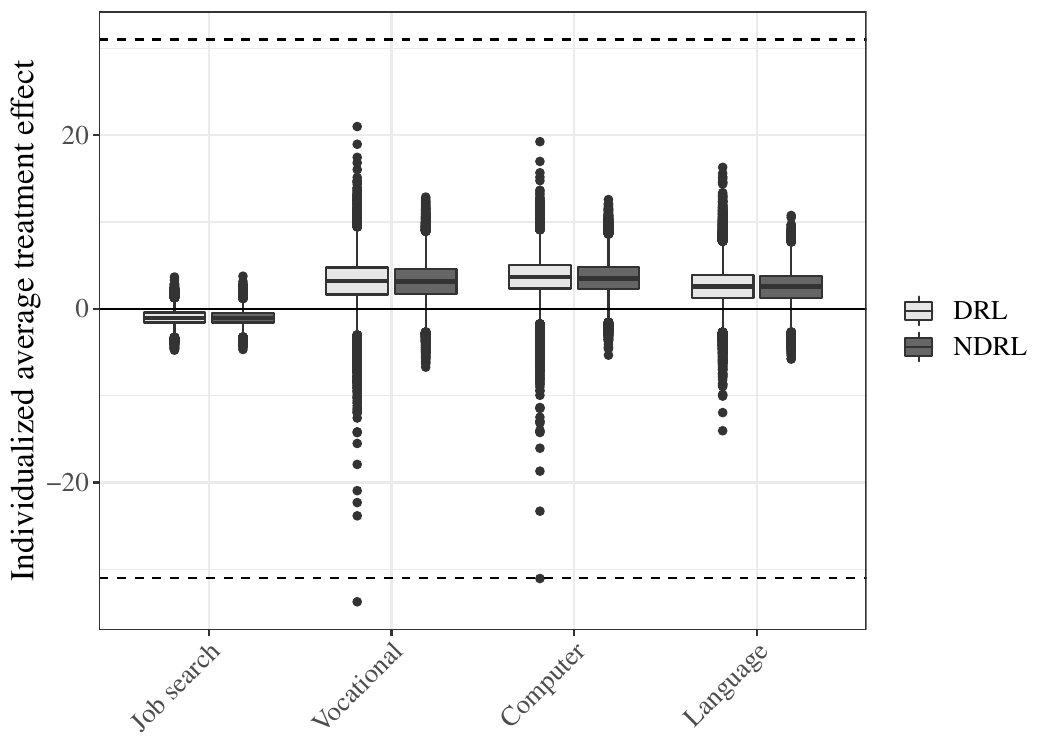}
    \subcaption*{\textit{Note:} The figure shows the distribution of IATEs for participating in the program labeled on the x-axis vs. non-participation estimated by the DR-learner (DRL) and the NDR-learner (NDRL). The dashed line indicates the possible range of the IATE of [-31,31] to illustrate that several DR-learner estimated IATEs lie outside this bound.}
    \label{fig:iate-box}
\end{figure}

We focus on the results based on the out-of-sample variant of the DR- and NDR-learner as the full sample variant leads to severe overfitting with predicted IATEs ranging from -265 to 161 that are up to eight times larger than what is possible given that the outcome is bounded between zero and 31.\footnote{See Appendix \ref{sec:app-iate} for results and discussion of the full sample.} However, Figure \ref{fig:iate-box} shows that the DR-learner produces impossible effect sizes even out-of-sample, which motivates the proposal of the NDR-learner as stabilised variant. Figure \ref{fig:iate-box} provides boxplots of the predicted IATEs and shows outliers lying below the smallest possible value of -31. However, the descriptive statistics provided in Table \ref{tab:iate_desc} and the joint and marginal distributions depicted in Figure \ref{fig:iates-dr-ndr} document that besides the outliers the distributions are quite similar and correlate with at least 0.90. Not surprisingly, the impact of normalised weights is much larger for the three smaller programs and nearly negligible for job search programs. Still, we base the following discussion for all programs on the more stable results of the NDR-learner.

We conduct a classification analysis as proposed by \citeA{Chernozhukov2018TheAverages} to understand which variables are most predictive of effect sizes. To this end, we split the predicted IATE distributions in quintiles and compare the covariate means of the observations falling into the fifth and first quintile. For comparability, we normalise all covariates to have mean zero and variance one. Table \ref{tab:clan-iate} shows the seven variables that have at least one absolute difference between the highest and lowest quintile that is larger than one standard deviation. For example, we observe that the group with the highest effects (the fifth quintile) of a job search program has a 1.32 standard deviations lower past income compared to the lowest IATE group (the first quintile). Also the other variables confirm the patterns that we document already in previous subsections. The effects of job search, vocational and computer training are higher for unskilled workers with lower previous labour market success and foreigners, while the opposite holds for language programs.\footnote{Table \ref{tab:app-clan} shows the classification analysis for all variables.}

\singlespacing
\begin{table}[t]
  \centering  \footnotesize
\begin{threeparttable}
  \caption{Classification analysis of IATEs} 
  \label{tab:clan-iate} 
\begin{tabular}{lcccccc} 
\toprule
\\[-1.8ex] & \multicolumn{1}{c}{Job search} & \multicolumn{1}{c}{Vocational} & \multicolumn{1}{c}{Computer} & \multicolumn{1}{c}{Language} \\ 
\\[-1.8ex] & \multicolumn{1}{c}{(1)} & \multicolumn{1}{c}{(2)} & \multicolumn{1}{c}{(3)} & \multicolumn{1}{c}{(4)} \\ 
\midrule
Past income & -1.32 & -0.84 & -1.17 & 1.01 \\ 
  [0.2em]
  Previous job: unskilled worker & 1.02 & 0.68 & 0.34 & -1.24 \\
  [0.2em]
  Mother tongue other than German, French, Italian & 0.69 & 0.68 & 0.00 & -1.17 \\ 
  [0.2em]
  Qualification: some degree & -0.88 & -0.65 & -0.41 & 1.15 \\ 
  [0.2em]
  Swiss citizen & -0.66 & -0.60 & 0.12 & 1.11 \\ 
  [0.2em]
  Fraction of months employed last 2 years & -1.06 & -0.37 & -0.47 & 0.30 \\ 
  [0.2em]
  Qualification: unskilled & 0.81 & 0.41 & 0.32 & -1.02 \\ 
    \bottomrule
    \end{tabular}%
         \begin{tablenotes} \item \textit{Note:} Table shows the differences in means of standardized covariates between the fifth and the first quintile of the respective estimated IATE distribution. \end{tablenotes}  
\end{threeparttable}
\end{table}
\doublespacing

\subsection{Optimal treatment assignment}

\begin{figure}[t] 
\centering
\caption{Optimal treatment assignment decision trees of depth two and three}\label{fig:op}
\begin{subfigure}{0.34\textwidth}
\includegraphics[width=\linewidth]{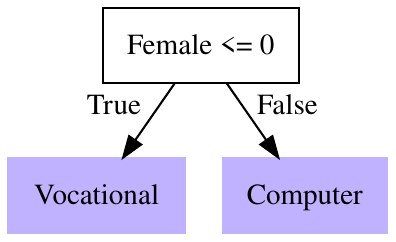}
\caption{Depth 1 \& 5 covariates} \label{fig:op1}
\end{subfigure} \hspace*{\fill}
\begin{subfigure}{0.59\textwidth}
\includegraphics[width=\linewidth]{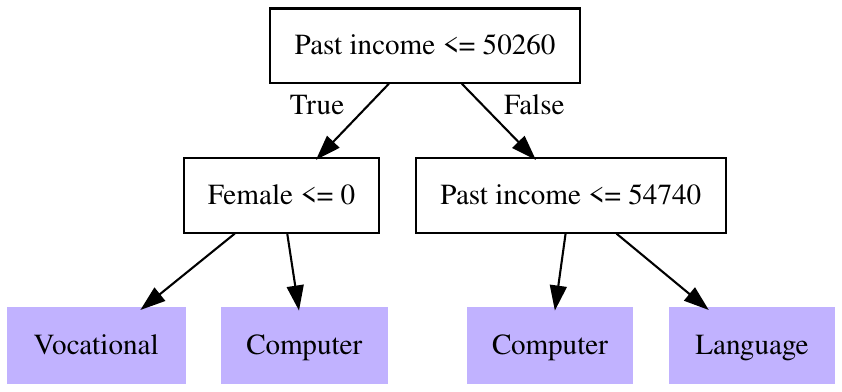}
\caption{Depth 2 \& 5 covariates} \label{fig:op2}
\end{subfigure}
\begin{subfigure}{0.34\textwidth}
\includegraphics[width=\linewidth]{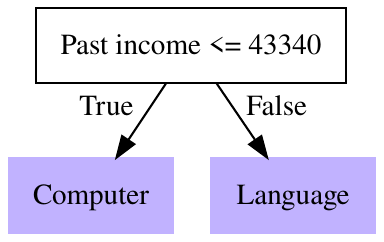}
\caption{Depth 1 \& 16 covariates} \label{fig:op3}
\end{subfigure} \hspace*{\fill}
\begin{subfigure}{0.59\textwidth}
\includegraphics[width=\linewidth]{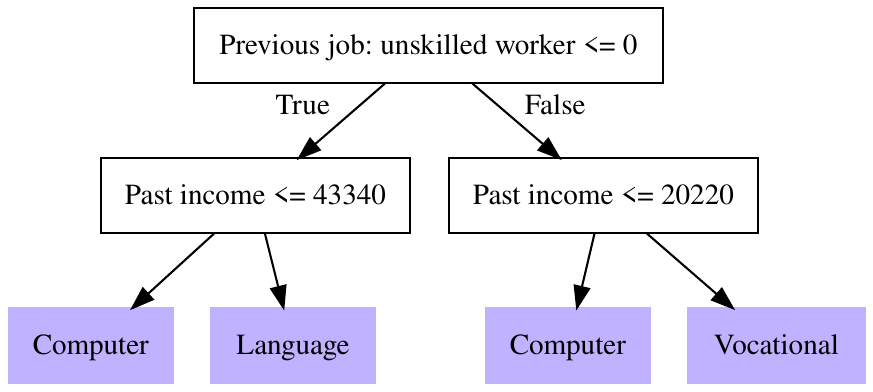}
\caption{Depth 2 \& 16 covariates} \label{fig:op4}
\end{subfigure}
\subcaption*{\textit{Notes:} Optimal assignment rules estimated following the procedure defined in Section \ref{sec:opt-pol}.} 
\end{figure}

The previous section documented substantial heterogeneities in the program effects. To leverage this heterogeneity for better targeting, we apply the DML based optimal policy algorithm of Section \ref{sec:opt-pol}. Figure \ref{fig:op1} shows the simplest decision tree with only one split for the five handpicked covariates. It would allocate men to vocational training and women to computer courses. This split is probably similar to what we would have suggested given the evidence presented in Table \ref{tab:cate-sg}. For a tree of depth two, such an eyeballing approach has its limits and the algorithmic approach provides a systematic way to arrive at an estimated optimal decision tree. The depth two tree in Figure \ref{fig:op2} splits first on past income and then recommends to send low earning men to vocational and low earning women to computer training, while high earners (more than CHF 55k) are recommended to participate in language training. In the absence of the possibility to split on gender, the depth one tree in Figure \ref{fig:op3} splits on past income roughly at the same value where the nonparametric GATEs of computer and language training intersect in Figure \ref{fig:cate-vverds}.\footnote{Appendix \ref{sec:app-op} provides the trees of depth three.}

Panel A of Table \ref{tab:op} summarises the results of the different trees. It shows the percentage of individuals that are placed in the different programs. Not surprisingly, all individuals are recommended to be placed into one of the three positively evaluated hard skill enhancing programs. 

One yet unsolved challenge is how to conduct statistical inference about the quality and stability of the decision trees. \citeA{Athey2021PolicyData} propose a form of cross-validation. To this end, we take the same folds that were used in the cross-fitting procedure to estimate the nuisance parameters. We build the decision tree in four folds and evaluate the value in the left out fold. First, we inspect how often the recommendations based on these trees coincide with the full sample policy rules. Figures \ref{fig:op-overlap} and \ref{fig:op-overlap-high} of Appendix \ref{sec:app-op} show that the cross-validated trees are not identical to the full sample ones. 

\singlespacing
\begin{table}[t]
  \centering  \footnotesize
\begin{threeparttable}
  \caption{Description of estimated optimal policies} 
  \label{tab:op} 
\begin{tabular}{lccccc} 
\toprule
\\[-1.8ex] & \multicolumn{1}{c}{No program} & \multicolumn{1}{c}{Job search} & \multicolumn{1}{c}{Vocational} & \multicolumn{1}{c}{Computer} & \multicolumn{1}{c}{Language}\\ 

\\[-1.8ex] & \multicolumn{1}{c}{(1)} & \multicolumn{1}{c}{(2)} & \multicolumn{1}{c}{(3)} & \multicolumn{1}{c}{(4)} & \multicolumn{1}{c}{(5)}\\ 
\midrule

\multicolumn{6}{l}{\textit{Panel A: Percent allocated to program}} \\
 [0.4em]
 Depth 1 \& 5 variables & 0 & 0 & 56 & 44 & 0 \\ 
 [0.4em]
 Depth 2 \& 5 variables  & 0 & 0 & 32 & 43 & 25 \\ 
 [0.4em]
 Depth 3 \& 5 variables  & 0 & 0 & 47 & 37 & 17 \\ 
  [0.4em]
 Depth 1 \& 16 variables & 0 & 0 & 0 & 54 & 46 \\ 
 [0.4em]
 Depth 2 \& 16 variables & 0 & 0 & 23 & 37 & 40 \\ 
 [0.4em]
 Depth 3 \& 16 variables & 0 & 0 & 42 & 30 & 27 \\ 
 [0.8em]
\multicolumn{6}{l}{\textit{Panel B: Cross-validated difference to APOs}} \\  
[0.4em]
Depth 1 \& 5 variables  & 3.75$^{***}$ & 4.77$^{***}$ & 0.49 & 0.32 & 1.21$^{**}$ \\ 
   & (0.40) & (0.42) & (0.44) & (0.48) & (0.49) \\ 
         [0.4em]
Depth 2 \& 5 variables  & 3.99$^{***}$ & 5.01$^{***}$ & 0.73 & 0.56 & 1.45$^{***}$ \\ 
   & (0.40) & (0.42) & (0.48) & (0.46) & (0.46) \\
         [0.4em]
 Depth 3 \& 5 variables & 3.50$^{***}$ & 4.52$^{***}$ & 0.24 & 0.07 & 0.96$^{**}$ \\ 
        & (0.41) & (0.43) & (0.45) & (0.49) & (0.47) \\ 
 [0.4em]
Depth 1 \& 16 variables  & 3.72$^{***}$ & 4.73$^{***}$ & 0.46 & 0.29 & 1.19$^{***}$ \\ 
   & (0.41) & (0.42) & (0.52) & (0.48) & (0.40) \\ 
         [0.4em]
Depth 2 \& 16 variables  & 3.63$^{***}$ & 4.65$^{***}$ & 0.37 & 0.20 & 1.10$^{***}$ \\ 
   & (0.43) & (0.44) & (0.47) & (0.53) & (0.42) \\ 
         [0.4em]
Depth 3 \& 16 variables & 3.94$^{***}$ & 4.96$^{***}$ & 0.68 & 0.51 & 1.41$^{***}$ \\ 
   & (0.42) & (0.43) & (0.47) & (0.49) & (0.46) \\ 
    \bottomrule
    \end{tabular}%
         \begin{tablenotes} \item \textit{Note:} Panel A shows the percentage of individuals being assigned to a specific program. Panel B shows a t-test of the difference of the cross-validated policy (standard errors in parentheses) and the APOs of the programs. $^{*}$p$<$0.1; $^{**}$p$<$0.05; $^{***}$p$<$0.01 \end{tablenotes}  
\end{threeparttable}
\end{table}
\doublespacing

\citeA{Zhou2018OfflineOptimization} propose another validation idea and test whether the optimal policy rules perform significantly better than sending all individuals to the same program. This is achieved by taking the difference of the APO score of the cross-validated policy rule and the APO score of the program $w$: $ \hat{\Delta}^{cv}_{i,w}(\pi) = \sum_{t=0}^{T} \mathds{1}(\hat{\pi}^{cv}(Z_i) = t) \hat{\Gamma}_{i,t} - \hat{\Gamma}_{i,w}$,
where $\hat{\pi}^{cv}(Z_i)$ is the policy rule that is estimated without individual $i$. A standard \textit{t}-test on the mean of $\hat{\Delta}^{cv}_{i,w}(\pi)$ tests then whether the cross-validated policy rules are significantly better than sending everybody to the same program. Note that the cross-validated policy rules do not necessarily coincide with the trees in the full sample and the cross-validation estimates not the value function for that specific tree. This requires to hold out a test set, which would be viable for an application with bigger programs.

The results are provided in Panel B of Table \ref{tab:op}. We can interpret the mean of $\hat{\Delta}^{cv}_{i,w}(\pi)$ as average treatment effect comparing a regime under the estimated assignment rule or a regime where everybody is sent to the program $w$. This effect is always positive indicating that the estimated rules can leverage the effect heterogeneities to improve the allocation. However, the cross-validated policy rules perform not significantly better than sending just everybody into vocational or computer programs. This would probably change if we could take costs or capacity constraints into account. However, we do not observe costs in this dataset and the optimal decision tree algorithm is currently not capable of incorporating capacity constraints in a systematic way. We leave both extensions for future research using a more detailed database on both costs and capacity constraints.

\section{Discussion and conclusions} \label{sec:conc}

This paper considers recent methodological developments based on Double Machine Learning (DML) through the lens of a standard program evaluation under unconfoundedness. DML based methods provide a convenient toolbox for a comprehensive program evaluation as different parameters of interest can be estimated using the same framework and a combination of standard statistical software. The application to an Active Labour Market Policy evaluation shows that the methods also produce plausible results in practice. The only exception is the DR-learner that required a modification, the newly introduced NDR-learner, before producing stable results for all individualised treatment effects. However, several conceptual and implementational issues remain open for investigation and refinement. 

In general, we know little about how to choose the estimator for the nuisance parameters. The pool of potential machine learning algorithms and their combinations is large and little is known, e.g., about the trade-off between high prediction performance and computation time in the causal setting. Also clear recommendations for the implementation of cross-fitting are missing. Another open question is how to deal with common support in general and for each estimand specifically. The literature on trimming rules is well developed for propensity score based methods estimating average effects. However, we are not only interested in average effects and the propensity score is not the only nuisance parameter of DML. It remains an open question whether the established trimming methods are also sensible for DML when common support becomes an issue.

The estimators for flexible heterogeneous treatment effects provide interesting new tools. However, it is currently not clear to what extent we can actually explore heterogeneity or to what extent we need to pre-define the heterogeneity of interest. The possibility to summarise pre-defined heterogeneity of interest using OLS, kernel or series regressions provide clearly valuable and easy to use options in applications. The instability of methods that aim for individualised heterogeneous effects shows that they should be used with caution and more research is required to investigate whether adjustments like the proposed NDR-learner are useful beyond the application of this paper.

The estimation of optimal treatment assignment rules is mostly unexplored in practice and many interesting issues in applications regarding inference, the implementation of different constraints, more flexible rules than decision trees, or the choice of variables that could or should enter the set of policy variables, which could be explored in future research.

The investigation of these DML specific questions but also the comparison with other more specialised causal machine learning methods for each estimand provides another interesting direction of future research. Such evidence would help to understand and guide which choices are critical in applications similar to the one in this paper.

\newpage

\bibliographystyle{apacite}
\renewcommand{\APACrefYearMonthDay}[3]{\APACrefYear{#1}}
\bibliography{references}

\begin{thebibliography}{}

\bibitem [\protect \citeauthoryear {%
Abadie%
\ \BBA {} Cattaneo%
}{%
Abadie%
\ \BBA {} Cattaneo%
}{%
{\protect \APACyear {2018}}%
}]{%
Abadie2018EconometricEvaluation}
\APACinsertmetastar {%
Abadie2018EconometricEvaluation}%
\begin{APACrefauthors}%
Abadie, A.%
\BCBT {}\ \BBA {} Cattaneo, M\BPBI D.%
\end{APACrefauthors}%
\unskip\
\newblock
\APACrefYearMonthDay{2018}{}{}.
\newblock
{\BBOQ}\APACrefatitle {{Econometric methods for program evaluation}}
  {{Econometric methods for program evaluation}}.{\BBCQ}
\newblock
\APACjournalVolNumPages{Annual Review of Economics}{10}{}{465--503}.
\PrintBackRefs{\CurrentBib}

\bibitem [\protect \citeauthoryear {%
Athey%
\ \BBA {} Imbens%
}{%
Athey%
\ \BBA {} Imbens%
}{%
{\protect \APACyear {2019}}%
}]{%
Athey2019MachineAbout}
\APACinsertmetastar {%
Athey2019MachineAbout}%
\begin{APACrefauthors}%
Athey, S.%
\BCBT {}\ \BBA {} Imbens, G.%
\end{APACrefauthors}%
\unskip\
\newblock
\APACrefYearMonthDay{2019}{}{}.
\newblock
{\BBOQ}\APACrefatitle {{Machine learning methods that economists should know
  about}} {{Machine learning methods that economists should know
  about}}.{\BBCQ}
\newblock
\APACjournalVolNumPages{Annual Review of Economics}{11}{}{685--725}.
\PrintBackRefs{\CurrentBib}

\bibitem [\protect \citeauthoryear {%
Athey%
\ \BBA {} Imbens%
}{%
Athey%
\ \BBA {} Imbens%
}{%
{\protect \APACyear {2016}}%
}]{%
Athey2016}
\APACinsertmetastar {%
Athey2016}%
\begin{APACrefauthors}%
Athey, S.%
\BCBT {}\ \BBA {} Imbens, G\BPBI W.%
\end{APACrefauthors}%
\unskip\
\newblock
\APACrefYearMonthDay{2016}{}{}.
\newblock
{\BBOQ}\APACrefatitle {{Recursive partitioning for heterogeneous causal
  effects}} {{Recursive partitioning for heterogeneous causal effects}}.{\BBCQ}
\newblock
\APACjournalVolNumPages{Proceedings of the National Academy of
  Sciences}{113}{27}{7353--7360}.
\PrintBackRefs{\CurrentBib}

\bibitem [\protect \citeauthoryear {%
Athey%
\ \BBA {} Imbens%
}{%
Athey%
\ \BBA {} Imbens%
}{%
{\protect \APACyear {2017}}%
}]{%
Athey2017}
\APACinsertmetastar {%
Athey2017}%
\begin{APACrefauthors}%
Athey, S.%
\BCBT {}\ \BBA {} Imbens, G\BPBI W.%
\end{APACrefauthors}%
\unskip\
\newblock
\APACrefYearMonthDay{2017}{}{}.
\newblock
{\BBOQ}\APACrefatitle {{The state of applied econometrics: causality and policy
  evaluation}} {{The state of applied econometrics: causality and policy
  evaluation}}.{\BBCQ}
\newblock
\APACjournalVolNumPages{Journal of Economic Perspectives}{31}{2}{3--32}.
\PrintBackRefs{\CurrentBib}

\bibitem [\protect \citeauthoryear {%
Athey%
, Imbens%
\BCBL {}\ \BBA {} Wager%
}{%
Athey%
\ \protect \BOthers {.}}{%
{\protect \APACyear {2018}}%
}]{%
Athey2018ApproximateDimensions}
\APACinsertmetastar {%
Athey2018ApproximateDimensions}%
\begin{APACrefauthors}%
Athey, S.%
, Imbens, G\BPBI W.%
\BCBL {}\ \BBA {} Wager, S.%
\end{APACrefauthors}%
\unskip\
\newblock
\APACrefYearMonthDay{2018}{}{}.
\newblock
{\BBOQ}\APACrefatitle {{Approximate residual balancing: Debiased inference of
  average treatment effects in high dimensions}} {{Approximate residual
  balancing: Debiased inference of average treatment effects in high
  dimensions}}.{\BBCQ}
\newblock
\APACjournalVolNumPages{Journal of the Royal Statistical Society: Series B
  (Statistical Methodology)}{80}{4}{597--632}.
\PrintBackRefs{\CurrentBib}

\bibitem [\protect \citeauthoryear {%
Athey%
, Tibshirani%
\BCBL {}\ \BBA {} Wager%
}{%
Athey%
\ \protect \BOthers {.}}{%
{\protect \APACyear {2019}}%
}]{%
Athey2017a}
\APACinsertmetastar {%
Athey2017a}%
\begin{APACrefauthors}%
Athey, S.%
, Tibshirani, J.%
\BCBL {}\ \BBA {} Wager, S.%
\end{APACrefauthors}%
\unskip\
\newblock
\APACrefYearMonthDay{2019}{}{}.
\newblock
{\BBOQ}\APACrefatitle {{Generalized random forests}} {{Generalized random
  forests}}.{\BBCQ}
\newblock
\APACjournalVolNumPages{Annals of Statistics}{47}{2}{1148 - 1178}.
\PrintBackRefs{\CurrentBib}

\bibitem [\protect \citeauthoryear {%
Athey%
\ \BBA {} Wager%
}{%
Athey%
\ \BBA {} Wager%
}{%
{\protect \APACyear {2021}}%
}]{%
Athey2021PolicyData}
\APACinsertmetastar {%
Athey2021PolicyData}%
\begin{APACrefauthors}%
Athey, S.%
\BCBT {}\ \BBA {} Wager, S.%
\end{APACrefauthors}%
\unskip\
\newblock
\APACrefYearMonthDay{2021}{}{}.
\newblock
{\BBOQ}\APACrefatitle {{Policy learning with observational data}} {{Policy
  learning with observational data}}.{\BBCQ}
\newblock
\APACjournalVolNumPages{Econometrica}{89}{1}{133--161}.
\PrintBackRefs{\CurrentBib}

\bibitem [\protect \citeauthoryear {%
Avagyan%
\ \BBA {} Vansteelandt%
}{%
Avagyan%
\ \BBA {} Vansteelandt%
}{%
{\protect \APACyear {2017}}%
}]{%
Avagyan2017HonestEstimation}
\APACinsertmetastar {%
Avagyan2017HonestEstimation}%
\begin{APACrefauthors}%
Avagyan, V.%
\BCBT {}\ \BBA {} Vansteelandt, S.%
\end{APACrefauthors}%
\unskip\
\newblock
\APACrefYearMonthDay{2017}{}{}.
\newblock
{\BBOQ}\APACrefatitle {{Honest data-adaptive inference for the average
  treatment effect under model misspecification using penalised bias-reduced
  double-robust estimation}} {{Honest data-adaptive inference for the average
  treatment effect under model misspecification using penalised bias-reduced
  double-robust estimation}}.{\BBCQ}
\newblock
\APACjournalVolNumPages{arXiv:1708.03787}{}{}{}.
\newblock
\begin{APACrefURL} \url{http://arxiv.org/abs/1708.03787} \end{APACrefURL}
\PrintBackRefs{\CurrentBib}

\bibitem [\protect \citeauthoryear {%
Baiardi%
\ \BBA {} Naghi%
}{%
Baiardi%
\ \BBA {} Naghi%
}{%
{\protect \APACyear {2021}}%
}]{%
Baiardi2021TheStudies}
\APACinsertmetastar {%
Baiardi2021TheStudies}%
\begin{APACrefauthors}%
Baiardi, A.%
\BCBT {}\ \BBA {} Naghi, A\BPBI A.%
\end{APACrefauthors}%
\unskip\
\newblock
\APACrefYearMonthDay{2021}{}{}.
\newblock
{\BBOQ}\APACrefatitle {{The value added of machine learning to causal
  inference: Evidence from revisited studies}} {{The value added of machine
  learning to causal inference: Evidence from revisited studies}}.{\BBCQ}
\newblock
\APACjournalVolNumPages{arXiv:2101.00878}{}{}{}.
\PrintBackRefs{\CurrentBib}

\bibitem [\protect \citeauthoryear {%
Bansak%
\ \protect \BOthers {.}}{%
Bansak%
\ \protect \BOthers {.}}{%
{\protect \APACyear {2018}}%
}]{%
Bansak2018ImprovingAssignment}
\APACinsertmetastar {%
Bansak2018ImprovingAssignment}%
\begin{APACrefauthors}%
Bansak, K.%
, Ferwerda, J.%
, Hainmueller, J.%
, Dillon, A.%
, Hangartner, D.%
, Lawrence, D.%
\BCBL {}\ \BBA {} Weinstein, J.%
\end{APACrefauthors}%
\unskip\
\newblock
\APACrefYearMonthDay{2018}{}{}.
\newblock
{\BBOQ}\APACrefatitle {{Improving refugee integration through data-driven
  algorithmic assignment}} {{Improving refugee integration through data-driven
  algorithmic assignment}}.{\BBCQ}
\newblock
\APACjournalVolNumPages{Science}{359}{6373}{325--329}.
\PrintBackRefs{\CurrentBib}

\bibitem [\protect \citeauthoryear {%
Belloni%
\ \BBA {} Chernozhukov%
}{%
Belloni%
\ \BBA {} Chernozhukov%
}{%
{\protect \APACyear {2013}}%
}]{%
Belloni2013LeastModels}
\APACinsertmetastar {%
Belloni2013LeastModels}%
\begin{APACrefauthors}%
Belloni, A.%
\BCBT {}\ \BBA {} Chernozhukov, V.%
\end{APACrefauthors}%
\unskip\
\newblock
\APACrefYearMonthDay{2013}{}{}.
\newblock
{\BBOQ}\APACrefatitle {{Least squares after model selection in high-dimensional
  sparse models}} {{Least squares after model selection in high-dimensional
  sparse models}}.{\BBCQ}
\newblock
\APACjournalVolNumPages{Bernoulli}{19}{2}{521--547}.
\PrintBackRefs{\CurrentBib}

\bibitem [\protect \citeauthoryear {%
Belloni%
, Chernozhukov%
, Fern{\'{a}}ndez-Val%
\BCBL {}\ \BBA {} Hansen%
}{%
Belloni%
\ \protect \BOthers {.}}{%
{\protect \APACyear {2017}}%
}]{%
Belloni2017}
\APACinsertmetastar {%
Belloni2017}%
\begin{APACrefauthors}%
Belloni, A.%
, Chernozhukov, V.%
, Fern{\'{a}}ndez-Val, I.%
\BCBL {}\ \BBA {} Hansen, C.%
\end{APACrefauthors}%
\unskip\
\newblock
\APACrefYearMonthDay{2017}{}{}.
\newblock
{\BBOQ}\APACrefatitle {{Program evaluation and causal inference with
  high-dimensional data}} {{Program evaluation and causal inference with
  high-dimensional data}}.{\BBCQ}
\newblock
\APACjournalVolNumPages{Econometrica}{85}{1}{233--298}.
\PrintBackRefs{\CurrentBib}

\bibitem [\protect \citeauthoryear {%
Belloni%
, Chernozhukov%
\BCBL {}\ \BBA {} Hansen%
}{%
Belloni%
\ \protect \BOthers {.}}{%
{\protect \APACyear {2014}}%
}]{%
Belloni2014InferenceControls}
\APACinsertmetastar {%
Belloni2014InferenceControls}%
\begin{APACrefauthors}%
Belloni, A.%
, Chernozhukov, V.%
\BCBL {}\ \BBA {} Hansen, C.%
\end{APACrefauthors}%
\unskip\
\newblock
\APACrefYearMonthDay{2014}{}{}.
\newblock
{\BBOQ}\APACrefatitle {{Inference on treatment effects after selection among
  high-dimensional controls}} {{Inference on treatment effects after selection
  among high-dimensional controls}}.{\BBCQ}
\newblock
\APACjournalVolNumPages{Review of Economic Studies}{81}{2}{608--650}.
\PrintBackRefs{\CurrentBib}

\bibitem [\protect \citeauthoryear {%
Bertrand%
, Cr{\'{e}}pon%
, Marguerie%
\BCBL {}\ \BBA {} Premand%
}{%
Bertrand%
\ \protect \BOthers {.}}{%
{\protect \APACyear {2017}}%
}]{%
Bertrand2017}
\APACinsertmetastar {%
Bertrand2017}%
\begin{APACrefauthors}%
Bertrand, M.%
, Cr{\'{e}}pon, B.%
, Marguerie, A.%
\BCBL {}\ \BBA {} Premand, P.%
\end{APACrefauthors}%
\unskip\
\newblock
\APACrefYearMonthDay{2017}{}{}.
\newblock
{\BBOQ}\APACrefatitle {{Contemporaneous and post-program impacts of a public
  works program: Evidence from C{\^{o}}te d'Ivoire}} {{Contemporaneous and
  post-program impacts of a public works program: Evidence from C{\^{o}}te
  d'Ivoire}}.{\BBCQ}
\newblock
\APACjournalVolNumPages{World Bank Working Paper}{}{}{}.
\PrintBackRefs{\CurrentBib}

\bibitem [\protect \citeauthoryear {%
Breiman%
}{%
Breiman%
}{%
{\protect \APACyear {2001}}%
}]{%
Breiman2001}
\APACinsertmetastar {%
Breiman2001}%
\begin{APACrefauthors}%
Breiman, L.%
\end{APACrefauthors}%
\unskip\
\newblock
\APACrefYearMonthDay{2001}{}{}.
\newblock
{\BBOQ}\APACrefatitle {{Random forests}} {{Random forests}}.{\BBCQ}
\newblock
\APACjournalVolNumPages{Machine Learning}{45}{1}{5--32}.
\PrintBackRefs{\CurrentBib}

\bibitem [\protect \citeauthoryear {%
Buja%
, Hastie%
\BCBL {}\ \BBA {} Tibshirani%
}{%
Buja%
\ \protect \BOthers {.}}{%
{\protect \APACyear {1989}}%
}]{%
Buja1989LinearModels}
\APACinsertmetastar {%
Buja1989LinearModels}%
\begin{APACrefauthors}%
Buja, A.%
, Hastie, T.%
\BCBL {}\ \BBA {} Tibshirani, R.%
\end{APACrefauthors}%
\unskip\
\newblock
\APACrefYearMonthDay{1989}{}{}.
\newblock
{\BBOQ}\APACrefatitle {{Linear smoothers and additive models}} {{Linear
  smoothers and additive models}}.{\BBCQ}
\newblock
\APACjournalVolNumPages{The Annals of Statistics}{17}{2}{453--510}.
\PrintBackRefs{\CurrentBib}

\bibitem [\protect \citeauthoryear {%
Busso%
, DiNardo%
\BCBL {}\ \BBA {} McCrary%
}{%
Busso%
\ \protect \BOthers {.}}{%
{\protect \APACyear {2014}}%
}]{%
busso2014new}
\APACinsertmetastar {%
busso2014new}%
\begin{APACrefauthors}%
Busso, M.%
, DiNardo, J.%
\BCBL {}\ \BBA {} McCrary, J.%
\end{APACrefauthors}%
\unskip\
\newblock
\APACrefYearMonthDay{2014}{}{}.
\newblock
{\BBOQ}\APACrefatitle {{New evidence on the finite sample properties of
  propensity score reweighting and matching estimators}} {{New evidence on the
  finite sample properties of propensity score reweighting and matching
  estimators}}.{\BBCQ}
\newblock
\APACjournalVolNumPages{Review of Economics and Statistics}{96}{5}{885--897}.
\PrintBackRefs{\CurrentBib}

\bibitem [\protect \citeauthoryear {%
Card%
, Kluve%
\BCBL {}\ \BBA {} Weber%
}{%
Card%
\ \protect \BOthers {.}}{%
{\protect \APACyear {2018}}%
}]{%
Card2018WhatEvaluations}
\APACinsertmetastar {%
Card2018WhatEvaluations}%
\begin{APACrefauthors}%
Card, D.%
, Kluve, J.%
\BCBL {}\ \BBA {} Weber, A.%
\end{APACrefauthors}%
\unskip\
\newblock
\APACrefYearMonthDay{2018}{}{}.
\newblock
{\BBOQ}\APACrefatitle {{What works? A meta analysis of recent active labor
  market program evaluations}} {{What works? A meta analysis of recent active
  labor market program evaluations}}.{\BBCQ}
\newblock
\APACjournalVolNumPages{Journal of the European Economic
  Association}{16}{3}{894--931}.
\PrintBackRefs{\CurrentBib}

\bibitem [\protect \citeauthoryear {%
Chernozhukov%
, Chetverikov%
\BCBL {}\ \protect \BOthers {.}}{%
Chernozhukov%
, Chetverikov%
\BCBL {}\ \protect \BOthers {.}}{%
{\protect \APACyear {2018}}%
}]{%
Chernozhukov2018}
\APACinsertmetastar {%
Chernozhukov2018}%
\begin{APACrefauthors}%
Chernozhukov, V.%
, Chetverikov, D.%
, Demirer, M.%
, Duflo, E.%
, Hansen, C.%
, Newey, W.%
\BCBL {}\ \BBA {} Robins, J.%
\end{APACrefauthors}%
\unskip\
\newblock
\APACrefYearMonthDay{2018}{}{}.
\newblock
{\BBOQ}\APACrefatitle {{Double/Debiased machine learning for treatment and
  structural parameters}} {{Double/Debiased machine learning for treatment and
  structural parameters}}.{\BBCQ}
\newblock
\APACjournalVolNumPages{The Econometrics Journal}{21}{1}{C1-C68}.
\PrintBackRefs{\CurrentBib}

\bibitem [\protect \citeauthoryear {%
Chernozhukov%
, Demirer%
, Duflo%
\BCBL {}\ \BBA {} Fernandez-Val%
}{%
Chernozhukov%
\ \protect \BOthers {.}}{%
{\protect \APACyear {2017}}%
}]{%
Chernozhukov2017GenericExperiments}
\APACinsertmetastar {%
Chernozhukov2017GenericExperiments}%
\begin{APACrefauthors}%
Chernozhukov, V.%
, Demirer, M.%
, Duflo, E.%
\BCBL {}\ \BBA {} Fernandez-Val, I.%
\end{APACrefauthors}%
\unskip\
\newblock
\APACrefYearMonthDay{2017}{}{}.
\newblock
{\BBOQ}\APACrefatitle {{Generic machine learning inference on heterogenous
  treatment effects in randomized experiments}} {{Generic machine learning
  inference on heterogenous treatment effects in randomized
  experiments}}.{\BBCQ}
\newblock
\APACjournalVolNumPages{arXiv:1712.04802}{}{}{}.
\newblock
\begin{APACrefURL} \url{http://arxiv.org/abs/1712.04802} \end{APACrefURL}
\PrintBackRefs{\CurrentBib}

\bibitem [\protect \citeauthoryear {%
Chernozhukov%
, Fernandez-Val%
\BCBL {}\ \BBA {} Luo%
}{%
Chernozhukov%
, Fernandez-Val%
\BCBL {}\ \BBA {} Luo%
}{%
{\protect \APACyear {2018}}%
}]{%
Chernozhukov2018TheAverages}
\APACinsertmetastar {%
Chernozhukov2018TheAverages}%
\begin{APACrefauthors}%
Chernozhukov, V.%
, Fernandez-Val, I.%
\BCBL {}\ \BBA {} Luo, Y.%
\end{APACrefauthors}%
\unskip\
\newblock
\APACrefYearMonthDay{2018}{}{}.
\newblock
{\BBOQ}\APACrefatitle {{The sorted effects method: Discovering heterogeneous
  effects beyond their averages}} {{The sorted effects method: Discovering
  heterogeneous effects beyond their averages}}.{\BBCQ}
\newblock
\APACjournalVolNumPages{Econometrica}{86}{6}{1911--1938}.
\PrintBackRefs{\CurrentBib}

\bibitem [\protect \citeauthoryear {%
Cockx%
, Lechner%
\BCBL {}\ \BBA {} Bollens%
}{%
Cockx%
\ \protect \BOthers {.}}{%
{\protect \APACyear {2020}}%
}]{%
Cockx2020PriorityBelgium}
\APACinsertmetastar {%
Cockx2020PriorityBelgium}%
\begin{APACrefauthors}%
Cockx, B.%
, Lechner, M.%
\BCBL {}\ \BBA {} Bollens, J.%
\end{APACrefauthors}%
\unskip\
\newblock
\APACrefYearMonthDay{2020}{}{}.
\newblock
{\BBOQ}\APACrefatitle {{Priority to unemployed immigrants? A causal machine
  learning evaluation of training in Belgium}} {{Priority to unemployed
  immigrants? A causal machine learning evaluation of training in
  Belgium}}.{\BBCQ}
\newblock
\APACjournalVolNumPages{CEPR Discussion Paper No. DP14270}{}{}{}.
\PrintBackRefs{\CurrentBib}

\bibitem [\protect \citeauthoryear {%
Colangelo%
\ \BBA {} Lee%
}{%
Colangelo%
\ \BBA {} Lee%
}{%
{\protect \APACyear {2019}}%
}]{%
Colangelo2019DoubleTreatments}
\APACinsertmetastar {%
Colangelo2019DoubleTreatments}%
\begin{APACrefauthors}%
Colangelo, K.%
\BCBT {}\ \BBA {} Lee, Y\BHBI Y.%
\end{APACrefauthors}%
\unskip\
\newblock
\APACrefYearMonthDay{2019}{}{}.
\newblock
{\BBOQ}\APACrefatitle {{Double debiased machine learning nonparametric
  inference with continuous treatments}} {{Double debiased machine learning
  nonparametric inference with continuous treatments}}.{\BBCQ}
\newblock
\APACjournalVolNumPages{arXiv:2004.03036}{}{}{}.
\PrintBackRefs{\CurrentBib}

\bibitem [\protect \citeauthoryear {%
Cr{\'{e}}pon%
\ \BBA {} van~den Berg%
}{%
Cr{\'{e}}pon%
\ \BBA {} van~den Berg%
}{%
{\protect \APACyear {2016}}%
}]{%
crepon2016active}
\APACinsertmetastar {%
crepon2016active}%
\begin{APACrefauthors}%
Cr{\'{e}}pon, B.%
\BCBT {}\ \BBA {} van~den Berg, G\BPBI J.%
\end{APACrefauthors}%
\unskip\
\newblock
\APACrefYearMonthDay{2016}{}{}.
\newblock
{\BBOQ}\APACrefatitle {{Active labor market policies}} {{Active labor market
  policies}}.{\BBCQ}
\newblock
\APACjournalVolNumPages{Annual Review of Economics}{8}{}{521--546}.
\PrintBackRefs{\CurrentBib}

\bibitem [\protect \citeauthoryear {%
Curth%
\ \BBA {} van~der Schaar%
}{%
Curth%
\ \BBA {} van~der Schaar%
}{%
{\protect \APACyear {2021}}%
}]{%
Curth2021NonparametricAlgorithms}
\APACinsertmetastar {%
Curth2021NonparametricAlgorithms}%
\begin{APACrefauthors}%
Curth, A.%
\BCBT {}\ \BBA {} van~der Schaar, M.%
\end{APACrefauthors}%
\unskip\
\newblock
\APACrefYearMonthDay{2021}{}{}.
\newblock
{\BBOQ}\APACrefatitle {{Nonparametric estimation of heterogeneous treatment
  effects: From theory to learning algorithms}} {{Nonparametric estimation of
  heterogeneous treatment effects: From theory to learning algorithms}}.{\BBCQ}
\newblock
\BIn{} \APACrefbtitle {Proceedings of The 24th International Conference on
  Artificial Intelligence and Statistics} {Proceedings of the 24th
  international conference on artificial intelligence and statistics}\
  (\BVOL~130, \BPGS\ 1810--1818).
\newblock
\APACaddressPublisher{}{PMLR}.
\PrintBackRefs{\CurrentBib}

\bibitem [\protect \citeauthoryear {%
Davis%
\ \BBA {} Heller%
}{%
Davis%
\ \BBA {} Heller%
}{%
{\protect \APACyear {2020}}%
}]{%
Davis2020RethinkingJobs}
\APACinsertmetastar {%
Davis2020RethinkingJobs}%
\begin{APACrefauthors}%
Davis, J\BPBI M\BPBI V.%
\BCBT {}\ \BBA {} Heller, S\BPBI B.%
\end{APACrefauthors}%
\unskip\
\newblock
\APACrefYearMonthDay{2020}{}{}.
\newblock
{\BBOQ}\APACrefatitle {{Rethinking the benefits of youth employment programs:
  The heterogeneous effects of summer jobs}} {{Rethinking the benefits of youth
  employment programs: The heterogeneous effects of summer jobs}}.{\BBCQ}
\newblock
\APACjournalVolNumPages{The Review of Economics and
  Statistics}{102}{4}{664--677}.
\PrintBackRefs{\CurrentBib}

\bibitem [\protect \citeauthoryear {%
Dudik%
, Langford%
\BCBL {}\ \BBA {} Li%
}{%
Dudik%
\ \protect \BOthers {.}}{%
{\protect \APACyear {2011}}%
}]{%
Dudik2011DoublyLearning}
\APACinsertmetastar {%
Dudik2011DoublyLearning}%
\begin{APACrefauthors}%
Dudik, M.%
, Langford, J.%
\BCBL {}\ \BBA {} Li, L.%
\end{APACrefauthors}%
\unskip\
\newblock
\APACrefYearMonthDay{2011}{}{}.
\newblock
{\BBOQ}\APACrefatitle {{Doubly robust policy evaluation and learning}} {{Doubly
  robust policy evaluation and learning}}.{\BBCQ}
\newblock
\APACjournalVolNumPages{arXiv:1103.4601}{}{}{}.
\newblock
\begin{APACrefURL} \url{http://arxiv.org/abs/1103.4601} \end{APACrefURL}
\PrintBackRefs{\CurrentBib}

\bibitem [\protect \citeauthoryear {%
Fan%
, Hsu%
, Lieli%
\BCBL {}\ \BBA {} Zhang%
}{%
Fan%
\ \protect \BOthers {.}}{%
{\protect \APACyear {2020}}%
}]{%
fan2020EstimationData}
\APACinsertmetastar {%
fan2020EstimationData}%
\begin{APACrefauthors}%
Fan, Q.%
, Hsu, Y\BHBI C.%
, Lieli, R\BPBI P.%
\BCBL {}\ \BBA {} Zhang, Y.%
\end{APACrefauthors}%
\unskip\
\newblock
\APACrefYearMonthDay{2020}{}{}.
\newblock
{\BBOQ}\APACrefatitle {{Estimation of conditional average treatment effects
  with high-dimensional data}} {{Estimation of conditional average treatment
  effects with high-dimensional data}}.{\BBCQ}
\newblock
\APACjournalVolNumPages{Journal of Business {\&} Economic
  Statistics}{}{}{published ahead of print 14 September 2020}.
\PrintBackRefs{\CurrentBib}

\bibitem [\protect \citeauthoryear {%
Farbmacher%
, K{\"{o}}gel%
\BCBL {}\ \BBA {} Spindler%
}{%
Farbmacher%
\ \protect \BOthers {.}}{%
{\protect \APACyear {2021}}%
}]{%
Farbmacher2021HeterogeneousCognition}
\APACinsertmetastar {%
Farbmacher2021HeterogeneousCognition}%
\begin{APACrefauthors}%
Farbmacher, H.%
, K{\"{o}}gel, H.%
\BCBL {}\ \BBA {} Spindler, M.%
\end{APACrefauthors}%
\unskip\
\newblock
\APACrefYearMonthDay{2021}{}{}.
\newblock
{\BBOQ}\APACrefatitle {{Heterogeneous effects of poverty on cognition}}
  {{Heterogeneous effects of poverty on cognition}}.{\BBCQ}
\newblock
\APACjournalVolNumPages{Labour Economics}{71}{}{102028}.
\PrintBackRefs{\CurrentBib}

\bibitem [\protect \citeauthoryear {%
Farrell%
}{%
Farrell%
}{%
{\protect \APACyear {2015}}%
}]{%
Farrell2015}
\APACinsertmetastar {%
Farrell2015}%
\begin{APACrefauthors}%
Farrell, M\BPBI H.%
\end{APACrefauthors}%
\unskip\
\newblock
\APACrefYearMonthDay{2015}{}{}.
\newblock
{\BBOQ}\APACrefatitle {{Robust inference on average treatment effects with
  possibly more covariates than observations}} {{Robust inference on average
  treatment effects with possibly more covariates than observations}}.{\BBCQ}
\newblock
\APACjournalVolNumPages{Journal of Econometrics}{189}{1}{1--23}.
\PrintBackRefs{\CurrentBib}

\bibitem [\protect \citeauthoryear {%
Farrell%
, Liang%
\BCBL {}\ \BBA {} Misra%
}{%
Farrell%
\ \protect \BOthers {.}}{%
{\protect \APACyear {2021}}%
}]{%
Farrell2021DeepInference}
\APACinsertmetastar {%
Farrell2021DeepInference}%
\begin{APACrefauthors}%
Farrell, M\BPBI H.%
, Liang, T.%
\BCBL {}\ \BBA {} Misra, S.%
\end{APACrefauthors}%
\unskip\
\newblock
\APACrefYearMonthDay{2021}{}{}.
\newblock
{\BBOQ}\APACrefatitle {{Deep neural networks for estimation and inference}}
  {{Deep neural networks for estimation and inference}}.{\BBCQ}
\newblock
\APACjournalVolNumPages{Econometrica}{89}{1}{181--213}.
\PrintBackRefs{\CurrentBib}

\bibitem [\protect \citeauthoryear {%
Foster%
\ \BBA {} Syrgkanis%
}{%
Foster%
\ \BBA {} Syrgkanis%
}{%
{\protect \APACyear {2019}}%
}]{%
Foster2019OrthogonalLearning}
\APACinsertmetastar {%
Foster2019OrthogonalLearning}%
\begin{APACrefauthors}%
Foster, D\BPBI J.%
\BCBT {}\ \BBA {} Syrgkanis, V.%
\end{APACrefauthors}%
\unskip\
\newblock
\APACrefYearMonthDay{2019}{}{}.
\newblock
{\BBOQ}\APACrefatitle {{Orthogonal statistical learning}} {{Orthogonal
  statistical learning}}.{\BBCQ}
\newblock
\APACjournalVolNumPages{arXiv:1901.09036}{}{}{}.
\newblock
\begin{APACrefURL} \url{http://arxiv.org/abs/1901.09036} \end{APACrefURL}
\PrintBackRefs{\CurrentBib}

\bibitem [\protect \citeauthoryear {%
Gerfin%
\ \BBA {} Lechner%
}{%
Gerfin%
\ \BBA {} Lechner%
}{%
{\protect \APACyear {2002}}%
}]{%
gerfin2002microeconometric}
\APACinsertmetastar {%
gerfin2002microeconometric}%
\begin{APACrefauthors}%
Gerfin, M.%
\BCBT {}\ \BBA {} Lechner, M.%
\end{APACrefauthors}%
\unskip\
\newblock
\APACrefYearMonthDay{2002}{}{}.
\newblock
{\BBOQ}\APACrefatitle {{A microeconometric evaluation of the active labour
  market policy in Switzerland}} {{A microeconometric evaluation of the active
  labour market policy in Switzerland}}.{\BBCQ}
\newblock
\APACjournalVolNumPages{Economic Journal}{112}{482}{854--893}.
\PrintBackRefs{\CurrentBib}

\bibitem [\protect \citeauthoryear {%
Glynn%
\ \BBA {} Quinn%
}{%
Glynn%
\ \BBA {} Quinn%
}{%
{\protect \APACyear {2009}}%
}]{%
Glynn2009AnEstimator}
\APACinsertmetastar {%
Glynn2009AnEstimator}%
\begin{APACrefauthors}%
Glynn, A\BPBI N.%
\BCBT {}\ \BBA {} Quinn, K\BPBI M.%
\end{APACrefauthors}%
\unskip\
\newblock
\APACrefYearMonthDay{2009}{}{}.
\newblock
{\BBOQ}\APACrefatitle {{An introduction to the augmented inverse propensity
  weighted estimator}} {{An introduction to the augmented inverse propensity
  weighted estimator}}.{\BBCQ}
\newblock
\APACjournalVolNumPages{Political Analysis}{18}{1}{36--56}.
\PrintBackRefs{\CurrentBib}

\bibitem [\protect \citeauthoryear {%
Gulyas%
\ \BBA {} Pytka%
}{%
Gulyas%
\ \BBA {} Pytka%
}{%
{\protect \APACyear {2019}}%
}]{%
Gulyas2019UnderstandingApproach}
\APACinsertmetastar {%
Gulyas2019UnderstandingApproach}%
\begin{APACrefauthors}%
Gulyas, A.%
\BCBT {}\ \BBA {} Pytka, K.%
\end{APACrefauthors}%
\unskip\
\newblock
\APACrefYearMonthDay{2019}{}{}.
\newblock
{\BBOQ}\APACrefatitle {{Understanding the sources of earnings losses after job
  displacement: A machine-learning approach}} {{Understanding the sources of
  earnings losses after job displacement: A machine-learning approach}}.{\BBCQ}
\newblock
\APACjournalVolNumPages{Discussion Paper Series – CRC TR 224 No. 131}{}{}{}.
\PrintBackRefs{\CurrentBib}

\bibitem [\protect \citeauthoryear {%
H{\'{a}}jek%
}{%
H{\'{a}}jek%
}{%
{\protect \APACyear {1971}}%
}]{%
Hajek1971CommentOne}
\APACinsertmetastar {%
Hajek1971CommentOne}%
\begin{APACrefauthors}%
H{\'{a}}jek, J.%
\end{APACrefauthors}%
\unskip\
\newblock
\APACrefYearMonthDay{1971}{}{}.
\newblock
{\BBOQ}\APACrefatitle {{Comment on “An essay on the logical foundations of
  survey sampling, part one”}} {{Comment on “An essay on the logical
  foundations of survey sampling, part one”}}.{\BBCQ}
\newblock
\BIn{} V\BPBI P.~Godambe\ \BBA {} D\BPBI A.~Sprott\ (\BEDS), \APACrefbtitle
  {Foundations of Statistical Inference} {Foundations of statistical
  inference}\ (\BPG~236).
\newblock
\APACaddressPublisher{Toronto}{Holt, Rinehart and Winston}.
\PrintBackRefs{\CurrentBib}

\bibitem [\protect \citeauthoryear {%
Hayfield%
\ \BBA {} Racine%
}{%
Hayfield%
\ \BBA {} Racine%
}{%
{\protect \APACyear {2008}}%
}]{%
Hayfield2008NonparametricPackage}
\APACinsertmetastar {%
Hayfield2008NonparametricPackage}%
\begin{APACrefauthors}%
Hayfield, T.%
\BCBT {}\ \BBA {} Racine, J\BPBI S.%
\end{APACrefauthors}%
\unskip\
\newblock
\APACrefYearMonthDay{2008}{}{}.
\newblock
{\BBOQ}\APACrefatitle {{Nonparametric econometrics: The np package}}
  {{Nonparametric econometrics: The np package}}.{\BBCQ}
\newblock
\APACjournalVolNumPages{Journal of Statistical Software}{27}{5}{}.
\PrintBackRefs{\CurrentBib}

\bibitem [\protect \citeauthoryear {%
Heiler%
\ \BBA {} Knaus%
}{%
Heiler%
\ \BBA {} Knaus%
}{%
{\protect \APACyear {2021}}%
}]{%
Heiler2021EffectTreatments}
\APACinsertmetastar {%
Heiler2021EffectTreatments}%
\begin{APACrefauthors}%
Heiler, P.%
\BCBT {}\ \BBA {} Knaus, M\BPBI C.%
\end{APACrefauthors}%
\unskip\
\newblock
\APACrefYearMonthDay{2021}{}{}.
\newblock
{\BBOQ}\APACrefatitle {{Effect or treatment heterogeneity? Policy evaluation
  with aggregated and disaggregated treatments}} {{Effect or treatment
  heterogeneity? Policy evaluation with aggregated and disaggregated
  treatments}}.{\BBCQ}
\newblock
\APACjournalVolNumPages{arXiv:2110.01427}{}{}{}.
\newblock
\begin{APACrefURL} \url{http://arxiv.org/abs/2110.01427} \end{APACrefURL}
\PrintBackRefs{\CurrentBib}

\bibitem [\protect \citeauthoryear {%
Hirano%
\ \BBA {} Porter%
}{%
Hirano%
\ \BBA {} Porter%
}{%
{\protect \APACyear {2009}}%
}]{%
Hirano2009AsymptoticsRules}
\APACinsertmetastar {%
Hirano2009AsymptoticsRules}%
\begin{APACrefauthors}%
Hirano, K.%
\BCBT {}\ \BBA {} Porter, J\BPBI R.%
\end{APACrefauthors}%
\unskip\
\newblock
\APACrefYearMonthDay{2009}{}{}.
\newblock
{\BBOQ}\APACrefatitle {{Asymptotics for statistical treatment rules}}
  {{Asymptotics for statistical treatment rules}}.{\BBCQ}
\newblock
\APACjournalVolNumPages{Econometrica}{77}{5}{1683--1701}.
\PrintBackRefs{\CurrentBib}

\bibitem [\protect \citeauthoryear {%
Holland%
}{%
Holland%
}{%
{\protect \APACyear {1986}}%
}]{%
Holland1986StatisticsInference}
\APACinsertmetastar {%
Holland1986StatisticsInference}%
\begin{APACrefauthors}%
Holland, P\BPBI W.%
\end{APACrefauthors}%
\unskip\
\newblock
\APACrefYearMonthDay{1986}{}{}.
\newblock
{\BBOQ}\APACrefatitle {{Statistics and causal inference}} {{Statistics and
  causal inference}}.{\BBCQ}
\newblock
\APACjournalVolNumPages{Journal of the American Statistical
  Association}{81}{396}{945--960}.
\PrintBackRefs{\CurrentBib}

\bibitem [\protect \citeauthoryear {%
Huber%
, Lechner%
\BCBL {}\ \BBA {} Mellace%
}{%
Huber%
\ \protect \BOthers {.}}{%
{\protect \APACyear {2017}}%
}]{%
Huber2017}
\APACinsertmetastar {%
Huber2017}%
\begin{APACrefauthors}%
Huber, M.%
, Lechner, M.%
\BCBL {}\ \BBA {} Mellace, G.%
\end{APACrefauthors}%
\unskip\
\newblock
\APACrefYearMonthDay{2017}{}{}.
\newblock
{\BBOQ}\APACrefatitle {{Why do tougher caseworkers increase employment? The
  role of program assignment as a causal mechanism}} {{Why do tougher
  caseworkers increase employment? The role of program assignment as a causal
  mechanism}}.{\BBCQ}
\newblock
\APACjournalVolNumPages{Review of Economics and Statistics}{99}{1}{180--183}.
\PrintBackRefs{\CurrentBib}

\bibitem [\protect \citeauthoryear {%
Imbens%
}{%
Imbens%
}{%
{\protect \APACyear {2000}}%
}]{%
Imbens2000TheFunctions}
\APACinsertmetastar {%
Imbens2000TheFunctions}%
\begin{APACrefauthors}%
Imbens, G\BPBI W.%
\end{APACrefauthors}%
\unskip\
\newblock
\APACrefYearMonthDay{2000}{}{}.
\newblock
{\BBOQ}\APACrefatitle {{The role of the propensity score in estimating
  dose-response functions}} {{The role of the propensity score in estimating
  dose-response functions}}.{\BBCQ}
\newblock
\APACjournalVolNumPages{Biometrika}{87}{3}{706--710}.
\PrintBackRefs{\CurrentBib}

\bibitem [\protect \citeauthoryear {%
Imbens%
}{%
Imbens%
}{%
{\protect \APACyear {2004}}%
}]{%
Imbens2004NonparametricReview}
\APACinsertmetastar {%
Imbens2004NonparametricReview}%
\begin{APACrefauthors}%
Imbens, G\BPBI W.%
\end{APACrefauthors}%
\unskip\
\newblock
\APACrefYearMonthDay{2004}{}{}.
\newblock
{\BBOQ}\APACrefatitle {{Nonparametric estimation of average treatment effects
  under exogeneity: A review}} {{Nonparametric estimation of average treatment
  effects under exogeneity: A review}}.{\BBCQ}
\newblock
\APACjournalVolNumPages{Review of Economics and Statistics}{86}{1}{4--29}.
\PrintBackRefs{\CurrentBib}

\bibitem [\protect \citeauthoryear {%
Imbens%
\ \BBA {} Rubin%
}{%
Imbens%
\ \BBA {} Rubin%
}{%
{\protect \APACyear {2015}}%
}]{%
Imbens2015CausalSciences}
\APACinsertmetastar {%
Imbens2015CausalSciences}%
\begin{APACrefauthors}%
Imbens, G\BPBI W.%
\BCBT {}\ \BBA {} Rubin, D\BPBI B.%
\end{APACrefauthors}%
\unskip\
\newblock
\APACrefYear{2015}.
\newblock
\APACrefbtitle {{Causal inference in statistics, social, and biomedical
  sciences}} {{Causal inference in statistics, social, and biomedical
  sciences}}.
\newblock
\APACaddressPublisher{}{Cambridge University Press}.
\PrintBackRefs{\CurrentBib}

\bibitem [\protect \citeauthoryear {%
Isphording%
}{%
Isphording%
}{%
{\protect \APACyear {2014}}%
}]{%
Isphording2014LanguageSuccess}
\APACinsertmetastar {%
Isphording2014LanguageSuccess}%
\begin{APACrefauthors}%
Isphording, I\BPBI E.%
\end{APACrefauthors}%
\unskip\
\newblock
\APACrefYearMonthDay{2014}{}{}.
\newblock
{\BBOQ}\APACrefatitle {{Language and labor market success}} {{Language and
  labor market success}}.{\BBCQ}
\newblock
\APACjournalVolNumPages{IZA Discussion Papers No. 8572}{}{8572}{}.
\PrintBackRefs{\CurrentBib}

\bibitem [\protect \citeauthoryear {%
Kallus%
}{%
Kallus%
}{%
{\protect \APACyear {2018}}%
}]{%
Kallus2018BalancedLearning}
\APACinsertmetastar {%
Kallus2018BalancedLearning}%
\begin{APACrefauthors}%
Kallus, N.%
\end{APACrefauthors}%
\unskip\
\newblock
\APACrefYearMonthDay{2018}{}{}.
\newblock
{\BBOQ}\APACrefatitle {{Balanced policy evaluation and learning}} {{Balanced
  policy evaluation and learning}}.{\BBCQ}
\newblock
\BIn{} \APACrefbtitle {Advances in Neural Information Processing Systems}
  {Advances in neural information processing systems}\ (\BPGS\ 8895--8906).
\PrintBackRefs{\CurrentBib}

\bibitem [\protect \citeauthoryear {%
Kallus%
, Mao%
\BCBL {}\ \BBA {} Uehara%
}{%
Kallus%
\ \protect \BOthers {.}}{%
{\protect \APACyear {2019}}%
}]{%
Kallus2019LocalizedBeyond}
\APACinsertmetastar {%
Kallus2019LocalizedBeyond}%
\begin{APACrefauthors}%
Kallus, N.%
, Mao, X.%
\BCBL {}\ \BBA {} Uehara, M.%
\end{APACrefauthors}%
\unskip\
\newblock
\APACrefYearMonthDay{2019}{}{}.
\newblock
{\BBOQ}\APACrefatitle {{Localized Debiased Machine Learning: Efficient
  Estimation of Quantile Treatment Effects, Conditional Value at Risk, and
  Beyond}} {{Localized Debiased Machine Learning: Efficient Estimation of
  Quantile Treatment Effects, Conditional Value at Risk, and Beyond}}.{\BBCQ}
\newblock
\APACjournalVolNumPages{arXiv:1912.12945}{}{}{}.
\newblock
\begin{APACrefURL} \url{http://arxiv.org/abs/1912.12945} \end{APACrefURL}
\PrintBackRefs{\CurrentBib}

\bibitem [\protect \citeauthoryear {%
Kang%
\ \BBA {} Schafer%
}{%
Kang%
\ \BBA {} Schafer%
}{%
{\protect \APACyear {2007}}%
}]{%
Kang2007}
\APACinsertmetastar {%
Kang2007}%
\begin{APACrefauthors}%
Kang, J\BPBI D\BPBI Y.%
\BCBT {}\ \BBA {} Schafer, J\BPBI L.%
\end{APACrefauthors}%
\unskip\
\newblock
\APACrefYearMonthDay{2007}{}{}.
\newblock
{\BBOQ}\APACrefatitle {{Demystifying double robustness: A comparison of
  alternative strategies for estimating a population mean from incomplete
  data}} {{Demystifying double robustness: A comparison of alternative
  strategies for estimating a population mean from incomplete data}}.{\BBCQ}
\newblock
\APACjournalVolNumPages{Statistical Science}{22}{4}{523--539}.
\PrintBackRefs{\CurrentBib}

\bibitem [\protect \citeauthoryear {%
Kennedy%
}{%
Kennedy%
}{%
{\protect \APACyear {2020}}%
}]{%
Kennedy2020OptimalEffects}
\APACinsertmetastar {%
Kennedy2020OptimalEffects}%
\begin{APACrefauthors}%
Kennedy, E\BPBI H.%
\end{APACrefauthors}%
\unskip\
\newblock
\APACrefYearMonthDay{2020}{}{}.
\newblock
{\BBOQ}\APACrefatitle {{Optimal doubly robust estimation of heterogeneous
  causal effects}} {{Optimal doubly robust estimation of heterogeneous causal
  effects}}.{\BBCQ}
\newblock
\APACjournalVolNumPages{arXiv:2004.14497}{}{}{}.
\newblock
\begin{APACrefURL} \url{http://arxiv.org/abs/2004.14497} \end{APACrefURL}
\PrintBackRefs{\CurrentBib}

\bibitem [\protect \citeauthoryear {%
Kennedy%
, Ma%
, McHugh%
\BCBL {}\ \BBA {} Small%
}{%
Kennedy%
\ \protect \BOthers {.}}{%
{\protect \APACyear {2017}}%
}]{%
Kennedy2017Non-parametricEffects}
\APACinsertmetastar {%
Kennedy2017Non-parametricEffects}%
\begin{APACrefauthors}%
Kennedy, E\BPBI H.%
, Ma, Z.%
, McHugh, M\BPBI D.%
\BCBL {}\ \BBA {} Small, D\BPBI S.%
\end{APACrefauthors}%
\unskip\
\newblock
\APACrefYearMonthDay{2017}{}{}.
\newblock
{\BBOQ}\APACrefatitle {{Non-parametric methods for doubly robust estimation of
  continuous treatment effects}} {{Non-parametric methods for doubly robust
  estimation of continuous treatment effects}}.{\BBCQ}
\newblock
\APACjournalVolNumPages{Journal of the Royal Statistical Society: Series B
  (Statistical Methodology)}{79}{}{1229--1245}.
\PrintBackRefs{\CurrentBib}

\bibitem [\protect \citeauthoryear {%
Kitagawa%
\ \BBA {} Tetenov%
}{%
Kitagawa%
\ \BBA {} Tetenov%
}{%
{\protect \APACyear {2018}}%
}]{%
Kitagawa2018WhoChoice}
\APACinsertmetastar {%
Kitagawa2018WhoChoice}%
\begin{APACrefauthors}%
Kitagawa, T.%
\BCBT {}\ \BBA {} Tetenov, A.%
\end{APACrefauthors}%
\unskip\
\newblock
\APACrefYearMonthDay{2018}{}{}.
\newblock
{\BBOQ}\APACrefatitle {{Who should be treated? Empirical welfare maximization
  methods for treatment choice}} {{Who should be treated? Empirical welfare
  maximization methods for treatment choice}}.{\BBCQ}
\newblock
\APACjournalVolNumPages{Econometrica}{86}{2}{591--616}.
\PrintBackRefs{\CurrentBib}

\bibitem [\protect \citeauthoryear {%
Knaus%
}{%
Knaus%
}{%
{\protect \APACyear {2021}}%
}]{%
Knaus2021ASkills}
\APACinsertmetastar {%
Knaus2021ASkills}%
\begin{APACrefauthors}%
Knaus, M\BPBI C.%
\end{APACrefauthors}%
\unskip\
\newblock
\APACrefYearMonthDay{2021}{}{}.
\newblock
{\BBOQ}\APACrefatitle {{A double machine learning approach to estimate the
  effects of musical practice on student’s skills}} {{A double machine
  learning approach to estimate the effects of musical practice on student’s
  skills}}.{\BBCQ}
\newblock
\APACjournalVolNumPages{Journal of the Royal Statistical Society. Series A:
  Statistics in Society}{184}{1}{282--300}.
\PrintBackRefs{\CurrentBib}

\bibitem [\protect \citeauthoryear {%
Knaus%
, Lechner%
\BCBL {}\ \BBA {} Strittmatter%
}{%
Knaus%
\ \protect \BOthers {.}}{%
{\protect \APACyear {2020}}%
}]{%
Knaus2020HeterogeneousApproach}
\APACinsertmetastar {%
Knaus2020HeterogeneousApproach}%
\begin{APACrefauthors}%
Knaus, M\BPBI C.%
, Lechner, M.%
\BCBL {}\ \BBA {} Strittmatter, A.%
\end{APACrefauthors}%
\unskip\
\newblock
\APACrefYearMonthDay{2020}{}{}.
\newblock
{\BBOQ}\APACrefatitle {{Heterogeneous employment effects of job search
  programmes: A machine learning approach}} {{Heterogeneous employment effects
  of job search programmes: A machine learning approach}}.{\BBCQ}
\newblock
\APACjournalVolNumPages{Journal of Human Resources}{0718-9615R}{}{published
  ahead of print 26 March 2020}.
\PrintBackRefs{\CurrentBib}

\bibitem [\protect \citeauthoryear {%
Knaus%
, Lechner%
\BCBL {}\ \BBA {} Strittmatter%
}{%
Knaus%
\ \protect \BOthers {.}}{%
{\protect \APACyear {2021}}%
}]{%
Knaus2021}
\APACinsertmetastar {%
Knaus2021}%
\begin{APACrefauthors}%
Knaus, M\BPBI C.%
, Lechner, M.%
\BCBL {}\ \BBA {} Strittmatter, A.%
\end{APACrefauthors}%
\unskip\
\newblock
\APACrefYearMonthDay{2021}{}{}.
\newblock
{\BBOQ}\APACrefatitle {{Machine Learning Estimation of Heterogeneous Causal
  Effects: Empirical Monte Carlo Evidence}} {{Machine Learning Estimation of
  Heterogeneous Causal Effects: Empirical Monte Carlo Evidence}}.{\BBCQ}
\newblock
\APACjournalVolNumPages{The Econometrics Journal}{24}{1}{134--161}.
\PrintBackRefs{\CurrentBib}

\bibitem [\protect \citeauthoryear {%
Knittel%
\ \BBA {} Stolper%
}{%
Knittel%
\ \BBA {} Stolper%
}{%
{\protect \APACyear {2019}}%
}]{%
Knittel2019UsingUse}
\APACinsertmetastar {%
Knittel2019UsingUse}%
\begin{APACrefauthors}%
Knittel, C\BPBI R.%
\BCBT {}\ \BBA {} Stolper, S.%
\end{APACrefauthors}%
\unskip\
\newblock
\APACrefYearMonthDay{2019}{}{}.
\newblock
{\BBOQ}\APACrefatitle {{Using machine learning to target treatment: The case of
  household energy use}} {{Using machine learning to target treatment: The case
  of household energy use}}.{\BBCQ}
\newblock
\APACjournalVolNumPages{NBER Working Paper No. 26531}{}{}{}.
\PrintBackRefs{\CurrentBib}

\bibitem [\protect \citeauthoryear {%
Kozbur%
}{%
Kozbur%
}{%
{\protect \APACyear {2020}}%
}]{%
Kozbur2020AnalysisSelection}
\APACinsertmetastar {%
Kozbur2020AnalysisSelection}%
\begin{APACrefauthors}%
Kozbur, D.%
\end{APACrefauthors}%
\unskip\
\newblock
\APACrefYearMonthDay{2020}{}{}.
\newblock
{\BBOQ}\APACrefatitle {{Analysis of testing-based forward model selection}}
  {{Analysis of testing-based forward model selection}}.{\BBCQ}
\newblock
\APACjournalVolNumPages{Econometrica}{88}{5}{2147--2173}.
\PrintBackRefs{\CurrentBib}

\bibitem [\protect \citeauthoryear {%
Kreif%
\ \BBA {} DiazOrdaz%
}{%
Kreif%
\ \BBA {} DiazOrdaz%
}{%
{\protect \APACyear {2019}}%
}]{%
Kreif2019MachineInference}
\APACinsertmetastar {%
Kreif2019MachineInference}%
\begin{APACrefauthors}%
Kreif, N.%
\BCBT {}\ \BBA {} DiazOrdaz, K.%
\end{APACrefauthors}%
\unskip\
\newblock
\APACrefYearMonthDay{2019}{}{}.
\newblock
{\BBOQ}\APACrefatitle {{Machine learning in policy evaluation: new tools for
  causal inference}} {{Machine learning in policy evaluation: new tools for
  causal inference}}.{\BBCQ}
\newblock
\APACjournalVolNumPages{arXiv:1903.00402}{}{}{}.
\newblock
\begin{APACrefURL} \url{http://arxiv.org/abs/1903.00402} \end{APACrefURL}
\PrintBackRefs{\CurrentBib}

\bibitem [\protect \citeauthoryear {%
K{\"{u}}nzel%
, Sekhon%
, Bickel%
\BCBL {}\ \BBA {} Yu%
}{%
K{\"{u}}nzel%
\ \protect \BOthers {.}}{%
{\protect \APACyear {2019}}%
}]{%
Kunzel2017}
\APACinsertmetastar {%
Kunzel2017}%
\begin{APACrefauthors}%
K{\"{u}}nzel, S\BPBI R.%
, Sekhon, J\BPBI S.%
, Bickel, P\BPBI J.%
\BCBL {}\ \BBA {} Yu, B.%
\end{APACrefauthors}%
\unskip\
\newblock
\APACrefYearMonthDay{2019}{}{}.
\newblock
{\BBOQ}\APACrefatitle {{Metalearners for estimating heterogeneous treatment
  effects using machine learning}} {{Metalearners for estimating heterogeneous
  treatment effects using machine learning}}.{\BBCQ}
\newblock
\APACjournalVolNumPages{Proceedings of the National Academy of
  Sciences}{116}{10}{4156--4165}.
\PrintBackRefs{\CurrentBib}

\bibitem [\protect \citeauthoryear {%
Lalive%
, van Ours%
\BCBL {}\ \BBA {} Zweim{\"{u}}ller%
}{%
Lalive%
\ \protect \BOthers {.}}{%
{\protect \APACyear {2008}}%
}]{%
Lalive2008TheUnemployment}
\APACinsertmetastar {%
Lalive2008TheUnemployment}%
\begin{APACrefauthors}%
Lalive, R.%
, van Ours, J.%
\BCBL {}\ \BBA {} Zweim{\"{u}}ller, J.%
\end{APACrefauthors}%
\unskip\
\newblock
\APACrefYearMonthDay{2008}{}{}.
\newblock
{\BBOQ}\APACrefatitle {{The impact of active labor market programs on the
  duration of unemployment}} {{The impact of active labor market programs on
  the duration of unemployment}}.{\BBCQ}
\newblock
\APACjournalVolNumPages{Economic Journal}{118}{525}{235--257}.
\PrintBackRefs{\CurrentBib}

\bibitem [\protect \citeauthoryear {%
Lechner%
}{%
Lechner%
}{%
{\protect \APACyear {1999}}%
}]{%
lechner1999earnings}
\APACinsertmetastar {%
lechner1999earnings}%
\begin{APACrefauthors}%
Lechner, M.%
\end{APACrefauthors}%
\unskip\
\newblock
\APACrefYearMonthDay{1999}{}{}.
\newblock
{\BBOQ}\APACrefatitle {{Earnings and employment effects of continuous
  gff-the-job training in East Germany after unification}} {{Earnings and
  employment effects of continuous gff-the-job training in East Germany after
  unification}}.{\BBCQ}
\newblock
\APACjournalVolNumPages{Journal of Business {\&} Economic
  Statistics}{17}{1}{74--90}.
\PrintBackRefs{\CurrentBib}

\bibitem [\protect \citeauthoryear {%
Lechner%
}{%
Lechner%
}{%
{\protect \APACyear {2001}}%
}]{%
Lechner2001}
\APACinsertmetastar {%
Lechner2001}%
\begin{APACrefauthors}%
Lechner, M.%
\end{APACrefauthors}%
\unskip\
\newblock
\APACrefYearMonthDay{2001}{}{}.
\newblock
{\BBOQ}\APACrefatitle {{Identification and estimation of causal effects of
  multiple treatments under the conditional independence assumption}}
  {{Identification and estimation of causal effects of multiple treatments
  under the conditional independence assumption}}.{\BBCQ}
\newblock
\BIn{} M.~Lechner\ \BBA {} E.~Pfeiffer\ (\BEDS), \APACrefbtitle {Econometric
  Evaluation of Labour Market Policies} {Econometric evaluation of labour
  market policies}\ (\BPGS\ 43--58).
\newblock
\APACaddressPublisher{Heidelberg}{Physica}.
\PrintBackRefs{\CurrentBib}

\bibitem [\protect \citeauthoryear {%
Lechner%
}{%
Lechner%
}{%
{\protect \APACyear {2018}}%
}]{%
Lechner2018}
\APACinsertmetastar {%
Lechner2018}%
\begin{APACrefauthors}%
Lechner, M.%
\end{APACrefauthors}%
\unskip\
\newblock
\APACrefYearMonthDay{2018}{}{}.
\newblock
{\BBOQ}\APACrefatitle {{Modified causal forests for estimating heterogeneous
  causal effects}} {{Modified causal forests for estimating heterogeneous
  causal effects}}.{\BBCQ}
\newblock
\APACjournalVolNumPages{arXiv:1812.09487}{}{}{}.
\newblock
\begin{APACrefURL} \url{https://arxiv.org/abs/1812.09487} \end{APACrefURL}
\PrintBackRefs{\CurrentBib}

\bibitem [\protect \citeauthoryear {%
Lechner%
\ \protect \BOthers {.}}{%
Lechner%
\ \protect \BOthers {.}}{%
{\protect \APACyear {2020}}%
}]{%
Lechner2020SwissDataset}
\APACinsertmetastar {%
Lechner2020SwissDataset}%
\begin{APACrefauthors}%
Lechner, M.%
, Knaus, M.%
, Huber, M.%
, Fr{\"{o}}lich, M.%
, Behncke, S.%
, Mellace, G.%
\BCBL {}\ \BBA {} Strittmatter, A.%
\end{APACrefauthors}%
\unskip\
\newblock
\APACrefYearMonthDay{2020}{}{}.
\newblock
\APACrefbtitle {{Swiss Active Labor Market Policy Evaluation [Dataset]}.}
  {{Swiss Active Labor Market Policy Evaluation [Dataset]}.}
\newblock
\APACaddressPublisher{}{Distributed by FORS, Lausanne}.
\newblock
\begin{APACrefURL} \url{https://doi.org/10.23662/FORS-DS-1203-1}
  \end{APACrefURL}
\PrintBackRefs{\CurrentBib}

\bibitem [\protect \citeauthoryear {%
Lechner%
\ \BBA {} Smith%
}{%
Lechner%
\ \BBA {} Smith%
}{%
{\protect \APACyear {2007}}%
}]{%
lechner2007value}
\APACinsertmetastar {%
lechner2007value}%
\begin{APACrefauthors}%
Lechner, M.%
\BCBT {}\ \BBA {} Smith, J.%
\end{APACrefauthors}%
\unskip\
\newblock
\APACrefYearMonthDay{2007}{}{}.
\newblock
{\BBOQ}\APACrefatitle {{What is the value added by caseworkers?}} {{What is the
  value added by caseworkers?}}{\BBCQ}
\newblock
\APACjournalVolNumPages{Labour Economics}{14}{2}{135--151}.
\PrintBackRefs{\CurrentBib}

\bibitem [\protect \citeauthoryear {%
Lunceford%
\ \BBA {} Davidian%
}{%
Lunceford%
\ \BBA {} Davidian%
}{%
{\protect \APACyear {2004}}%
}]{%
Lunceford2004StratificationStudy}
\APACinsertmetastar {%
Lunceford2004StratificationStudy}%
\begin{APACrefauthors}%
Lunceford, J\BPBI K.%
\BCBT {}\ \BBA {} Davidian, M.%
\end{APACrefauthors}%
\unskip\
\newblock
\APACrefYearMonthDay{2004}{}{}.
\newblock
{\BBOQ}\APACrefatitle {{Stratification and weighting via the propensity score
  in estimation of causal treatment effects: A comparative study}}
  {{Stratification and weighting via the propensity score in estimation of
  causal treatment effects: A comparative study}}.{\BBCQ}
\newblock
\APACjournalVolNumPages{Statistics in Medicine}{23}{19}{2937--2960}.
\PrintBackRefs{\CurrentBib}

\bibitem [\protect \citeauthoryear {%
Luo%
\ \BBA {} Spindler%
}{%
Luo%
\ \BBA {} Spindler%
}{%
{\protect \APACyear {2016}}%
}]{%
Luo2016High-DimensionalConvergence}
\APACinsertmetastar {%
Luo2016High-DimensionalConvergence}%
\begin{APACrefauthors}%
Luo, Y.%
\BCBT {}\ \BBA {} Spindler, M.%
\end{APACrefauthors}%
\unskip\
\newblock
\APACrefYearMonthDay{2016}{}{}.
\newblock
{\BBOQ}\APACrefatitle {{High-dimensional L2-boosting: Rate of Convergence}}
  {{High-dimensional L2-boosting: Rate of Convergence}}.{\BBCQ}
\newblock
\APACjournalVolNumPages{arXiv:1602.08927}{}{}{}.
\newblock
\begin{APACrefURL} \url{http://arxiv.org/abs/1602.08927} \end{APACrefURL}
\PrintBackRefs{\CurrentBib}

\bibitem [\protect \citeauthoryear {%
Manski%
}{%
Manski%
}{%
{\protect \APACyear {2004}}%
}]{%
Manski2004StatisticalPopulations}
\APACinsertmetastar {%
Manski2004StatisticalPopulations}%
\begin{APACrefauthors}%
Manski, C\BPBI F.%
\end{APACrefauthors}%
\unskip\
\newblock
\APACrefYearMonthDay{2004}{}{}.
\newblock
{\BBOQ}\APACrefatitle {{Statistical treatment rules for heterogeneous
  populations}} {{Statistical treatment rules for heterogeneous
  populations}}.{\BBCQ}
\newblock
\APACjournalVolNumPages{Econometrica}{72}{4}{1221--1246}.
\PrintBackRefs{\CurrentBib}

\bibitem [\protect \citeauthoryear {%
Nie%
\ \BBA {} Wager%
}{%
Nie%
\ \BBA {} Wager%
}{%
{\protect \APACyear {2021}}%
}]{%
Nie2021}
\APACinsertmetastar {%
Nie2021}%
\begin{APACrefauthors}%
Nie, X.%
\BCBT {}\ \BBA {} Wager, S.%
\end{APACrefauthors}%
\unskip\
\newblock
\APACrefYearMonthDay{2021}{}{}.
\newblock
{\BBOQ}\APACrefatitle {{Quasi-oracle estimation of heterogeneous treatment
  effects}} {{Quasi-oracle estimation of heterogeneous treatment
  effects}}.{\BBCQ}
\newblock
\APACjournalVolNumPages{Biometrika}{108}{2}{299--319}.
\PrintBackRefs{\CurrentBib}

\bibitem [\protect \citeauthoryear {%
Ning%
, Peng%
\BCBL {}\ \BBA {} Imai%
}{%
Ning%
\ \protect \BOthers {.}}{%
{\protect \APACyear {2020}}%
}]{%
Ning2020RobustScore}
\APACinsertmetastar {%
Ning2020RobustScore}%
\begin{APACrefauthors}%
Ning, Y.%
, Peng, S.%
\BCBL {}\ \BBA {} Imai, K.%
\end{APACrefauthors}%
\unskip\
\newblock
\APACrefYearMonthDay{2020}{}{}.
\newblock
{\BBOQ}\APACrefatitle {{Robust estimation of causal effects via
  high-dimensional covariate balancing propensity score}} {{Robust estimation
  of causal effects via high-dimensional covariate balancing propensity
  score}}.{\BBCQ}
\newblock
\APACjournalVolNumPages{Biometrika}{107}{3}{533--554}.
\PrintBackRefs{\CurrentBib}

\bibitem [\protect \citeauthoryear {%
Oprescu%
, Syrgkanis%
\BCBL {}\ \BBA {} Wu%
}{%
Oprescu%
\ \protect \BOthers {.}}{%
{\protect \APACyear {2019}}%
}]{%
Oprescu2019OrthogonalInference}
\APACinsertmetastar {%
Oprescu2019OrthogonalInference}%
\begin{APACrefauthors}%
Oprescu, M.%
, Syrgkanis, V.%
\BCBL {}\ \BBA {} Wu, Z\BPBI S.%
\end{APACrefauthors}%
\unskip\
\newblock
\APACrefYearMonthDay{2019}{}{}.
\newblock
{\BBOQ}\APACrefatitle {{Orthogonal random forest for causal inference}}
  {{Orthogonal random forest for causal inference}}.{\BBCQ}
\newblock
\APACjournalVolNumPages{36th International Conference on Machine Learning, ICML
  2019}{2019-June}{}{8655--8696}.
\PrintBackRefs{\CurrentBib}

\bibitem [\protect \citeauthoryear {%
Racine%
\ \BBA {} Nie%
}{%
Racine%
\ \BBA {} Nie%
}{%
{\protect \APACyear {2021}}%
}]{%
Racine2021Crs:Splines}
\APACinsertmetastar {%
Racine2021Crs:Splines}%
\begin{APACrefauthors}%
Racine, J\BPBI S.%
\BCBT {}\ \BBA {} Nie, Z.%
\end{APACrefauthors}%
\unskip\
\newblock
\APACrefYearMonthDay{2021}{}{}.
\newblock
\APACrefbtitle {{crs: Categorical Regression Splines}.} {{crs: Categorical
  Regression Splines}.}
\newblock
\begin{APACrefURL} \url{https://cran.r-project.org/package=crs}
  \end{APACrefURL}
\PrintBackRefs{\CurrentBib}

\bibitem [\protect \citeauthoryear {%
Robins%
, Rotnitzky%
\BCBL {}\ \BBA {} Zhao%
}{%
Robins%
\ \protect \BOthers {.}}{%
{\protect \APACyear {1994}}%
}]{%
Robins1994}
\APACinsertmetastar {%
Robins1994}%
\begin{APACrefauthors}%
Robins, J\BPBI M.%
, Rotnitzky, A.%
\BCBL {}\ \BBA {} Zhao, L\BPBI P.%
\end{APACrefauthors}%
\unskip\
\newblock
\APACrefYearMonthDay{1994}{}{}.
\newblock
{\BBOQ}\APACrefatitle {{Estimation of regression coefficients when some
  regressors are not always observed}} {{Estimation of regression coefficients
  when some regressors are not always observed}}.{\BBCQ}
\newblock
\APACjournalVolNumPages{Journal of the American Statistical
  Association}{89}{427}{846--866}.
\PrintBackRefs{\CurrentBib}

\bibitem [\protect \citeauthoryear {%
Robins%
, Rotnitzky%
\BCBL {}\ \BBA {} Zhao%
}{%
Robins%
\ \protect \BOthers {.}}{%
{\protect \APACyear {1995}}%
}]{%
Robins1995AnalysisData}
\APACinsertmetastar {%
Robins1995AnalysisData}%
\begin{APACrefauthors}%
Robins, J\BPBI M.%
, Rotnitzky, A.%
\BCBL {}\ \BBA {} Zhao, L\BPBI P.%
\end{APACrefauthors}%
\unskip\
\newblock
\APACrefYearMonthDay{1995}{}{}.
\newblock
{\BBOQ}\APACrefatitle {{Analysis of semiparametric regression models for
  repeated outcomes in the presence of missing data}} {{Analysis of
  semiparametric regression models for repeated outcomes in the presence of
  missing data}}.{\BBCQ}
\newblock
\APACjournalVolNumPages{Journal of the American Statistical
  Association}{90}{429}{106--121}.
\PrintBackRefs{\CurrentBib}

\bibitem [\protect \citeauthoryear {%
Robins%
, Sued%
, Lei-Gomez%
\BCBL {}\ \BBA {} Rotnitzky%
}{%
Robins%
\ \protect \BOthers {.}}{%
{\protect \APACyear {2007}}%
}]{%
Robins2007Comment:Variable}
\APACinsertmetastar {%
Robins2007Comment:Variable}%
\begin{APACrefauthors}%
Robins, J\BPBI M.%
, Sued, M.%
, Lei-Gomez, Q.%
\BCBL {}\ \BBA {} Rotnitzky, A.%
\end{APACrefauthors}%
\unskip\
\newblock
\APACrefYearMonthDay{2007}{}{}.
\newblock
{\BBOQ}\APACrefatitle {{Comment: Performance of double-robust estimators when
  "inverse probability" weights are highly variable}} {{Comment: Performance of
  double-robust estimators when "inverse probability" weights are highly
  variable}}.{\BBCQ}
\newblock
\APACjournalVolNumPages{Statistical Science}{22}{4}{544--559}.
\PrintBackRefs{\CurrentBib}

\bibitem [\protect \citeauthoryear {%
Rubin%
}{%
Rubin%
}{%
{\protect \APACyear {1974}}%
}]{%
Rubin1974}
\APACinsertmetastar {%
Rubin1974}%
\begin{APACrefauthors}%
Rubin, D\BPBI B.%
\end{APACrefauthors}%
\unskip\
\newblock
\APACrefYearMonthDay{1974}{}{}.
\newblock
{\BBOQ}\APACrefatitle {{Estimating causal effects of treatments in randomized
  and nonrandomized studies.}} {{Estimating causal effects of treatments in
  randomized and nonrandomized studies.}}{\BBCQ}
\newblock
\APACjournalVolNumPages{Journal of Educational Psychology}{66}{5}{688--701}.
\PrintBackRefs{\CurrentBib}

\bibitem [\protect \citeauthoryear {%
Semenova%
\ \BBA {} Chernozhukov%
}{%
Semenova%
\ \BBA {} Chernozhukov%
}{%
{\protect \APACyear {2021}}%
}]{%
Semenova2021DebiasedFunctions}
\APACinsertmetastar {%
Semenova2021DebiasedFunctions}%
\begin{APACrefauthors}%
Semenova, V.%
\BCBT {}\ \BBA {} Chernozhukov, V.%
\end{APACrefauthors}%
\unskip\
\newblock
\APACrefYearMonthDay{2021}{}{}.
\newblock
{\BBOQ}\APACrefatitle {{Debiased machine learning of conditional average
  treatment effects and other causal functions}} {{Debiased machine learning of
  conditional average treatment effects and other causal functions}}.{\BBCQ}
\newblock
\APACjournalVolNumPages{The Econometrics Journal}{24}{2}{264--289}.
\PrintBackRefs{\CurrentBib}

\bibitem [\protect \citeauthoryear {%
Smucler%
, Rotnitzky%
\BCBL {}\ \BBA {} Robins%
}{%
Smucler%
\ \protect \BOthers {.}}{%
{\protect \APACyear {2019}}%
}]{%
Smucler2019AContrasts}
\APACinsertmetastar {%
Smucler2019AContrasts}%
\begin{APACrefauthors}%
Smucler, E.%
, Rotnitzky, A.%
\BCBL {}\ \BBA {} Robins, J\BPBI M.%
\end{APACrefauthors}%
\unskip\
\newblock
\APACrefYearMonthDay{2019}{}{}.
\newblock
{\BBOQ}\APACrefatitle {{A unifying approach for doubly-robust L1 regularized
  estimation of causal contrasts}} {{A unifying approach for doubly-robust L1
  regularized estimation of causal contrasts}}.{\BBCQ}
\newblock
\APACjournalVolNumPages{arXiv:1904.03737}{}{}{}.
\newblock
\begin{APACrefURL} \url{http://arxiv.org/abs/1904.03737} \end{APACrefURL}
\PrintBackRefs{\CurrentBib}

\bibitem [\protect \citeauthoryear {%
Stoye%
}{%
Stoye%
}{%
{\protect \APACyear {2009}}%
}]{%
Stoye2009MinimaxSamples}
\APACinsertmetastar {%
Stoye2009MinimaxSamples}%
\begin{APACrefauthors}%
Stoye, J.%
\end{APACrefauthors}%
\unskip\
\newblock
\APACrefYearMonthDay{2009}{}{}.
\newblock
{\BBOQ}\APACrefatitle {{Minimax regret treatment choice with finite samples}}
  {{Minimax regret treatment choice with finite samples}}.{\BBCQ}
\newblock
\APACjournalVolNumPages{Journal of Econometrics}{151}{1}{70--81}.
\PrintBackRefs{\CurrentBib}

\bibitem [\protect \citeauthoryear {%
Stoye%
}{%
Stoye%
}{%
{\protect \APACyear {2012}}%
}]{%
Stoye2012MinimaxExperiments}
\APACinsertmetastar {%
Stoye2012MinimaxExperiments}%
\begin{APACrefauthors}%
Stoye, J.%
\end{APACrefauthors}%
\unskip\
\newblock
\APACrefYearMonthDay{2012}{}{}.
\newblock
{\BBOQ}\APACrefatitle {{Minimax regret treatment choice with covariates or with
  limited validity of experiments}} {{Minimax regret treatment choice with
  covariates or with limited validity of experiments}}.{\BBCQ}
\newblock
\APACjournalVolNumPages{Journal of Econometrics}{166}{1}{138--156}.
\PrintBackRefs{\CurrentBib}

\bibitem [\protect \citeauthoryear {%
Strittmatter%
}{%
Strittmatter%
}{%
{\protect \APACyear {2018}}%
}]{%
Strittmatter2018WhatEvaluation}
\APACinsertmetastar {%
Strittmatter2018WhatEvaluation}%
\begin{APACrefauthors}%
Strittmatter, A.%
\end{APACrefauthors}%
\unskip\
\newblock
\APACrefYearMonthDay{2018}{}{}.
\newblock
{\BBOQ}\APACrefatitle {{What is the value added by using causal machine
  learning methods in a welfare experiment evaluation?}} {{What is the value
  added by using causal machine learning methods in a welfare experiment
  evaluation?}}{\BBCQ}
\newblock
\APACjournalVolNumPages{arXiv:1812.06533}{}{}{}.
\newblock
\begin{APACrefURL} \url{http://arxiv.org/abs/1812.06533} \end{APACrefURL}
\PrintBackRefs{\CurrentBib}

\bibitem [\protect \citeauthoryear {%
Sverdrup%
, Kanodia%
, Zhou%
, Athey%
\BCBL {}\ \BBA {} Wager%
}{%
Sverdrup%
\ \protect \BOthers {.}}{%
{\protect \APACyear {2020}}%
}]{%
Sverdrup2020Policytree:Trees}
\APACinsertmetastar {%
Sverdrup2020Policytree:Trees}%
\begin{APACrefauthors}%
Sverdrup, E.%
, Kanodia, A.%
, Zhou, Z.%
, Athey, S.%
\BCBL {}\ \BBA {} Wager, S.%
\end{APACrefauthors}%
\unskip\
\newblock
\APACrefYearMonthDay{2020}{}{}.
\newblock
{\BBOQ}\APACrefatitle {{policytree: Policy learning via doubly robust empirical
  welfare maximization over trees}} {{policytree: Policy learning via doubly
  robust empirical welfare maximization over trees}}.{\BBCQ}
\newblock
\APACjournalVolNumPages{Journal of Open Source Software}{5}{50}{2232}.
\PrintBackRefs{\CurrentBib}

\bibitem [\protect \citeauthoryear {%
Syrgkanis%
\ \BBA {} Zampetakis%
}{%
Syrgkanis%
\ \BBA {} Zampetakis%
}{%
{\protect \APACyear {2020}}%
}]{%
Syrgkanis2020EstimationDimensions}
\APACinsertmetastar {%
Syrgkanis2020EstimationDimensions}%
\begin{APACrefauthors}%
Syrgkanis, V.%
\BCBT {}\ \BBA {} Zampetakis, M.%
\end{APACrefauthors}%
\unskip\
\newblock
\APACrefYearMonthDay{2020}{}{}.
\newblock
{\BBOQ}\APACrefatitle {{Estimation and inference with trees and forests in high
  dimensions}} {{Estimation and inference with trees and forests in high
  dimensions}}.{\BBCQ}
\newblock
\BIn{} \APACrefbtitle {Conference on Learning Theory} {Conference on learning
  theory}\ (\BPGS\ 3453--3454).
\newblock
\APACaddressPublisher{}{PMLR}.
\newblock
\begin{APACrefURL} \url{http://arxiv.org/abs/2007.03210} \end{APACrefURL}
\PrintBackRefs{\CurrentBib}

\bibitem [\protect \citeauthoryear {%
Tan%
}{%
Tan%
}{%
{\protect \APACyear {2020}}%
}]{%
Tan2020Model-assistedData}
\APACinsertmetastar {%
Tan2020Model-assistedData}%
\begin{APACrefauthors}%
Tan, Z.%
\end{APACrefauthors}%
\unskip\
\newblock
\APACrefYearMonthDay{2020}{}{}.
\newblock
{\BBOQ}\APACrefatitle {{Model-assisted inference for treatment effects using
  regularized calibrated estimation with high-dimensional data}}
  {{Model-assisted inference for treatment effects using regularized calibrated
  estimation with high-dimensional data}}.{\BBCQ}
\newblock
\APACjournalVolNumPages{The Annals of Statistics}{48}{2}{811--837}.
\PrintBackRefs{\CurrentBib}

\bibitem [\protect \citeauthoryear {%
Tian%
, Alizadeh%
, Gentles%
\BCBL {}\ \BBA {} Tibshirani%
}{%
Tian%
\ \protect \BOthers {.}}{%
{\protect \APACyear {2014}}%
}]{%
Tian2014}
\APACinsertmetastar {%
Tian2014}%
\begin{APACrefauthors}%
Tian, L.%
, Alizadeh, A\BPBI A.%
, Gentles, A\BPBI J.%
\BCBL {}\ \BBA {} Tibshirani, R.%
\end{APACrefauthors}%
\unskip\
\newblock
\APACrefYearMonthDay{2014}{}{}.
\newblock
{\BBOQ}\APACrefatitle {{A simple method for estimating interactions between a
  treatment and a large number of covariates}} {{A simple method for estimating
  interactions between a treatment and a large number of covariates}}.{\BBCQ}
\newblock
\APACjournalVolNumPages{Journal of the American Statistical
  Association}{109}{508}{1517--1532}.
\PrintBackRefs{\CurrentBib}

\bibitem [\protect \citeauthoryear {%
van~der Laan%
\ \BBA {} Rubin%
}{%
van~der Laan%
\ \BBA {} Rubin%
}{%
{\protect \APACyear {2006}}%
}]{%
vanderLaan2006TargetedLearning}
\APACinsertmetastar {%
vanderLaan2006TargetedLearning}%
\begin{APACrefauthors}%
van~der Laan, M\BPBI J.%
\BCBT {}\ \BBA {} Rubin, D.%
\end{APACrefauthors}%
\unskip\
\newblock
\APACrefYearMonthDay{2006}{}{}.
\newblock
{\BBOQ}\APACrefatitle {{Targeted maximum likelihood learning}} {{Targeted
  maximum likelihood learning}}.{\BBCQ}
\newblock
\APACjournalVolNumPages{International Journal of Biostatistics}{2}{1}{}.
\PrintBackRefs{\CurrentBib}

\bibitem [\protect \citeauthoryear {%
Wager%
\ \BBA {} Athey%
}{%
Wager%
\ \BBA {} Athey%
}{%
{\protect \APACyear {2018}}%
}]{%
Wager2017}
\APACinsertmetastar {%
Wager2017}%
\begin{APACrefauthors}%
Wager, S.%
\BCBT {}\ \BBA {} Athey, S.%
\end{APACrefauthors}%
\unskip\
\newblock
\APACrefYearMonthDay{2018}{}{}.
\newblock
{\BBOQ}\APACrefatitle {{Estimation and inference of heterogeneous treatment
  effects using random forests}} {{Estimation and inference of heterogeneous
  treatment effects using random forests}}.{\BBCQ}
\newblock
\APACjournalVolNumPages{Journal of the American Statistical
  Association}{113}{523}{1228--1242}.
\PrintBackRefs{\CurrentBib}

\bibitem [\protect \citeauthoryear {%
Wager%
\ \BBA {} Walther%
}{%
Wager%
\ \BBA {} Walther%
}{%
{\protect \APACyear {2015}}%
}]{%
Wager2015AdaptiveForests}
\APACinsertmetastar {%
Wager2015AdaptiveForests}%
\begin{APACrefauthors}%
Wager, S.%
\BCBT {}\ \BBA {} Walther, G.%
\end{APACrefauthors}%
\unskip\
\newblock
\APACrefYearMonthDay{2015}{}{}.
\newblock
{\BBOQ}\APACrefatitle {{Adaptive concentration of regression trees, with
  application to random forests}} {{Adaptive concentration of regression trees,
  with application to random forests}}.{\BBCQ}
\newblock
\APACjournalVolNumPages{arXiv:1503.06388}{}{}{}.
\newblock
\begin{APACrefURL} \url{http://arxiv.org/abs/1503.06388} \end{APACrefURL}
\PrintBackRefs{\CurrentBib}

\bibitem [\protect \citeauthoryear {%
Wunsch%
}{%
Wunsch%
}{%
{\protect \APACyear {2016}}%
}]{%
Wunsch2016HowWorkers}
\APACinsertmetastar {%
Wunsch2016HowWorkers}%
\begin{APACrefauthors}%
Wunsch, C.%
\end{APACrefauthors}%
\unskip\
\newblock
\APACrefYearMonthDay{2016}{}{}.
\newblock
{\BBOQ}\APACrefatitle {{How to minimize lock-in effects of programs for
  unemployed workers}} {{How to minimize lock-in effects of programs for
  unemployed workers}}.{\BBCQ}
\newblock
\APACjournalVolNumPages{IZA World of Labor}{}{}{}.
\PrintBackRefs{\CurrentBib}

\bibitem [\protect \citeauthoryear {%
Zhou%
, Athey%
\BCBL {}\ \BBA {} Wager%
}{%
Zhou%
\ \protect \BOthers {.}}{%
{\protect \APACyear {2018}}%
}]{%
Zhou2018OfflineOptimization}
\APACinsertmetastar {%
Zhou2018OfflineOptimization}%
\begin{APACrefauthors}%
Zhou, Z.%
, Athey, S.%
\BCBL {}\ \BBA {} Wager, S.%
\end{APACrefauthors}%
\unskip\
\newblock
\APACrefYearMonthDay{2018}{}{}.
\newblock
{\BBOQ}\APACrefatitle {{Offline multi-action policy learning: Generalization
  and optimization}} {{Offline multi-action policy learning: Generalization and
  optimization}}.{\BBCQ}
\newblock
\APACjournalVolNumPages{arXiv:1810.04778}{}{}{}.
\newblock
\begin{APACrefURL} \url{http://arxiv.org/abs/1810.04778} \end{APACrefURL}
\PrintBackRefs{\CurrentBib}

\bibitem [\protect \citeauthoryear {%
Zimmert%
\ \BBA {} Lechner%
}{%
Zimmert%
\ \BBA {} Lechner%
}{%
{\protect \APACyear {2019}}%
}]{%
Zimmert2019NonparametricConfounding}
\APACinsertmetastar {%
Zimmert2019NonparametricConfounding}%
\begin{APACrefauthors}%
Zimmert, M.%
\BCBT {}\ \BBA {} Lechner, M.%
\end{APACrefauthors}%
\unskip\
\newblock
\APACrefYearMonthDay{2019}{}{}.
\newblock
{\BBOQ}\APACrefatitle {{Nonparametric estimation of causal heterogeneity under
  high-dimensional confounding}} {{Nonparametric estimation of causal
  heterogeneity under high-dimensional confounding}}.{\BBCQ}
\newblock
\APACjournalVolNumPages{arXiv:1908.08779}{}{}{}.
\newblock
\begin{APACrefURL} \url{http://arxiv.org/abs/1908.08779} \end{APACrefURL}
\PrintBackRefs{\CurrentBib}

\end{thebibliography}

\newpage
\begin{appendices}
\counterwithin{figure}{section}
\counterwithin{table}{section}

\huge \noindent \textbf{Appendices}


\normalsize

\section{Identification and Neyman orthogonality} \label{sec:app-a}
\subsection{Doubly robust identification}  \label{sec:app-ident}

To revisit identification and identification double robustness of Equation 2.3 under Assumption 2.1, rewrite the conditional average potential outcome in the following way, where $\mu_w(x) = E[Y_i(w) \mid X_i=x]$, $\mu(w,x) = E[Y_i \mid W_i=w, X_i=x]$ and $e_w(x) = E \left[ D_i(w) \middle|  X_i = x \right] = P \left[ W_i = w \middle|  X_i = x \right] \overset{1b}{>} 0$:
\begin{align}
    \mu_w(x) & = E \left[ \mu(w,x) + \dfrac{D_i(w) (Y_i - \mu(w,x))}{e_w(x)} \middle|  X_i = x \right] \nonumber \\
    & = E \left[Y_i(w) - Y_i(w) + \mu(w,x) + \dfrac{D_i(w) (Y_i - \mu(w,x))}{e_w(x)} \middle|  X_i = x \right] \nonumber\\
    & \overset{\mathrm{1c}}{=} E \left[Y_i(w) - Y_i(w) + \mu(w,x) + \dfrac{D_i(w) (Y_i(w) - \mu(w,x))}{e_w(x)} \middle|  X_i = x \right] \nonumber\\
    & =  E \left[Y_i(w) \mid  X_i = x \right] + E \left[ \left(Y_i(w) - \mu(w,x) \right) \left(\dfrac{D_i(w) - e_w(x)}{e_w(x)} \right) \middle|  X_i = x \right] \nonumber\\
    & = \mu_w(x) + E \left[ \left(Y_i(w) - \mu(w,x) \right) \left(\dfrac{D_i(w) - e_w(x)}{e_w(x)} \right) \middle|  X_i = x \right]  \label{eq:id3-app}
\end{align}

The conditional average potential outcome is thus identified if the second part of Equation \ref{eq:id3-app} equals zero. This happens under three scenarios:

1. Correct propensity score and correct outcome regression:
\begin{align*}
   E & \left[ \left(Y_i(w) - \mu(w,x) \right) \left(\dfrac{D_i(w) - e_w(x)}{e_w(x)} \right) \middle|  X_i = x \right] \\
   & \overset{\mathrm{1a}}{=}   E \left[ \left(Y_i(w) - \mu(w,x) \right)  \middle|  X_i = x \right] E \left[\left(\dfrac{D_i(w) - e_w(x)}{e_w(x)} \right) \middle|  X_i = x \right] \\
    & =    \left( E \left[Y_i(w) \mid  X_i = x \right] - \mu(w,x) \right)   \left(\dfrac{E \left[ D_i(w) \mid  X_i = x \right] - e_w(x)}{e_w(x)} \right)  \\
    & =    \left( \mu_w(x) - \mu(w,x) \right)   \left(\dfrac{e_w(x) - e_w(x)}{e_w(x)} \right)  \\
    & \overset{\mathrm{1a}}{=}  \underbrace{\left( \mu_w(x) - \mu_w(x) \right)}_{=0}   
    \underbrace{\left(\dfrac{e_w(x) - e_w(x)}{e_w(x)} \right)}_{=0} = 0
\end{align*}

2. Correct propensity score but instead of correct outcome regression $\mu(w,x)$, use some function $g(x)$:
\begin{align*}
   E & \left[ \left(Y_i(w) - g(x) \right) \left(\dfrac{D_i(w) - e_w(x)}{e_w(x)} \right) \middle|  X_i = x \right] \\
   & \overset{\mathrm{1a}}{=}   E \left[ \left(Y_i(w) - g(x) \right)  \middle|  X_i = x \right] E \left[\left(\dfrac{D_i(w) - e_w(x)}{e_w(x)} \right) \middle|  X_i = x \right] \\
    & =    \left( E \left[Y_i(w) \mid  X_i = x \right] - g(x) \right)   \left(\dfrac{E \left[ D_i(w) \mid  X_i = x \right] - e_w(x)}{e_w(x)} \right)  \\
    & =    \left( \mu_w(x) - g(x) \right)   
    \underbrace{\left(\dfrac{e_w(x) - e_w(x)}{e_w(x)} \right)}_{=0}  = 0
\end{align*}

3. Correct outcome regression but instead of correct propensity score $e_w(x)$, use some function $h(x)$:
\begin{align*}
   E & \left[ \left(Y_i(w) - \mu(w,x) \right) \left(\dfrac{D_i(w) - h(x)}{h(x)} \right) \middle|  X_i = x \right] \\
   & \overset{\mathrm{1a}}{=}   E \left[ \left(Y_i(w) - \mu(w,x) \right)  \middle|  X_i = x \right] E \left[\left(\dfrac{D_i(w) - h(x)}{h(x)} \right) \middle|  X_i = x \right] \\
    & =    \left( E \left[Y_i(w) \mid  X_i = x \right] - \mu(w,x) \right)   \left(\dfrac{E \left[ D_i(w) \mid  X_i = x \right] - h(x)}{h(x))} \right)  \\
    & =    \left( \mu_w(x) - \mu(w,x) \right)   \left(\dfrac{e_w(x) - h(x)}{h(x)} \right)  \\
    & \overset{\mathrm{1a}}{=}  \underbrace{\left( \mu_w(x) - \mu_w(x) \right)}_{=0}   
    \left(\dfrac{e_w(x) - h(x)}{h(x)} \right) = 0 \\
\end{align*}

\subsection{Neyman orthogonality}  \label{sec:app-neyman}

We revisit Neyman orthogonality of the APO score as the other scores follow similarly. The score defining the APO is
\begin{align}
    &E \Bigg[ \underbrace{\mu(w,X_i) + \frac{D_i(w) (Y_i - \mu(w,X_i))}{e_w(X_i)} - \gamma_w }_{\psi(Y_i,W_i,\mu(w,X_i),e_w(X_i))} \Bigg]= 0 \\
    \Rightarrow~~& \gamma_w = E \Bigg[ \mu(w,X_i) + \frac{D_i(w) (Y_i - \mu(w,X_i))}{e_w(X_i)} \Bigg]
\end{align}

Neyman-orthogonality of a score $\psi(\cdot)$ means that the Gateaux derivative with respect to the nuisance parameters is zero in expectation at the true nuisance parameters (NP). In our case this means that
\begin{align}\label{eq:neyman}
\partial_{r} E[\psi(Y_i,W_i,\mu + r(\tilde{\mu} - \mu),e + r(\tilde{e} - e))|X_i=x]|_{r = 0} = 0
\end{align}
where we suppress the dependencies of NPs and denote by, e.g., $\tilde{\mu}$ a value of the outcome nuisance that is different to the true value $\mu$. We can show that Equation \ref{eq:neyman} holds with the following steps:

First, add perturbations to the true nuisance parameters in the score 
\begin{align*}
\psi&(Y_i,W_i,\mu + r(\tilde{\mu} - \mu),e + r(\tilde{e} - e)) \\
&= (\mu + r(\tilde{\mu} - \mu)) + \frac{D_i(w) Y_i}{e + r(\tilde{e} - e)}  - \frac{D_i(w)(\mu + r(\tilde{\mu} - \mu))}{e + r(\tilde{e} - e)} - \gamma_w
\end{align*}

Second, take the conditional expectation
\begin{align*}
E&\left[\psi(Y_i,W_i,\mu + r(\tilde{\mu} - \mu),e + r(\tilde{e} - e)) \middle| X_i = x \right] \\
&= E\left[(\mu + r(\tilde{\mu} - \mu)) + \frac{D_i(w) Y_i}{e + r(\tilde{e} - e)}  - \frac{D_i(w)(\mu + r(\tilde{\mu} - \mu))}{e  + r(\tilde{e} - e)}  - \hat{\gamma}_w \middle| X_i = x \right] \\
&= (\mu + r(\tilde{\mu} - \mu)) + \frac{e \mu}{e + r(\tilde{e} - e)}  - \frac{e(\mu + r(\tilde{\mu} - \mu))}{e  + r(\tilde{e} - e)} - \gamma_w
\end{align*}
where we use that $E[D_i(w) Y_i \mid X_i = x] = E[D_i(w) \sum_w D_i(w) Y_i(w) \mid X_i = x] = E[D_i(w) Y_i(w) \mid X_i = x] \overset{1a}{=} e \mu$.

Third, take the derivative with respect to $r$
\begin{align*}
\partial_{r} E&\left[\psi(Y_i,W_i,\mu + r(\tilde{\mu} - \mu),e + r(\tilde{e} - e)) \middle| X_i = x \right]  \\
&= (\tilde{\mu} - \mu) - \frac{e \mu (\tilde{e} - e)}{(e + r(\tilde{e} - e))^2}  - \frac{e(\tilde{\mu} - \mu)(e + r(\tilde{e} - e)) - e(\mu + r(\tilde{\mu} - \mu)) (\tilde{e} - e)}{(e + r(\tilde{e} - e))^2}
\end{align*}

Finally, evaluate at the true nuisance values, i.e. set $r=0$ to show that \ref{eq:neyman} holds
\begin{align*}
\partial_{r} & E[\psi(Y_i,W_i,\mu + r(\tilde{\mu} - \mu),e + r(\tilde{e} - e))|X_i=x]|_{r = 0} \\ 
&= (\tilde{\mu} - \mu) - \frac{e \mu (\tilde{e} - e)}{(e + 0(\tilde{e} - e))^2}  - \frac{e(\tilde{\mu} - \mu)(e + 0(\tilde{e} - e)) - e(\mu + 0(\tilde{\mu} - \mu)) (\tilde{e} - e)}{(e + 0(\tilde{e} - e))^2} \\
&= (\tilde{\mu} - \mu) - \frac{e \mu (\tilde{e} - e)}{e^2}  - \frac{e(\tilde{\mu} - \mu)e - e\mu (\tilde{e} - e)}{e^2} \\
&= (\tilde{\mu} - \mu) - \frac{e \mu (\tilde{e} - e)}{e^2}  - \frac{e^2}{e^2} (\tilde{\mu} - \mu) + \frac{e\mu (\tilde{e} - e)}{e^2} \\
&= 0 
\end{align*}

\newpage

\section{DR- and NDR-learner}  \label{sec:app-drl}

This Appendix describes the algorithms that are applied to estimate out-of-sample IATEs using the DR- and NDR-learner. It mostly follows Algorithm 1 of \citeA{Kennedy2020OptimalEffects} and adapts it to the situation that we are interested in estimating IATEs for all observations without using them in the estimation step.

\begin{algorithm}[DR-learner] \label{alg:drlearner}
Let $(S_1^N,S_2^N,S_3^N,S_4^N)$ denote four independent samples of $N$ observations of $O_i=(X_i,W_i,Y_i)$. 
\begin{enumerate}
\item[Step 1.] Nuisance training:
\begin{enumerate}
\item Construct a model $\hat{e}_w(x)$ of the propensity scores $e_w(x)$ using $S_1^N$. 
\item Construct a model $(\hat{\mu}(w,x),\hat{\mu}(w',x))$ of the regression functions $(\mu(w,x), \mu(w',x))$ using $S_2^N$. 
\end{enumerate}
\item[Step 2.]  Pseudo-outcome regression: Construct the pseudo-outcome for every observation $i$ in subsample $S_3^N$ using the models of step 1
\begin{equation*}
\hat{\Delta}_{i,w,w'} = \hat{\mu}(w,X_i) - \hat{\mu}(w',X_i) + \dfrac{ D_i(w) }{\hat{e}_w(X_i)}  \Tilde{Y_i}(w,X_i)  - \dfrac{D_i(w') }{\hat{e}_{w'}(X_i)} \Tilde{Y_i}(w',X_i),
\end{equation*}
regress it on covariates $X_i$ in $S_3^N$, and use the model to predict IATEs in $S_4^N$, $\hat{\tau}_{w,w'}^{4,1}(x)$.
\item[Step 3.]  Cross-fitting: Repeat steps 1--2 twice, first using $S_2^N$ for the propensity score, $S_3^N$ for the outcome regression and $S_1^N$ as subsample to obtain IATE predictions in $S_4^N$ $\hat{\tau}_{w,w'}^{4,2}(x)$, and then using $S_3^N$ for the propensity score, $S_1^N$ for the outcome regression and $S_2^N$ as subsample to obtain IATE predictions in $S_4^N$, $\hat{\tau}_{w,w'}^{4,3}(x)$. 
\item[Step 4.]  Prediction: Predict IATEs in $S_4^N$ as the average of the three predictions $\hat{\tau}_{w,w'}^{drl}(x) = 1/3 \hat{\tau}_{w,w'}^{4,1}(x) + 1/3 \hat{\tau}_{w,w'}^{4,2}(x) + 1/3 \hat{\tau}_{w,w'}^{4,3}(x)$.
\item[Step 5.]  Iteration: Repeat steps 1--4 three times. First, with $S_1^N$, $S_2^N$ and $S_4^N$ to predict IATEs for $S_3^N$, second with $S_1^N$, $S_3^N$ and $S_4^N$ to predict IATEs for $S_2^N$ and finally with $S_2^N$, $S_3^N$ and $S_4^N$ to predict IATEs for $S_1^N$.
\end{enumerate}
\end{algorithm}

\newpage

The NDR-learner follows the same basic steps but modifies step two:

\begin{algorithm}[NDR-learner] \label{alg:ndrlearner}
Let $(S_1^N,S_2^N,S_3^N,S_4^N)$ denote four independent samples of $N$ observations of $O_i=(X_i,W_i,Y_i)$. 
\begin{enumerate}
\item[Step 1.] Nuisance training:
\begin{enumerate}
\item Construct a model $\hat{e}_w(x)$ of the propensity scores $e_w(x)$ using $S_1^N$. 
\item Construct a model $(\hat{\mu}(w,x),\hat{\mu}(w',x))$ of the regression functions $(\mu(w,x), \mu(w',x))$ using $S_2^N$. 
\end{enumerate}
\item[Step 2a.]  Pseudo-outcome regression: Construct the pseudo-outcome for every observation $i$ in subsample $S_3^N$ using the models of step 1
\begin{equation*}
\hat{\Delta}_{i,w,w'} = \hat{\mu}(w,X_i) - \hat{\mu}(w',X_i) + \dfrac{ D_i(w) }{\hat{e}_w(X_i)}  \Tilde{Y_i}(w,X_i)  - \dfrac{D_i(w') }{\hat{e}_{w'}(X_i)} \Tilde{Y_i}(w',X_i),
\end{equation*}
regress it on covariates $X_i$ in $S_3^N$, and use the model to predict IATEs in $S_4^N$.
\item[Step 2b.]  Normalisation: For every observation $j$ in $S_4^N$: (i) extract the weights underlying its prediction $\alpha_i(X_j)$ and (ii) use it to calculate the normalised DR-learner given in Equation 3.9, where the sum goes over observations in $S_3^N$, to obtain $\hat{\tau}_{w,w'}^{4,1}(X_j)$.
\item[Step 3.]  Cross-fitting: Repeat steps 1--2 twice, first using $S_2^N$ for the propensity score, $S_3^N$ for the outcome regression and $S_1^N$ as subsample to obtain IATE predictions in $S_4^N$ $\hat{\tau}_{w,w'}^{4,2}(x)$, and then using $S_3^N$ for the propensity score, $S_1^N$ for the outcome regression and $S_2^N$ as subsample to obtain IATE predictions in $S_4^N$, $\hat{\tau}_{w,w'}^{4,3}(x)$. 
\item[Step 4.]  Prediction: Predict IATEs in $S_4^N$ as the average of the three predictions $\hat{\tau}_{w,w'}^{drl}(x) = 1/3 \hat{\tau}_{w,w'}^{4,1}(x) + 1/3 \hat{\tau}_{w,w'}^{4,2}(x) + 1/3 \hat{\tau}_{w,w'}^{4,3}(x)$.
\item[Step 5.]  Iteration: Repeat steps 1--4 three times. First, with $S_1^N$, $S_2^N$ and $S_4^N$ to predict IATEs for $S_3^N$, second with $S_1^N$, $S_3^N$ and $S_4^N$ to predict IATEs for $S_2^N$ and finally with $S_2^N$, $S_3^N$ and $S_4^N$ to predict IATEs for $S_1^N$.
\end{enumerate}
\end{algorithm}

\newpage
\section{Results}  \label{sec:app-res}

\subsection{Descriptives}

Table \ref{tab:desc-app} provides the means of all confounders by program participation. It documents that especially measures of past labour market success like past income are associated with program participation.

\singlespacing
\begin{table}[h]
  \centering  \scriptsize
\begin{threeparttable}[t]
  \caption{Means of control variables by program} 
  \label{tab:desc-app} 
\begin{tabular}{lccccc} 
\toprule
\\[-1.8ex] & \multicolumn{1}{c}{No} & \multicolumn{1}{c}{JS} & \multicolumn{1}{c}{Voc} & \multicolumn{1}{c}{Comp} & \multicolumn{1}{c}{Lang}\\ 
\\[-1.8ex] & \multicolumn{1}{c}{(1)} & \multicolumn{1}{c}{(2)} & \multicolumn{1}{c}{(3)} & \multicolumn{1}{c}{(4)} & \multicolumn{1}{c}{(5)}\\ 
\midrule
Age & 36.6 & 37.3 & 37.5 & 39.1 & 35.3 \\ 
  Mother tongue in canton's language & 0.10 & 0.12 & 0.11 & 0.11 & 0.04 \\ 
  Lives in big city & 0.19 & 0.19 & 0.21 & 0.11 & 0.23 \\ 
  Lives in medium city & 0.12 & 0.13 & 0.12 & 0.15 & 0.15 \\ 
  Lives in no city & 0.68 & 0.68 & 0.67 & 0.73 & 0.63 \\ 
  Caseworker age & 44.1 & 44.1 & 44.8 & 44.6 & 44.6 \\ 
  Caseworker cooperative & 0.48 & 0.50 & 0.41 & 0.42 & 0.45 \\ 
  Caseworker education: above vocational training & 0.46 & 0.45 & 0.44 & 0.48 & 0.48 \\ 
  Caseworker education: tertiary track & 0.19 & 0.21 & 0.17 & 0.16 & 0.21 \\ 
  Caseworker female & 0.43 & 0.47 & 0.39 & 0.44 & 0.47 \\ 
  Missing caseworker characteristics & 0.05 & 0.05 & 0.04 & 0.05 & 0.05 \\ 
  Caseworker has own unemployemnt experience & 0.62 & 0.63 & 0.64 & 0.61 & 0.63 \\ 
  Caseworker tenure & 5.48 & 5.44 & 5.73 & 5.83 & 5.61 \\ 
  Caseworker education: vocational degree & 0.26 & 0.27 & 0.22 & 0.25 & 0.22 \\ 
  Fraction of months employed last 2 years & 0.81 & 0.84 & 0.83 & 0.84 & 0.72 \\ 
  Number of employment spells last 5 years & 1.21 & 0.97 & 0.93 & 0.86 & 0.78 \\ 
  Employability & 1.93 & 1.98 & 1.93 & 1.97 & 1.85 \\ 
  Female & 0.44 & 0.44 & 0.33 & 0.60 & 0.55 \\ 
  Foreigner with temporary permit & 0.13 & 0.11 & 0.12 & 0.04 & 0.44 \\ 
  Foreigner with permanent permit & 0.23 & 0.22 & 0.18 & 0.17 & 0.23 \\ 
  Cantonal GDP p.c. & 0.52 & 0.53 & 0.51 & 0.53 & 0.54 \\ 
  Married & 0.47 & 0.46 & 0.48 & 0.45 & 0.72 \\ 
  Mother tongue other than German, French, Italian & 0.33 & 0.29 & 0.31 & 0.18 & 0.64 \\ 
  Past income & 42528.0 & 46693.1 & 48653.8 & 43212.8 & 37300.5 \\ 
  Previous job: manager & 0.08 & 0.08 & 0.10 & 0.09 & 0.07 \\ 
  Missing sector & 0.18 & 0.15 & 0.15 & 0.16 & 0.29 \\ 
  Previous job in primary sector & 0.09 & 0.06 & 0.09 & 0.05 & 0.05 \\ 
  Previous job in secondary sector & 0.12 & 0.14 & 0.15 & 0.13 & 0.12 \\ 
  Previous job in tertiary sector & 0.61 & 0.65 & 0.61 & 0.67 & 0.54 \\ 
  Previous job: self-employed & 0.01 & 0.00 & 0.00 & 0.00 & 0.00 \\ 
  Previous job: skilled worker & 0.60 & 0.65 & 0.65 & 0.75 & 0.43 \\ 
  Previous job: unskilled worker & 0.29 & 0.24 & 0.22 & 0.15 & 0.48 \\ 
  Qualification: semiskilled & 0.16 & 0.14 & 0.17 & 0.14 & 0.15 \\ 
  Qualification: some degree & 0.58 & 0.62 & 0.63 & 0.72 & 0.38 \\ 
  Qualification: unskilled & 0.23 & 0.20 & 0.17 & 0.12 & 0.40 \\ 
  Qualification: skilled without degree & 0.03 & 0.03 & 0.02 & 0.02 & 0.07 \\ 
  Swiss citizen & 0.64 & 0.67 & 0.70 & 0.79 & 0.34 \\ 
  Allocation of unemployed to caseworkers: by industry & 0.60 & 0.67 & 0.58 & 0.51 & 0.64 \\ 
  Allocation of unemployed to caseworkers: by occupation & 0.51 & 0.57 & 0.46 & 0.45 & 0.57 \\ 
  Allocation of unemployed to caseworkers: by age & 0.04 & 0.04 & 0.04 & 0.06 & 0.05 \\ 
  Allocation of unemployed to caseworkers: by employability & 0.09 & 0.07 & 0.10 & 0.08 & 0.06 \\ 
  Allocation of unemployed to caseworkers: by region & 0.13 & 0.09 & 0.09 & 0.13 & 0.11 \\ 
  Allocation of unemployed to caseworkers: other & 0.09 & 0.07 & 0.08 & 0.10 & 0.09 \\ 
  Number of unemployment spells last 2 years & 0.57 & 0.39 & 0.52 & 0.37 & 0.43 \\ 
  Cantonal unemployment rate (in \%) & 3.52 & 3.59 & 3.41 & 3.36 & 3.63 \\ 
    \bottomrule
    \end{tabular}%
         \begin{tablenotes} \item \textit{Note:} Program specific means. \end{tablenotes}  
\end{threeparttable}
\end{table}
\doublespacing
\clearpage
\subsection{Nuisance parameters} \label{sec:app-res-np}

Nuisance parameters are only a tool to remove confounding but it is still informative to investigate which variables are most predictive of treatment probabilities and outcome. This is less straightforward for flexible tools like random forests than for the well-known regression outputs of parametric models. We conduct a classification analysis as proposed by \citeA{Chernozhukov2018TheAverages}. To this end, we split the predicted nuisance parameter distributions in quintiles and compare the covariate means of the observations falling into the fifth and first quintile. For comparability, we normalise all covariates to have mean zero and variance one and order the variables by their largest absolute difference between the highest and lowest quintile.

Table \ref{tab:app-clan-ps} shows that measures of citizenship, qualification and previous labour market success are important predictors of program selection. In line with intuition the former seems to drive a large part of the selection into language courses. Also Table \ref{tab:app-clan-out} showing the classification analysis for outcome predictions shows intuitive patterns. Again measures of citizenship, qualification and previous labour market success seem predictive for future employment with suggested correlations pointing in the expected directions. For example, Swiss citizens, individuals with a degree and high past income are overrepresented in the upper quintile, while individuals with a non-Swiss mother tongue and no qualification are underepresented in the upper quintile.

\begin{figure}[htp] 
\centering
\caption{Distribution of propensity scores} \label{fig:ps-dist}

\begin{subfigure}{0.45\textwidth}
\includegraphics[width=\linewidth]{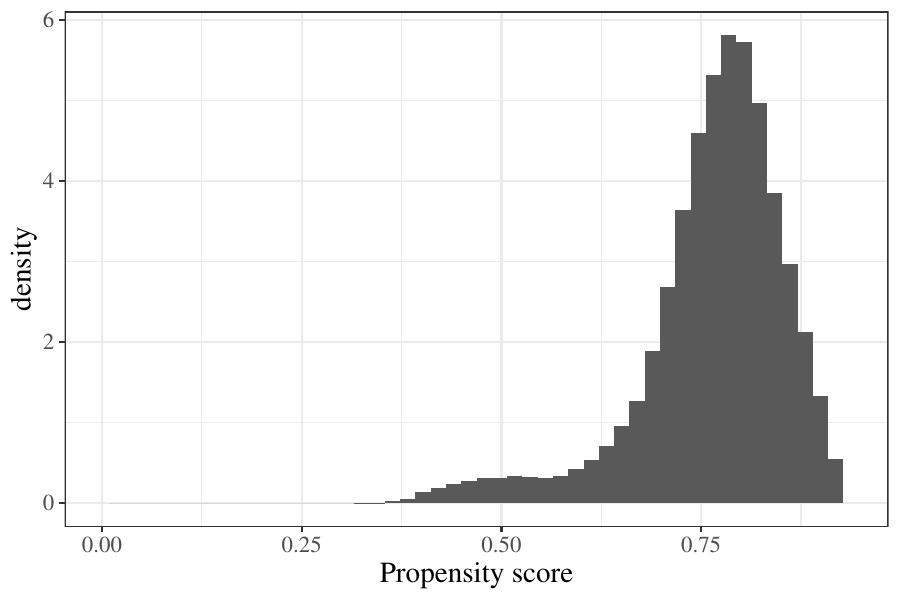}
\caption{No program} 
\end{subfigure}\hspace*{\fill}
\begin{subfigure}{0.45\textwidth}
\includegraphics[width=\linewidth]{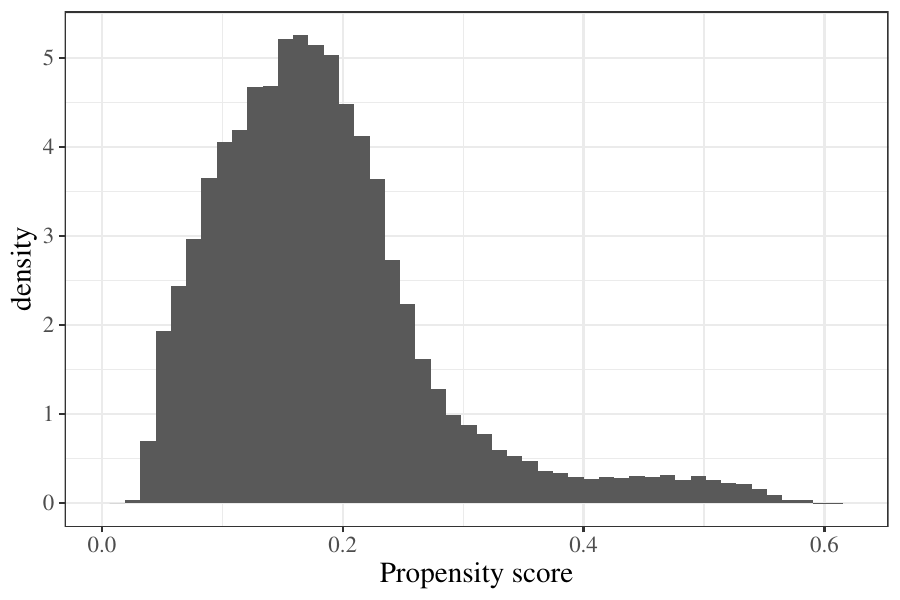}
\caption{Job search} 
\end{subfigure}

\medskip
\begin{subfigure}{0.45\textwidth}
\includegraphics[width=\linewidth]{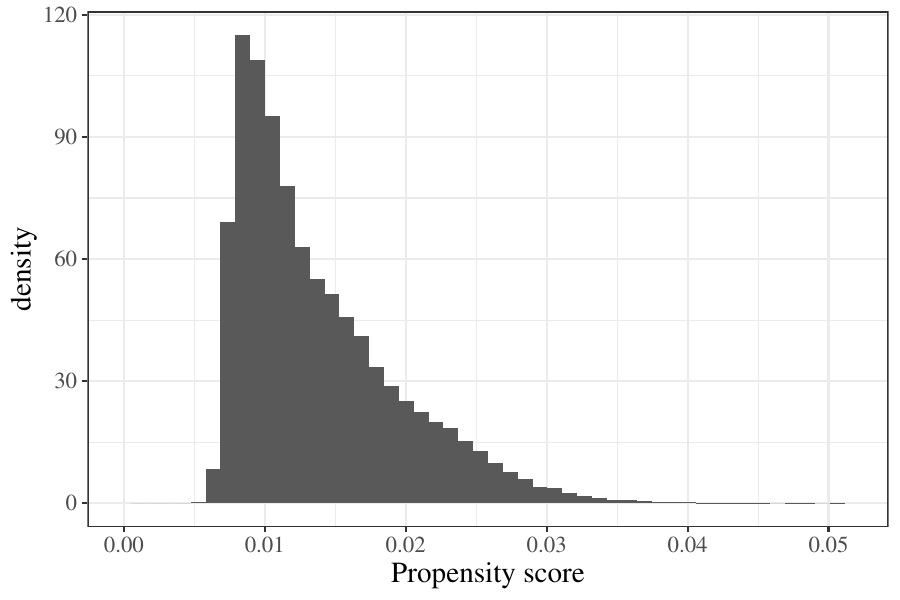}
\caption{Vocational}
\end{subfigure}\hspace*{\fill}
\begin{subfigure}{0.45\textwidth}
\includegraphics[width=\linewidth]{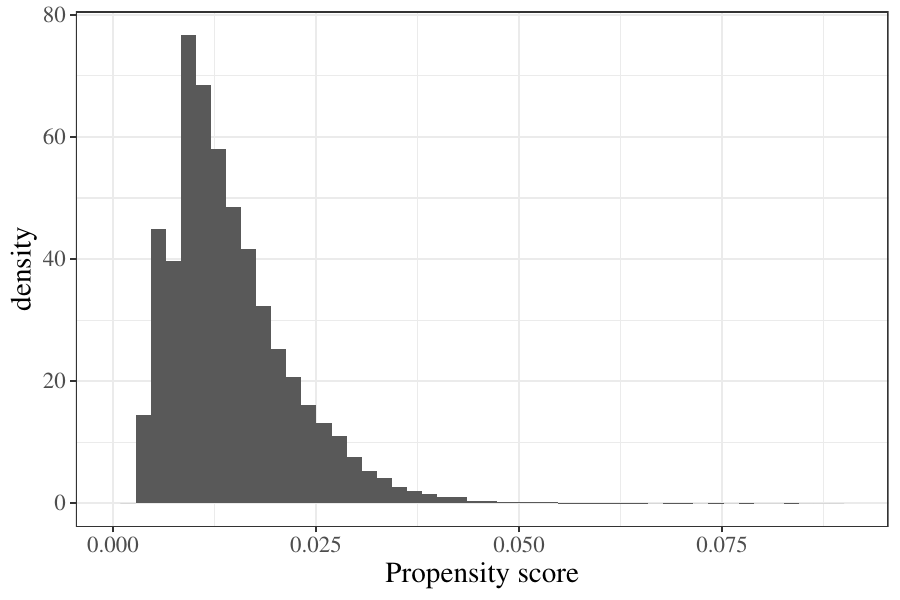}
\caption{Computer}
\end{subfigure}

\medskip
\begin{subfigure}{0.45\textwidth}
\includegraphics[width=\linewidth]{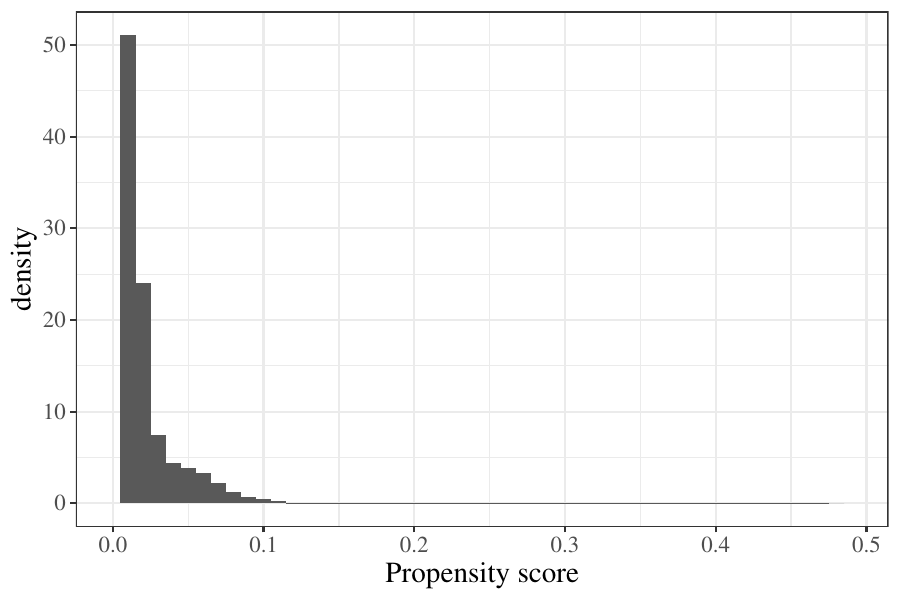}
\caption{Language}
\end{subfigure}

\end{figure}

Finally, we investigate the propensity score distributions for all programs. Figure \ref{fig:ps-dist} shows that propensity scores are quite variable. This indicates that selection into programs is not negligible. Further, Table \ref{tab:ps_desc} shows that some of the propensity scores get quite small with the smallest one being 0.003 for a computer training participant. This is not surprising given that already the unconditional participation probabilities for computer and vocational training are only about 0.015. However, the small propensity score \textit{per se} is not an indicator of poor overlap. The residual with the smallest propensity score receives a weight of $\sim1/ 0.003 = 333$, which is only 0.5\% of the total weights. Note that we could easily increase the smallest propensity score by randomly removing a large fraction of non-participants and participants of the job search program. This would discard valuable information and shows that the mere focus on the smallest propensity score can be misleading in cases with imbalanced treatment group sizes. More importantly, we observe overlap in the sense that all treatment groups contain individuals with similarly low propensity scores. Thus, overlap seems not to be a major issue in our application, at least for the low dimensional parameters of interest.

\singlespacing
\begin{table}[htp]
  \centering   \scriptsize
\begin{threeparttable}[t]
  \caption{Classification analysis of propensity scores} 
  \label{tab:app-clan-ps} 
\begin{tabular}{lccccccc} 
\toprule
\\[-1.8ex] & \multicolumn{1}{c}{No program} & \multicolumn{1}{c}{Job search} & \multicolumn{1}{c}{Vocational} & \multicolumn{1}{c}{Computer} & \multicolumn{1}{c}{Language} \\ 
\\[-1.8ex] & \multicolumn{1}{c}{(1)} & \multicolumn{1}{c}{(2)} & \multicolumn{1}{c}{(3)} & \multicolumn{1}{c}{(4)} & \multicolumn{1}{c}{(5)} \\ 
\midrule
Foreigner with temporary permit & -0.22 & -0.34 & -0.32 & -1.25 & 1.93 \\ 
  Swiss citizen & 0.08 & 0.22 & 0.43 & 1.59 & -1.86 \\ 
  Mother tongue other than German, French, Italian & 0.10 & -0.42 & -0.33 & -1.57 & 1.76 \\ 
  Previous job: unskilled worker & 0.31 & -0.43 & -0.61 & -1.52 & 0.99 \\ 
  Past income & -0.72 & 0.78 & 1.42 & 0.28 & -0.07 \\ 
  Previous job: skilled worker & -0.20 & 0.31 & 0.30 & 1.31 & -0.90 \\ 
  Qualification: some degree & -0.18 & 0.31 & 0.53 & 1.26 & -0.93 \\ 
  Qualification: unskilled & 0.04 & -0.16 & -0.55 & -1.16 & 0.86 \\ 
  Female & -0.08 & -0.06 & -1.07 & 1.00 & 0.41 \\ 
  Married & -0.23 & -0.03 & -0.17 & -0.60 & 1.07 \\ 
  Cantonal unemployment rate (in \%) & -0.71 & 0.73 & -0.90 & -0.53 & -0.14 \\ 
  Foreigner with permanent permit & 0.08 & 0.03 & -0.23 & -0.81 & 0.56 \\ 
  Age & -0.34 & 0.33 & 0.35 & 0.79 & -0.25 \\ 
  Cantonal GDP p.c. & -0.62 & 0.65 & -0.73 & -0.16 & -0.14 \\ 
  Number of employment spells last 5 years & 0.73 & -0.53 & -0.41 & -0.46 & -0.31 \\ 
  Allocation of unemployed to caseworkers: by occupation & -0.49 & 0.50 & -0.64 & -0.21 & 0.08 \\ 
  Allocation of unemployed to caseworkers: by region & 0.53 & -0.53 & 0.02 & 0.06 & 0.03 \\ 
  Fraction of months employed last 2 years & -0.26 & 0.47 & 0.53 & 0.40 & -0.45 \\ 
  Allocation of unemployed to caseworkers: by industry & -0.45 & 0.45 & -0.49 & -0.53 & 0.01 \\ 
  Previous job: manager & -0.21 & 0.18 & 0.52 & 0.17 & 0.09 \\ 
  Employability & -0.44 & 0.49 & 0.14 & 0.36 & -0.25 \\ 
  Previous job in tertiary sector & -0.20 & 0.27 & 0.01 & 0.45 & -0.31 \\ 
  Caseworker cooperative & -0.08 & 0.11 & -0.43 & -0.21 & -0.02 \\ 
  Missing sector & 0.15 & -0.29 & -0.41 & -0.35 & 0.43 \\ 
  Lives in big city & 0.07 & -0.08 & -0.04 & -0.42 & 0.09 \\ 
  Number of unemployment spells last 2 years & 0.39 & -0.33 & -0.06 & -0.40 & 0.02 \\ 
  Caseworker female & -0.30 & 0.27 & -0.39 & 0.14 & -0.03 \\ 
  Qualification: skilled without degree & -0.08 & -0.02 & -0.10 & -0.27 & 0.36 \\ 
  Lives in no city & -0.01 & 0.05 & -0.02 & 0.33 & -0.12 \\ 
  Previous job in primary sector & 0.30 & -0.26 & 0.30 & -0.26 & -0.02 \\ 
  Caseworker tenure & -0.02 & 0.00 & 0.24 & 0.12 & 0.05 \\ 
  Qualification: semiskilled & 0.24 & -0.22 & -0.04 & -0.24 & 0.10 \\ 
  Previous job in secondary sector & -0.13 & 0.15 & 0.21 & -0.05 & -0.02 \\ 
  Caseworker age & -0.05 & 0.03 & 0.12 & 0.08 & 0.19 \\ 
  Allocation of unemployed to caseworkers: by employability & 0.18 & -0.19 & 0.11 & 0.01 & -0.01 \\ 
  Caseworker education: tertiary track & -0.18 & 0.18 & -0.12 & -0.17 & 0.00 \\ 
  Caseworker has own unemployemnt experience & -0.14 & 0.16 & 0.07 & -0.02 & 0.02 \\ 
  Caseworker education: vocational degree & -0.10 & 0.14 & -0.15 & 0.07 & -0.14 \\ 
  Mother tongue in canton's language & -0.01 & 0.10 & -0.03 & 0.03 & 0.14 \\ 
  Allocation of unemployed to caseworkers: other & 0.05 & -0.04 & -0.13 & -0.00 & -0.07 \\ 
  Caseworker education: above vocational training & 0.12 & -0.13 & 0.04 & 0.11 & 0.02 \\ 
  Allocation of unemployed to caseworkers: by age & -0.02 & 0.01 & -0.06 & 0.08 & -0.00 \\ 
  Lives in medium city & -0.07 & 0.02 & 0.06 & 0.03 & 0.06 \\ 
  Previous job: self-employed & 0.01 & 0.01 & 0.02 & 0.04 & -0.07 \\ 
  Missing caseworker characteristics & 0.06 & -0.06 & 0.05 & -0.02 & 0.00 \\ 
    \bottomrule
    \end{tabular}%
         \begin{tablenotes} \item \textit{Note:} Table shows the differences in means of normalized covariates between the fifth and the first quintile of the respective propensity score distribution. Variables are ordered according to the largest absolute difference. \end{tablenotes}  
\end{threeparttable}
\end{table}
\doublespacing

\singlespacing
\begin{table}[htp]
  \centering   \scriptsize
\begin{threeparttable}[t]
  \caption{Classification analysis of outcome predictions} 
  \label{tab:app-clan-out} 
\begin{tabular}{lccccccc} 
\toprule
\\[-1.8ex] & \multicolumn{1}{c}{No program} & \multicolumn{1}{c}{Job search} & \multicolumn{1}{c}{Vocational} & \multicolumn{1}{c}{Computer} & \multicolumn{1}{c}{Language} \\ 
\\[-1.8ex] & \multicolumn{1}{c}{(1)} & \multicolumn{1}{c}{(2)} & \multicolumn{1}{c}{(3)} & \multicolumn{1}{c}{(4)} & \multicolumn{1}{c}{(5)} \\ 
\midrule
Mother tongue other than German, French, Italian & -1.44 & -1.66 & -1.16 & -2.06 & -1.89 \\ 
  Qualification: some degree & 1.80 & 1.96 & 1.57 & 1.64 & 1.85 \\ 
  Swiss citizen & 1.37 & 1.60 & 1.13 & 1.94 & 1.83 \\ 
  Previous job: unskilled worker & -1.72 & -1.76 & -1.55 & -1.65 & -1.87 \\ 
  Past income & 1.49 & 1.38 & 1.16 & 0.87 & 1.73 \\ 
  Qualification: unskilled & -1.43 & -1.57 & -1.53 & -1.36 & -1.52 \\ 
  Previous job: skilled worker & 1.23 & 1.27 & 1.21 & 1.34 & 1.36 \\ 
  Foreigner with permanent permit & -0.88 & -1.11 & -0.71 & -1.34 & -1.15 \\ 
  Number of unemployment spells last 2 years & -0.93 & -0.80 & -1.25 & -0.57 & -0.56 \\ 
  Married & -0.98 & -1.14 & -0.60 & -1.19 & -1.04 \\ 
  Foreigner with temporary permit & -0.84 & -0.89 & -0.72 & -1.08 & -1.16 \\ 
  Fraction of months employed last 2 years & 1.01 & 0.78 & 0.91 & 0.66 & 0.87 \\ 
  Employability & 0.93 & 0.84 & 0.46 & 0.49 & 0.62 \\ 
  Age & -0.72 & -0.76 & -0.25 & -0.34 & -0.29 \\ 
  Cantonal unemployment rate (in \%) & -0.10 & -0.05 & -0.75 & -0.02 & -0.04 \\ 
  Lives in big city & -0.24 & -0.22 & -0.74 & -0.36 & -0.30 \\ 
  Cantonal GDP p.c. & -0.04 & 0.02 & -0.74 & 0.04 & 0.06 \\ 
  Missing sector & -0.50 & -0.48 & -0.58 & -0.53 & -0.71 \\ 
  Number of employment spells last 5 years & -0.54 & -0.42 & -0.68 & -0.45 & -0.41 \\ 
  Lives in no city & 0.26 & 0.23 & 0.68 & 0.43 & 0.33 \\ 
  Previous job: manager & 0.54 & 0.52 & 0.40 & 0.32 & 0.67 \\ 
  Qualification: semiskilled & -0.63 & -0.67 & -0.26 & -0.49 & -0.59 \\ 
  Female & -0.42 & -0.32 & -0.44 & 0.23 & -0.60 \\ 
  Previous job in tertiary sector & 0.31 & 0.36 & 0.14 & 0.50 & 0.51 \\ 
  Mother tongue in canton's language & -0.21 & -0.25 & -0.22 & -0.45 & -0.33 \\ 
  Allocation of unemployed to caseworkers: by occupation & 0.18 & 0.22 & 0.32 & 0.39 & 0.25 \\ 
  Qualification: skilled without degree & -0.35 & -0.38 & -0.22 & -0.34 & -0.35 \\ 
  Previous job in secondary sector & 0.08 & 0.05 & 0.31 & 0.01 & 0.07 \\ 
  Caseworker age & 0.02 & 0.03 & 0.27 & -0.17 & 0.07 \\ 
  Caseworker female & 0.01 & 0.04 & -0.20 & 0.26 & -0.02 \\ 
  Caseworker tenure & -0.05 & -0.06 & -0.11 & -0.25 & -0.06 \\ 
  Previous job in primary sector & 0.05 & -0.04 & 0.18 & -0.17 & -0.01 \\ 
  Lives in medium city & -0.08 & -0.05 & -0.08 & -0.17 & -0.12 \\ 
  Caseworker education: vocational degree & 0.11 & 0.08 & 0.15 & 0.07 & 0.08 \\ 
  Allocation of unemployed to caseworkers: by employability & 0.04 & 0.03 & 0.15 & 0.05 & 0.04 \\ 
  Allocation of unemployed to caseworkers: by industry & 0.11 & 0.12 & 0.08 & 0.11 & 0.09 \\ 
  Caseworker education: above vocational training & 0.04 & 0.04 & 0.11 & 0.12 & 0.06 \\ 
  Missing caseworker characteristics & -0.04 & -0.05 & -0.11 & 0.02 & -0.06 \\ 
  Allocation of unemployed to caseworkers: by region & 0.04 & -0.04 & 0.09 & -0.02 & -0.03 \\ 
  Caseworker cooperative & -0.03 & -0.04 & -0.07 & 0.04 & -0.05 \\ 
  Allocation of unemployed to caseworkers: by age & 0.00 & -0.01 & 0.04 & 0.06 & 0.02 \\ 
  Caseworker education: tertiary track & 0.04 & 0.03 & 0.02 & -0.05 & -0.00 \\ 
  Caseworker has own unemployemnt experience & 0.03 & 0.04 & -0.00 & -0.03 & 0.03 \\ 
  Previous job: self-employed & -0.03 & -0.01 & -0.03 & -0.02 & 0.02 \\ 
  Allocation of unemployed to caseworkers: other & -0.02 & -0.03 & -0.01 & -0.02 & 0.01 \\ 
    \bottomrule
    \end{tabular}%
         \begin{tablenotes} \item \textit{Note:} Table shows the differences in means of normalized covariates between the fifth and the first quintile of the respective outcome prediction distribution. Variables are ordered according to the largest absolute difference. \end{tablenotes}  
\end{threeparttable}
\end{table}
\doublespacing

\onehalfspacing
\begin{table}[htp]
  \centering  \footnotesize
\begin{threeparttable}
  \caption{Summary statistics of propensity score distributions} 
  \label{tab:ps_desc} 
  \begin{tabular}{rccccc}
\toprule
 & No program & Job search & Vocational & Computer & Language \\ 
\midrule
Mean & 0.764 & 0.184 & 0.014 & 0.015 & 0.024 \\ 
  SD & 0.093 & 0.094 & 0.006 & 0.008 & 0.030 \\ 
  Minimum & 0.321 & 0.027 & 0.005 & 0.003 & 0.005 \\ 
  Q1 & 0.439 & 0.045 & 0.007 & 0.004 & 0.006 \\ 
  Q25 & 0.727 & 0.120 & 0.009 & 0.009 & 0.009 \\ 
  Q50 & 0.779 & 0.170 & 0.012 & 0.013 & 0.015 \\ 
  Q75 & 0.824 & 0.224 & 0.017 & 0.018 & 0.025 \\ 
  Q99 & 0.910 & 0.515 & 0.031 & 0.038 & 0.109 \\ 
  Maximum & 0.938 & 0.622 & 0.052 & 0.091 & 0.490 \\ 
    \bottomrule
\end{tabular}
\begin{tablenotes} \item \textit{Note:} The table provides summary statistics of the program specific propensity score distributions. The rows denoted by Q show the respective quantiles. \end{tablenotes}  
\end{threeparttable}
\end{table}
\doublespacing

\subsection{Average treatment effects}
\begin{figure}[h]
    \centering
        \caption{Average potential outcomes}
    \includegraphics[width=0.65\linewidth]{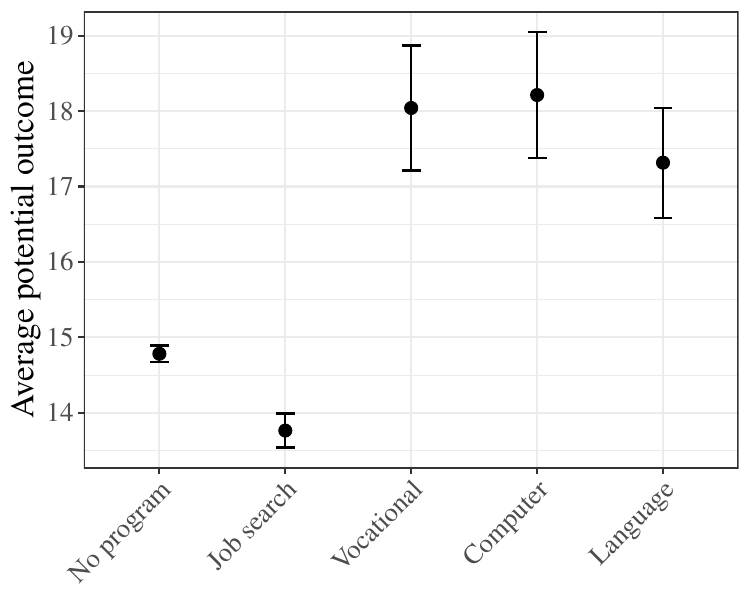}
    \label{fig:apo}
    \subcaption*{\textit{Notes:} Average potential outcomes with 95\% confidence intervals. Numeric results in Panel A of Table \ref{tab:avg-eff}.}
\end{figure}

\begin{figure}[t!] 

\caption{Average treatment effects over time} 
\label{fig:ate31}

\begin{subfigure}{0.49\textwidth}
\includegraphics[width=\linewidth]{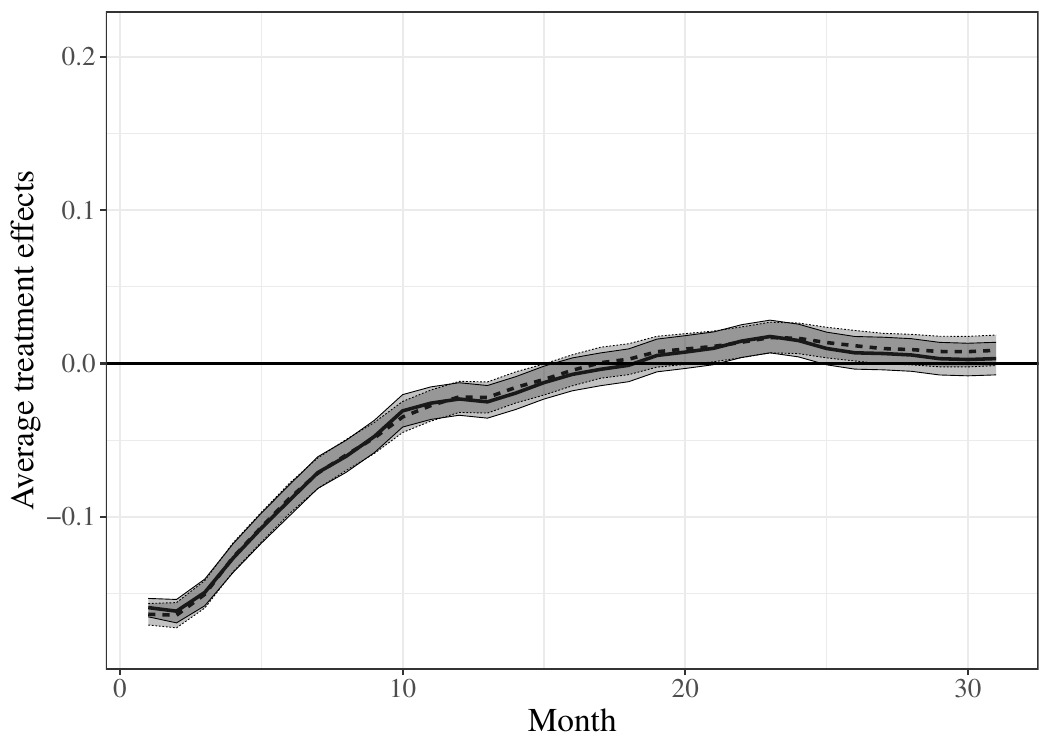}
\caption{Job search} \label{fig:a}
\end{subfigure}\hspace*{\fill}
\begin{subfigure}{0.49\textwidth}
\includegraphics[width=\linewidth]{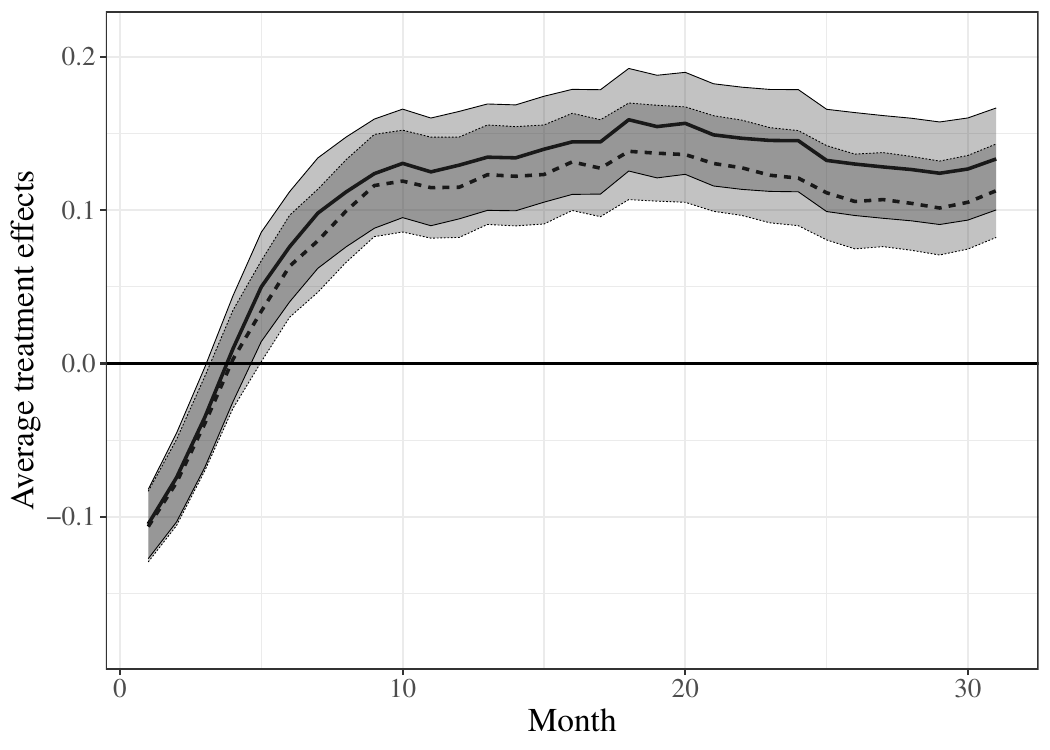}
\caption{Vocational} \label{fig:b}
\end{subfigure}

\medskip
\begin{subfigure}{0.49\textwidth}
\includegraphics[width=\linewidth]{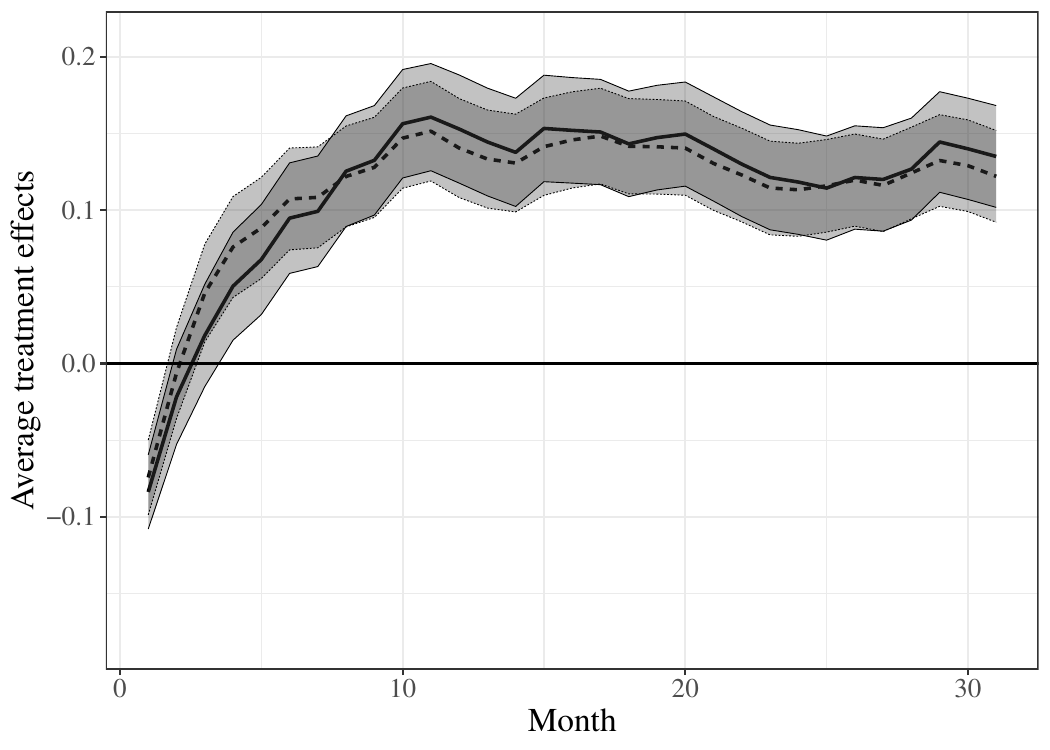}
\caption{Computer} \label{fig:c}
\end{subfigure}\hspace*{\fill}
\begin{subfigure}{0.49\textwidth}
\includegraphics[width=\linewidth]{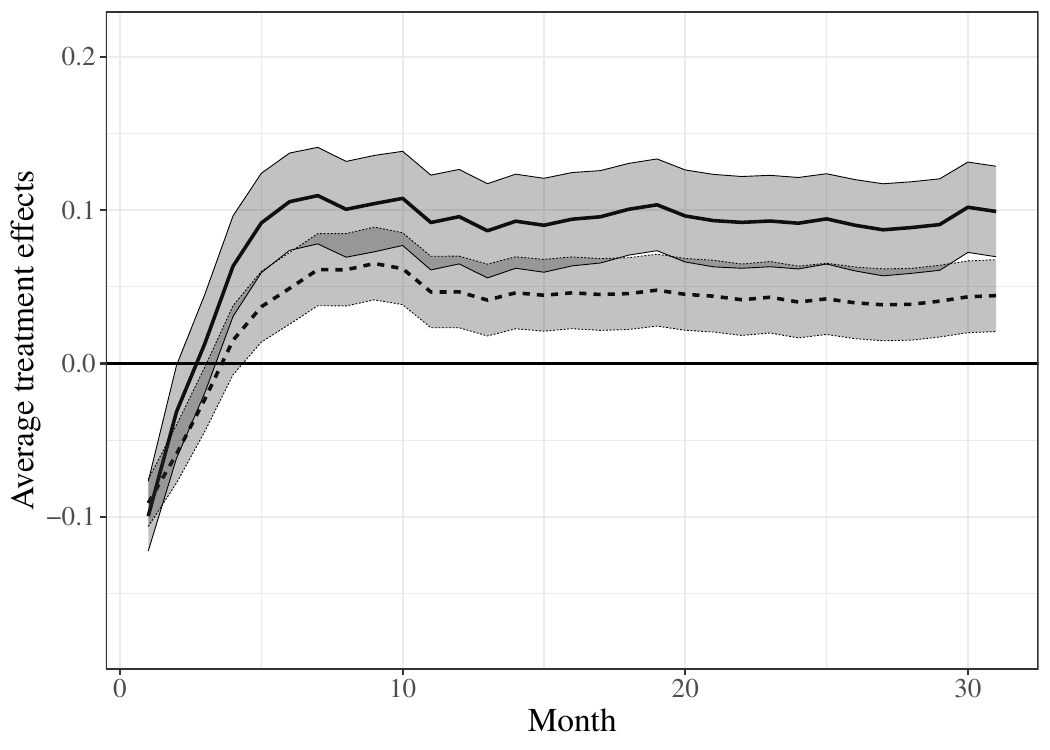}
\caption{Language} \label{fig:d}
\end{subfigure}
\subcaption*{\textit{Notes:} Solid lines show ATE, dashed lines ATET of the respective programs compared to nonparticipation on employment probability in the 31 months after program start. Grey area depicts the 95\% confidence intervals.}
\end{figure}

Figure \ref{fig:apo} shows the APO estimates and Figure \ref{fig:ate31} shows the effects of program participation on the employment probabilities over time. The latter documents that all program show the well-known lock-in effect within the first months after program start \cite<e.g.>{Wunsch2016HowWorkers}. However, participants of the hard skill programs catch up and show a sustained increase in employment rates of up to 10 percentage points.

\begin{table}[t]
  \centering  \footnotesize
\begin{threeparttable}
  \caption{Average effects} 
  \label{tab:avg-eff} 
\begin{tabular}{lcc} 
\toprule
 & \multicolumn{1}{c}{Estimate} & \multicolumn{1}{c}{Standard error} \\ 
 & \multicolumn{1}{c}{(1)} & \multicolumn{1}{c}{(2)}  \\ 
\midrule
\multicolumn{3}{l}{\textit{Panel A:} APO}\\
No program & 14.78 & 0.06  \\ 
  Job search & 13.76 & 0.12  \\ 
  Vocational & 18.04 & 0.42  \\ 
  Computer & 18.21 & 0.43  \\ 
  Language & 17.32 & 0.37  \\ 
[0.5em]  
\multicolumn{3}{l}{\textit{Panel B:} ATE}\\
Job search - no program & -1.02$^{***}$ & 0.13 \\ 
  Vocational - no program & 3.26$^{***}$ & 0.43 \\ 
  Computer - no program & 3.43$^{***}$ & 0.43 \\ 
  Language - no program & 2.54$^{***}$ & 0.38 \\ 
  [0.5em]  
\multicolumn{3}{l}{\textit{Panel C:} ATET}\\
Job search - no program & -0.98$^{***}$ & 0.12 \\ 
  Vocational - no program & 2.79$^{***}$ & 0.39 \\ 
Computer - no program & 3.47$^{***}$ & 0.40 \\ 
Language - no program & 1.08$^{***}$ & 0.29 \\
    \bottomrule
    \end{tabular}%
         \begin{tablenotes} \item \textit{Note:} Table shows DML based point estimates and standard errors of average effects.  $^{*}$p$<$0.1; $^{**}$p$<$0.05; $^{***}$p$<$0.01 \end{tablenotes}  
\end{threeparttable}
\end{table}

\clearpage
\subsection{GATEs}

\begin{figure}[h] 
\caption{Effect heterogeneity regarding age} 
\label{fig:cate-age}
\begin{subfigure}{0.399\textwidth}
\includegraphics[width=\linewidth]{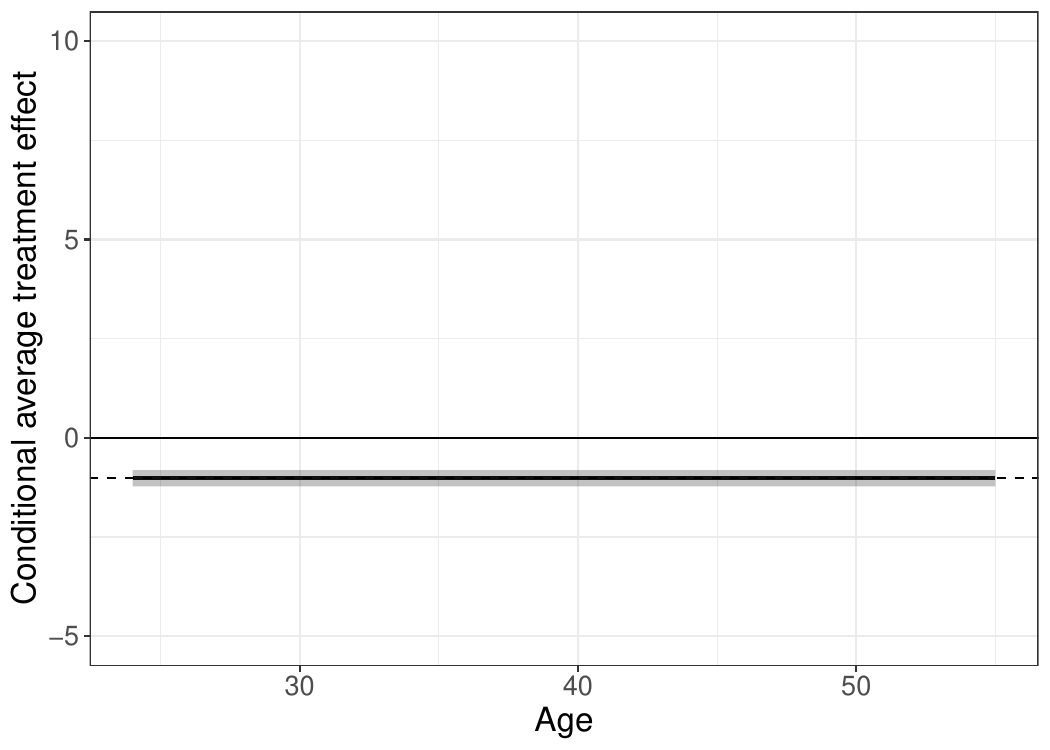}
\caption{Job search (Kernel)}
\end{subfigure}\hspace*{\fill}
\begin{subfigure}{0.399\textwidth}
\includegraphics[width=\linewidth]{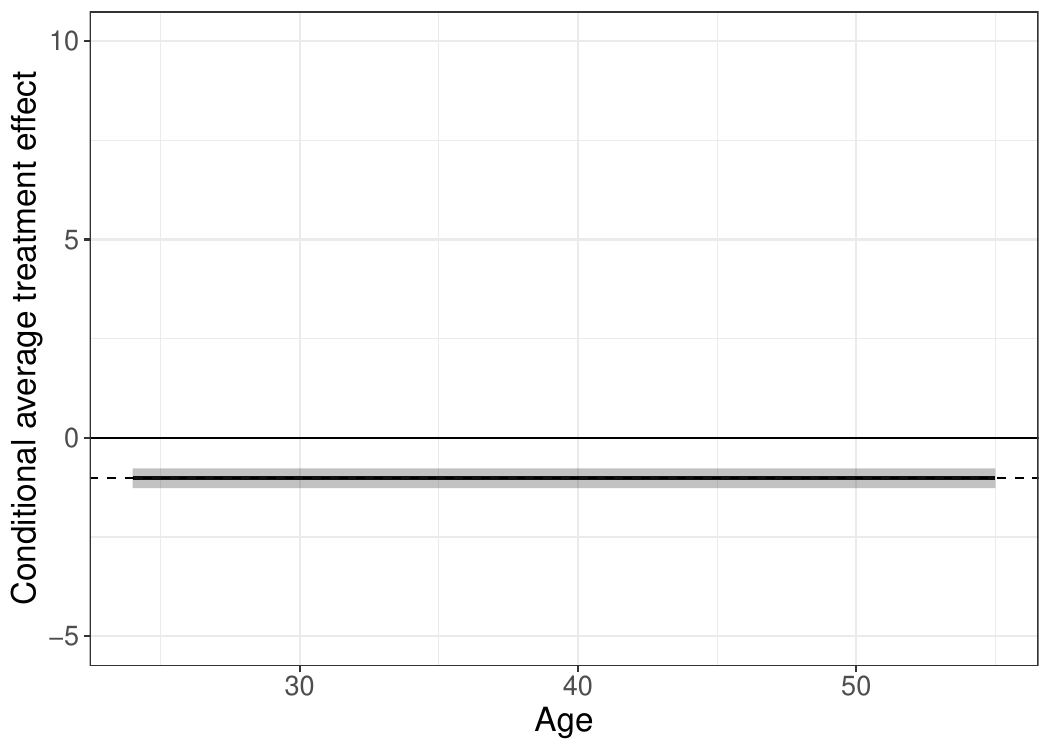}
\caption{Job search (Spline)}

\end{subfigure}

\medskip
\begin{subfigure}{0.399\textwidth}
\includegraphics[width=\linewidth]{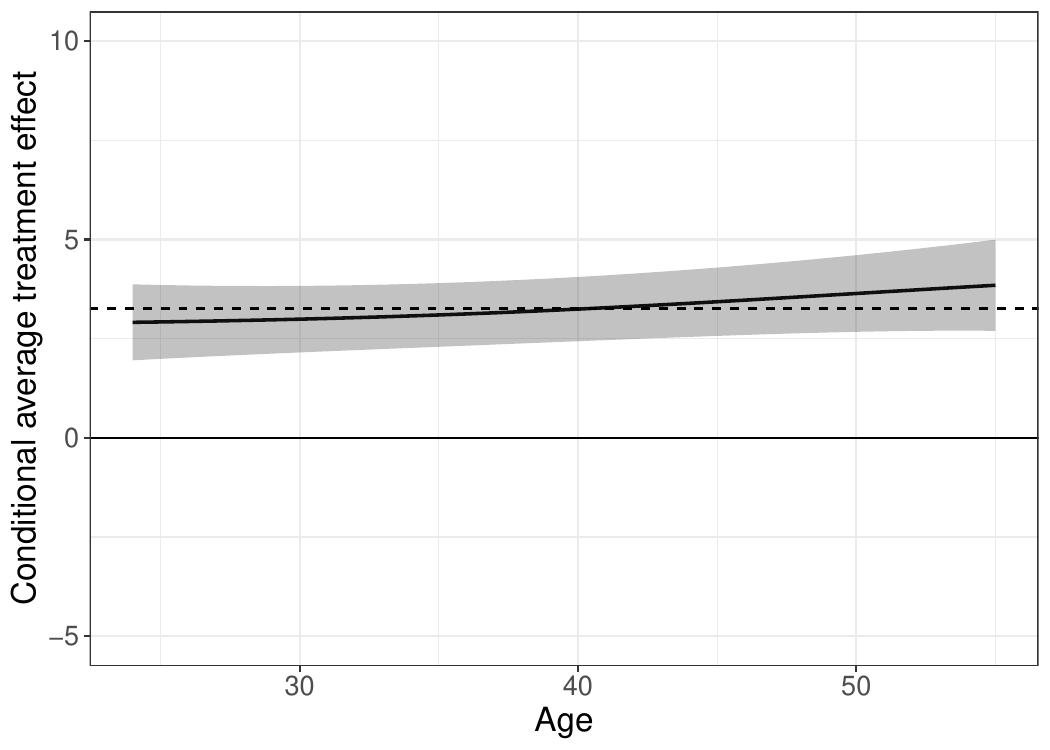}
\caption{Vocational (Kernel)}

\end{subfigure}\hspace*{\fill}
\begin{subfigure}{0.399\textwidth}
\includegraphics[width=\linewidth]{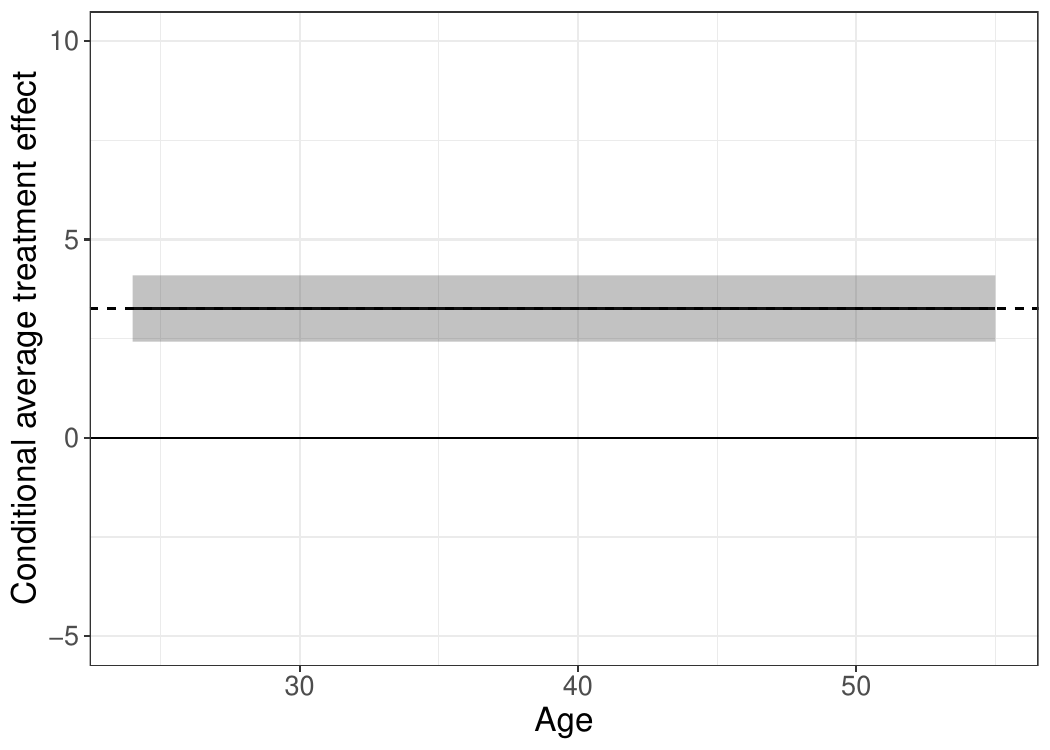}
\caption{Vocational (Spline)}
\end{subfigure}

\medskip

\begin{subfigure}{0.399\textwidth}
\includegraphics[width=\linewidth]{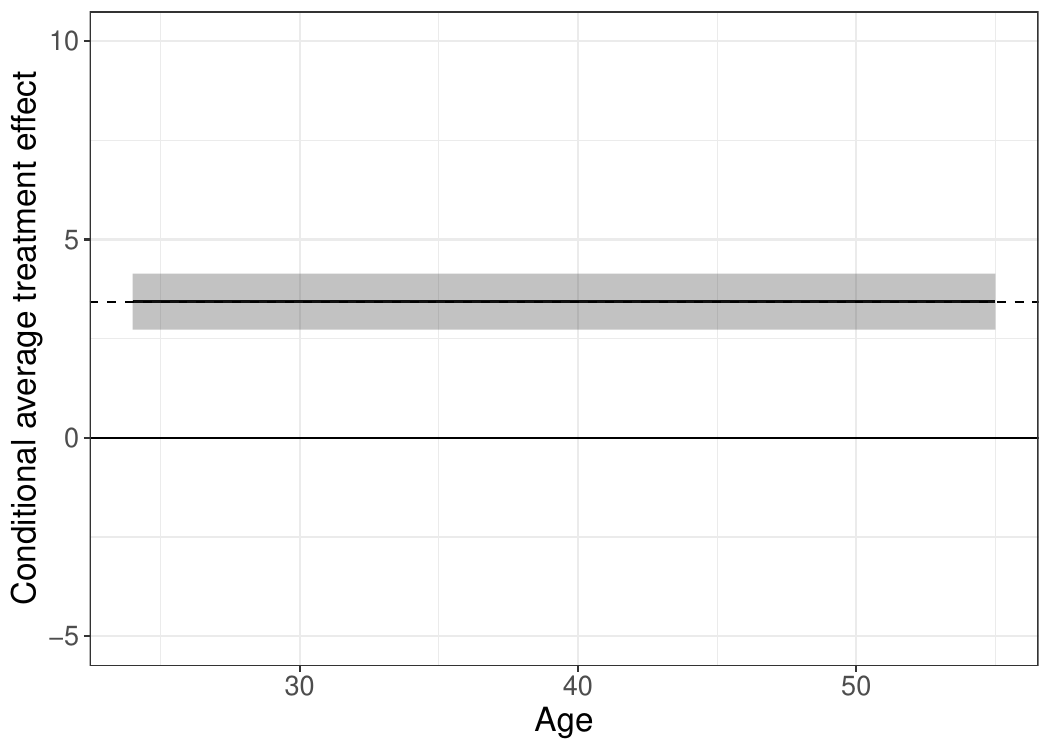}
\caption{Computer (Kernel)}
\end{subfigure}\hspace*{\fill}
\begin{subfigure}{0.399\textwidth}
\includegraphics[width=\linewidth]{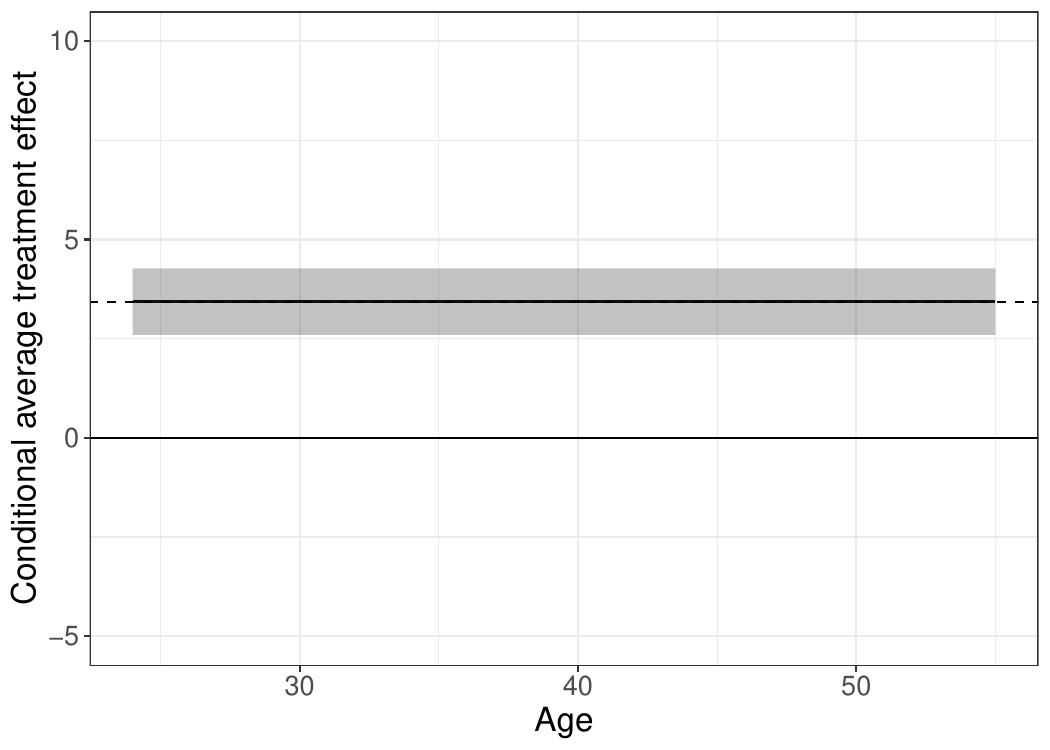}
\caption{Computer (Spline)}
\end{subfigure}

\medskip
\begin{subfigure}{0.399\textwidth}
\includegraphics[width=\linewidth]{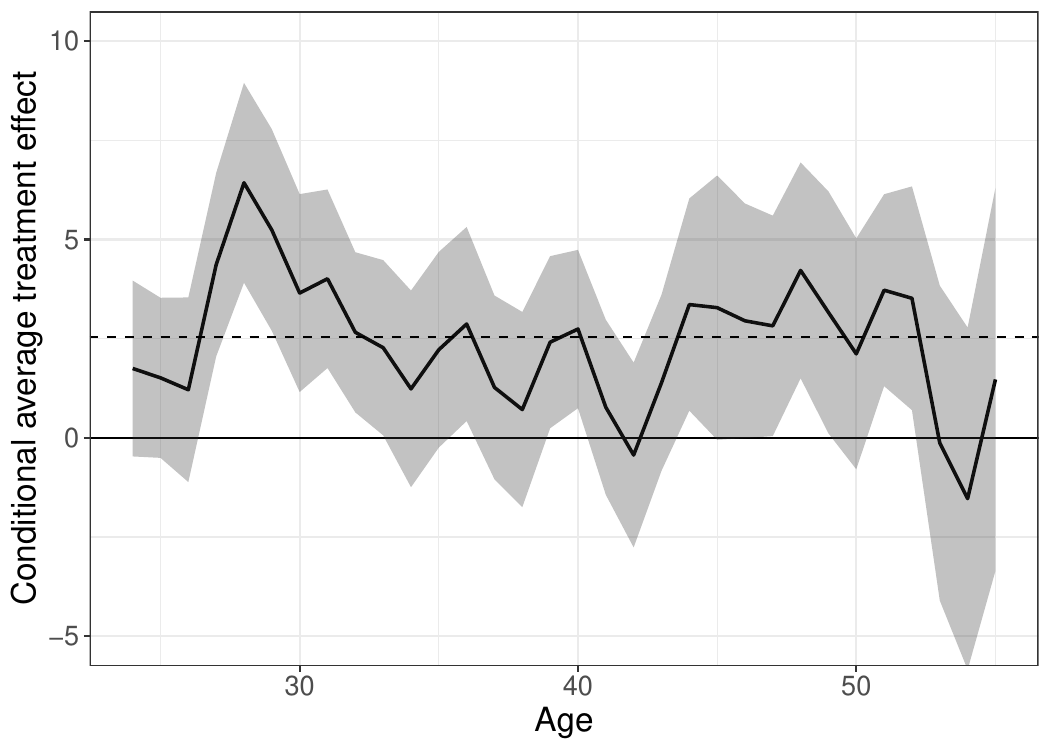}
\caption{Language (Kernel)}
\end{subfigure}\hspace*{\fill}
\begin{subfigure}{0.399\textwidth}
\includegraphics[width=\linewidth]{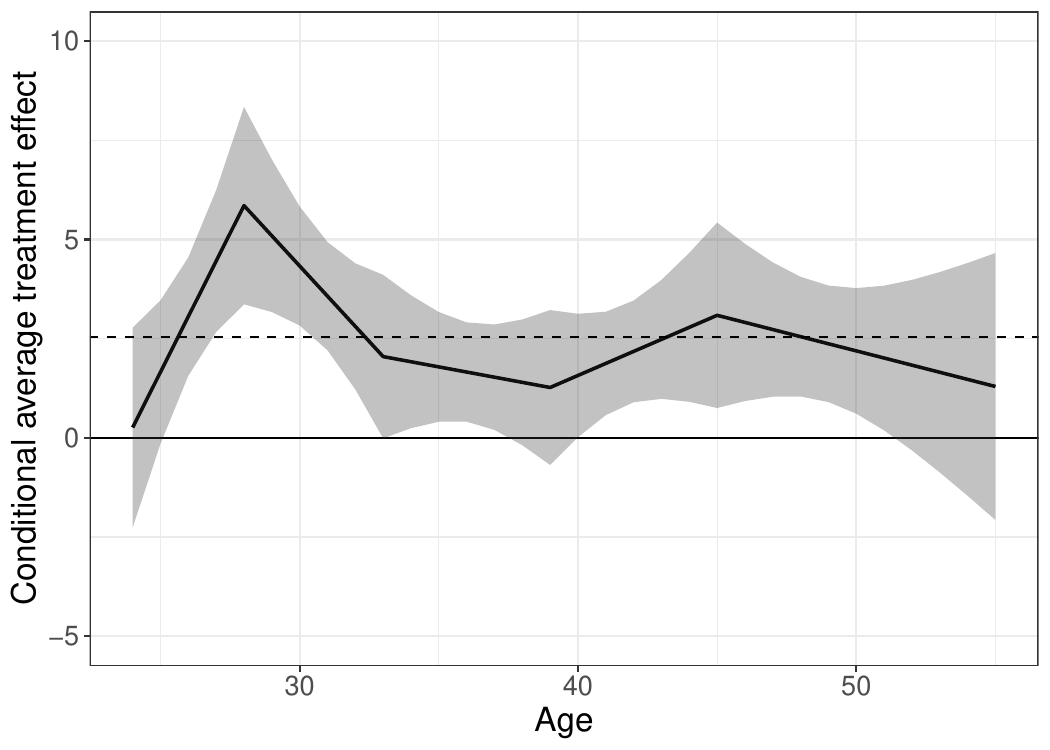}
\caption{Language (Spline)}
\end{subfigure}

\subcaption*{Dotted line indicates point estimate of the respective average treatment effect. Grey area shows 95\%-confidence interval.}
\end{figure}

Figure \ref{fig:cate-age} shows that the kernel and spline regressions detect no substantial heterogeneity for individuals of different age. All programs besides language courses show basically flat curves or at least the 95\% confidence intervals includes the ATE. Only language courses show a larger positive effect for job seekers in the late twenties. Both kernel and spline regressions agree regarding this but the relatively erratic curves prevent deriving strong conclusions from these results.

\clearpage

\subsection{IATEs}\label{sec:app-iate}

\begin{figure}[h]
    \centering
    \caption{Boxplot of IATEs estimated by DR- and NDR-learner}
    \includegraphics[width=0.7\textwidth]{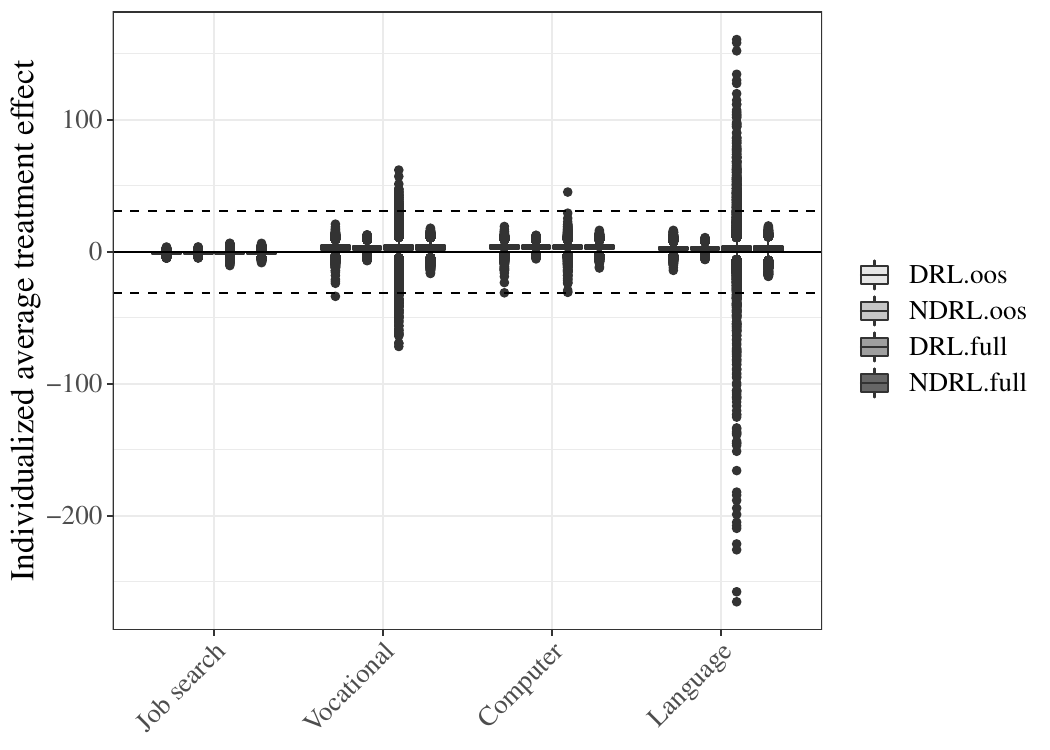}
    \subcaption*{\textit{Note:} The figure shows the distribution of IATEs for participating in the program labeled on the x-axis vs. non-participation estimated by the DR-learner (DRL) and the NDR-learner (NDRL). The first two boxplots of a group are obtained using the out-of-sample (oos) procedure of Appendix \ref{sec:app-drl} and the other two from the full sample. The dashed line indicates the possible range of the IATE of [-31,31] to illustrate that several DR-learner estimated IATEs lie outside this bound.}
    \label{fig:app-iate-box}
\end{figure}

Figure \ref{fig:app-iate-box} documents the extreme IATE predictions obtained using the full sample. Especially the DR-learner produces very extreme estimates for vocational training ranging from -265 to 161. Also in this extreme case, the NDR-learner mitigates the problem substantially. However, Table \ref{tab:iate_desc} documents that it still produces implausibly high values ranging from -19 to 20. The out-of-sample prediction of IATEs is thus preferred and discussed in the main text.

\begin{figure}[h] 
\centering
\caption{Joint and marginal distributions of IATEs}
\begin{subfigure}{0.49\textwidth}
\includegraphics[width=\linewidth]{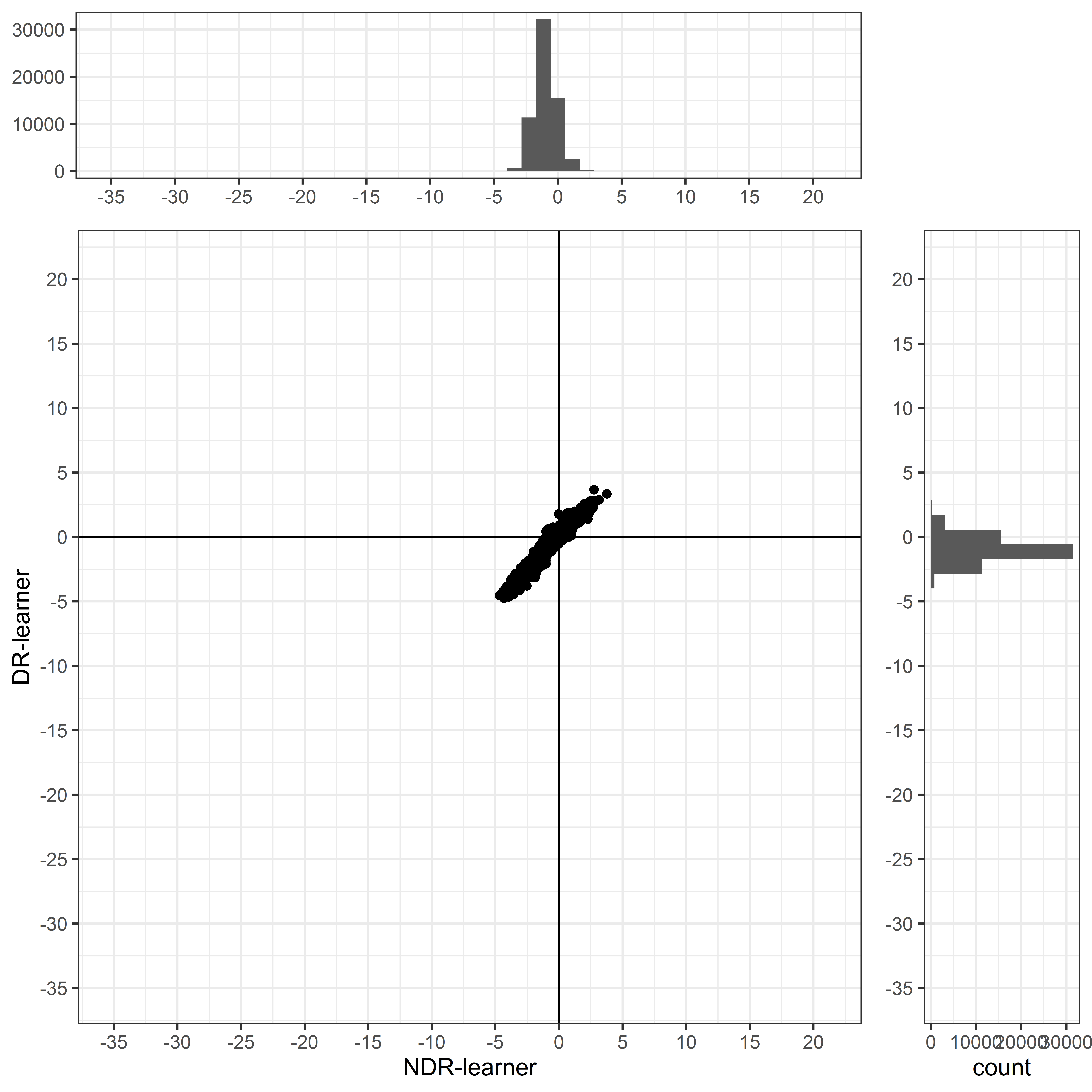}
\caption{Job search}
\end{subfigure} \hspace*{\fill}
\begin{subfigure}{0.49\textwidth}
\includegraphics[width=\linewidth]{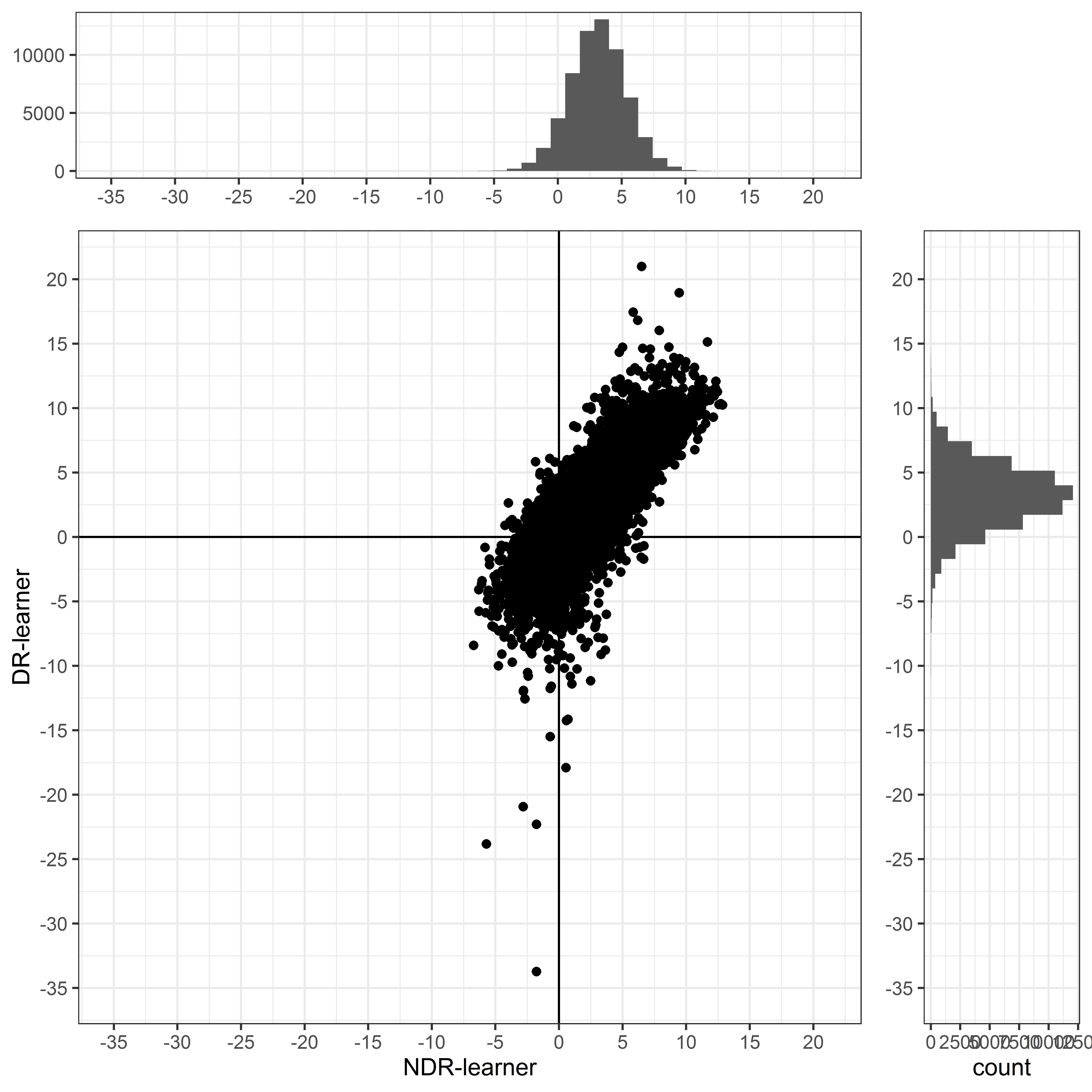}
\caption{Vocational}
\end{subfigure}
\begin{subfigure}{0.49\textwidth}
\includegraphics[width=\linewidth]{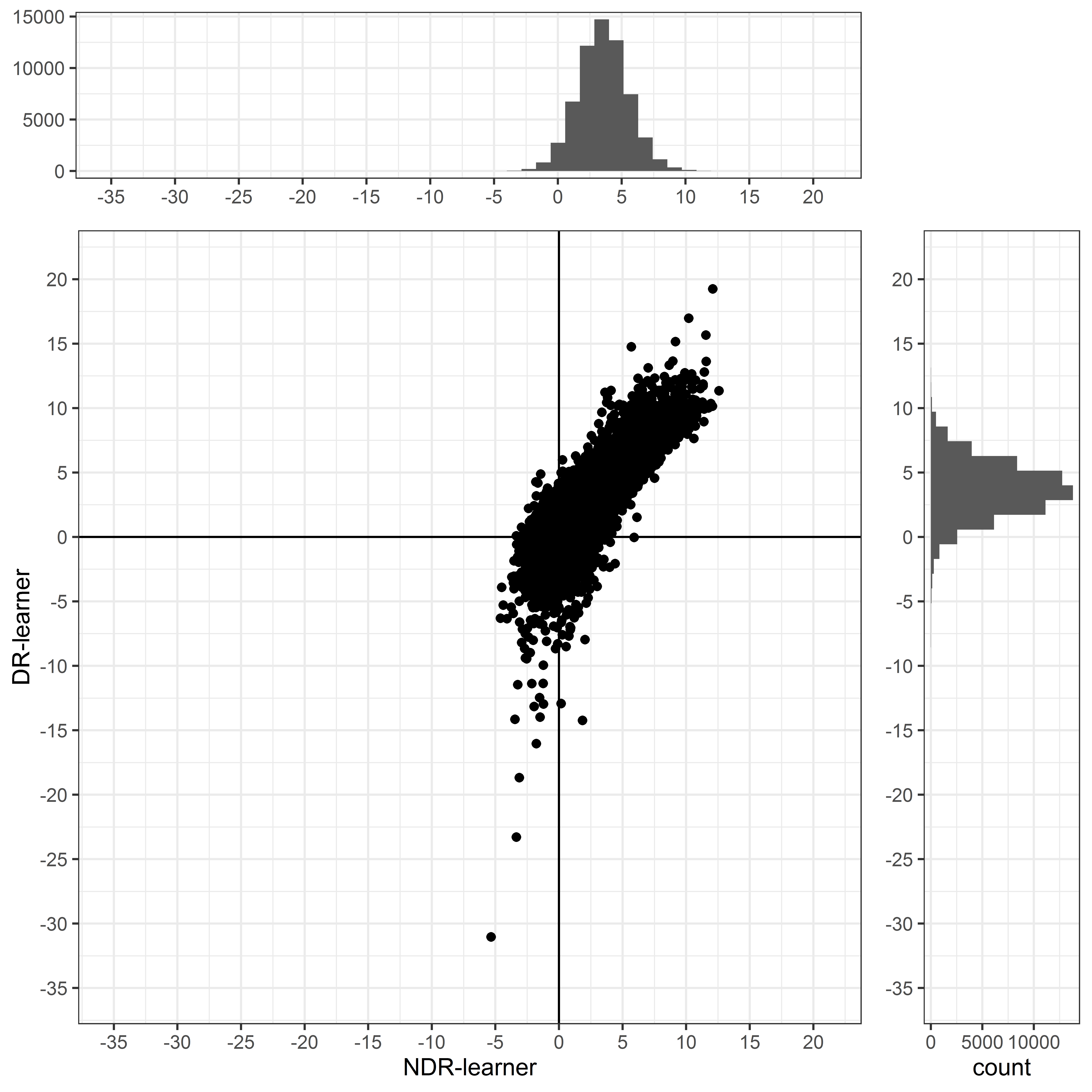}
\caption{Computer}
\end{subfigure} \hspace*{\fill}
\begin{subfigure}{0.49\textwidth}
\includegraphics[width=\linewidth]{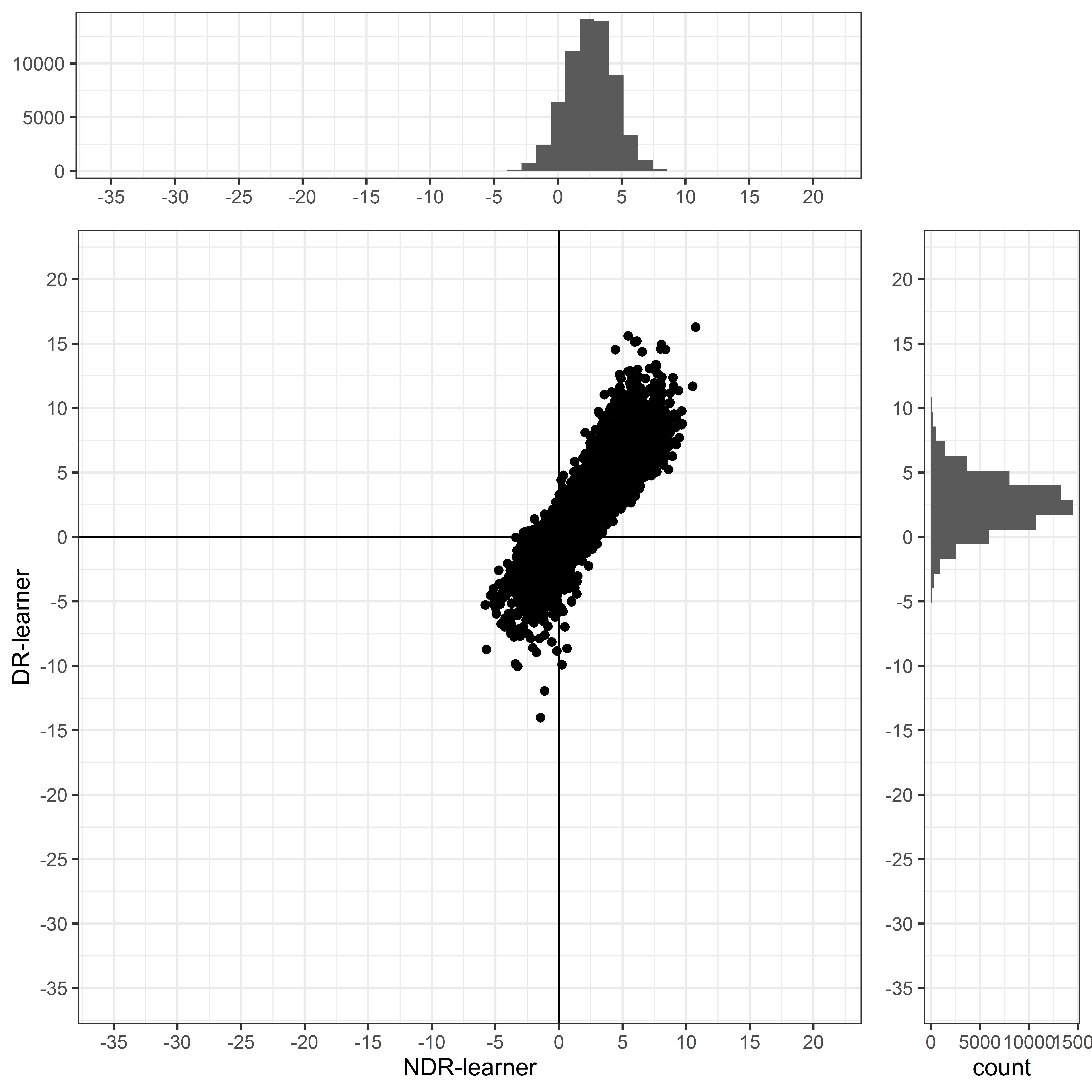}
\caption{Language}
\end{subfigure}
\subcaption*{\textit{Notes:} Figures show the joint and marginal distributions of IATEs estimated by DR-learner and NDR-learner.}  \label{fig:iates-dr-ndr}
\end{figure}

Table \ref{tab:iate_desc} and Figure \ref{fig:iates-dr-ndr} provide a detailed comparison of the IATEs estimated by DR- and NDR-learner. We see that the differences are mainly driven by few outliers as indicated by the much larger kurtosis for the DR-learner IATEs. However, most of the estimates are quite similar as the correlations of at least 0.90 provided in the last row of panel A in Table \ref{tab:iate_desc} and the scatter plots in Figure \ref{fig:iates-dr-ndr} document.

\onehalfspacing
\begin{table}[htp]
  \centering  \footnotesize
\begin{threeparttable}
  \caption{Summary statistics of IATE distributions} 
  \label{tab:iate_desc} 
\begin{tabular}{lcccccccc} 
\toprule
& \multicolumn{2}{c}{Job search} & \multicolumn{2}{c}{Vocational} & \multicolumn{2}{c}{Computer} & \multicolumn{2}{c}{Language}\\ 
\midrule
 & DRL & NDRL & DRL & NDRL & DRL & NDRL & DRL & NDRL \\ 
  \midrule
  \multicolumn{9}{l}{\textit{Panel A: Out-of-sample}} \\
Mean & -0.97 & -0.99 & 3.17 & 3.13 & 3.68 & 3.54 & 2.55 & 2.50 \\ 
  SD & 0.90 & 0.86 & 2.46 & 2.20 & 2.13 & 1.95 & 2.12 & 1.88 \\ 
  Minimum & -4.75 & -4.68 & \textbf{-33.72} & -6.71 & \textbf{-31.04} & -5.33 & -14.03 & -5.79 \\ 
  Q1 & -2.96 & -2.91 & -3.04 & -2.14 & -1.61 & -1.01 & -2.67 & -1.95 \\ 
  Q25 & -1.57 & -1.57 & 1.65 & 1.68 & 2.34 & 2.25 & 1.24 & 1.21 \\ 
  Q50 & -1.03 & -1.04 & 3.22 & 3.14 & 3.69 & 3.51 & 2.57 & 2.57 \\ 
  Q75 & -0.42 & -0.45 & 4.77 & 4.58 & 5.04 & 4.80 & 3.85 & 3.80 \\ 
  Q99 & 1.34 & 1.22 & 8.78 & 8.36 & 8.65 & 8.31 & 7.90 & 6.74 \\ 
  Maximum & 3.66 & 3.76 & 20.99 & 12.86 & 19.24 & 12.58 & 16.28 & 10.75 \\ 
  Kurtosis & 3.35 & 3.35 & 5.63 & 3.22 & 6.37 & 3.28 & 4.28 & 2.98 \\ 
  Correlation &  \multicolumn{2}{c}{0.99}  & \multicolumn{2}{c}{0.90} & \multicolumn{2}{c}{0.92}  & \multicolumn{2}{c}{0.93} \\ 
     [0.4em]
    \multicolumn{9}{l}{\textit{Panel B: Full sample}} \\
Mean & -1.03 & -1.04 & 3.15 & 3.08 & 3.46 & 3.32 & 2.50 & 2.40 \\ 
  SD & 1.20 & 1.14 & 4.07 & 3.03 & 2.31 & 2.17 & 7.51 & 3.42 \\ 
  Minimum & -10.42 & -8.21 & \textbf{-71.74} & -16.35 & -30.67 & -12.26 & \textbf{-265.23} & -18.58 \\ 
  Q1 & -3.89 & -3.71 & -6.18 & -4.78 & -2.06 & -2.05 & -11.07 & -6.46 \\ 
  Q25 & -1.80 & -1.79 & 1.27 & 1.22 & 2.08 & 1.94 & 0.41 & 0.23 \\ 
  Q50 & -1.07 & -1.08 & 3.24 & 3.14 & 3.48 & 3.34 & 2.53 & 2.52 \\ 
  Q75 & -0.29 & -0.33 & 5.19 & 5.04 & 4.88 & 4.73 & 4.61 & 4.70 \\ 
  Q99 & 1.98 & 1.82 & 11.57 & 10.14 & 8.79 & 8.45 & 16.98 & 10.05 \\ 
  Maximum & 6.53 & 6.45 & \textbf{62.01} & 17.92 & 45.33 & 16.30 & \textbf{160.92} & 19.63 \\ 
  Kurtosis & 4.07 & 3.86 & 45.54 & 4.02 & 13.18 & 3.80 & 228.23 & 3.96 \\ 
  Correlation &  \multicolumn{2}{c}{0.99}  & \multicolumn{2}{c}{0.86} & \multicolumn{2}{c}{0.94}  & \multicolumn{2}{c}{0.67} \\ 
    \bottomrule
    \end{tabular}%
\begin{tablenotes} \item \textit{Note:} The table provides summary statistics for the distributions of IATEs estimated by the DR-learner (DRL) and NDR-learner (NDRL). The rows denoted by Q show the respective quantiles. Correlation is calculated between the DR-learner and the NDR-learner. Bold numbers indicate values that are outside the possible range. \end{tablenotes}  
\end{threeparttable}
\end{table}
\doublespacing

Table \ref{tab:app-clan} shows the full results of the classification analysis. In line with previous results of \citeA{Knaus2020HeterogeneousApproach}, we observe that the caseworker characteristics play a negligible role in explaining variation in IATEs. All caseworker related variables are in the lower half of the table.

\singlespacing
\begin{table}[htp]
  \centering   \scriptsize
\begin{threeparttable}[t]
  \caption{Classification analysis of IATEs} 
  \label{tab:app-clan} 
\begin{tabular}{lcccccc} 
\toprule
\\[-1.8ex] & \multicolumn{1}{c}{Job search} & \multicolumn{1}{c}{Vocational} & \multicolumn{1}{c}{Computer} & \multicolumn{1}{c}{Language} \\ 
\\[-1.8ex] & \multicolumn{1}{c}{(1)} & \multicolumn{1}{c}{(2)} & \multicolumn{1}{c}{(3)} & \multicolumn{1}{c}{(4)} \\ 
\midrule
Past income & -1.32 & -0.84 & -1.17 & 1.01 \\ 
  Previous job: unskilled worker & 1.02 & 0.68 & 0.34 & -1.24 \\ 
  Mother tongue other than German, French, Italian & 0.69 & 0.68 & 0.00 & -1.17 \\ 
  Qualification: some degree & -0.88 & -0.65 & -0.41 & 1.15 \\ 
  Swiss citizen & -0.66 & -0.60 & 0.12 & 1.11 \\ 
  Fraction of months employed last 2 years & -1.06 & -0.37 & -0.47 & 0.30 \\ 
  Qualification: unskilled & 0.81 & 0.41 & 0.32 & -1.02 \\ 
  Previous job: skilled worker & -0.85 & -0.41 & -0.09 & 0.94 \\ 
  Missing sector & 0.89 & 0.05 & 0.21 & -0.53 \\ 
  Female & 0.62 & 0.01 & 0.88 & -0.48 \\ 
  Cantonal GDP p.c. & 0.31 & -0.78 & 0.01 & 0.26 \\ 
  Foreigner with temporary permit & 0.55 & 0.35 & 0.09 & -0.75 \\ 
  Cantonal unemployment rate (in \%) & 0.41 & -0.72 & 0.02 & 0.13 \\ 
  Married & 0.32 & 0.53 & 0.11 & -0.69 \\ 
  Foreigner with permanent permit & 0.31 & 0.40 & -0.21 & -0.67 \\ 
  Previous job in tertiary sector & -0.45 & -0.31 & 0.03 & 0.58 \\ 
  Employability & -0.45 & -0.55 & -0.55 & 0.22 \\ 
  Number of employment spells last 5 years & 0.53 & 0.22 & 0.04 & -0.08 \\ 
  Number of unemployment spells last 2 years & 0.47 & 0.11 & 0.13 & -0.15 \\ 
  Previous job: manager & -0.27 & -0.37 & -0.36 & 0.40 \\ 
  Lives in no city & -0.40 & 0.26 & 0.10 & 0.06 \\ 
  Lives in big city & 0.27 & -0.38 & -0.08 & -0.09 \\ 
  Age & 0.01 & 0.35 & 0.37 & -0.01 \\ 
  Qualification: semiskilled & 0.20 & 0.37 & 0.19 & -0.30 \\ 
  Caseworker age & 0.10 & 0.33 & -0.00 & 0.08 \\ 
  Previous job in primary sector & -0.33 & 0.21 & -0.28 & -0.19 \\ 
  Allocation of unemployed to caseworkers: by occupation & 0.17 & 0.02 & 0.23 & 0.30 \\ 
  Caseworker female & 0.07 & -0.26 & 0.29 & -0.09 \\ 
  Allocation of unemployed to caseworkers: by region & -0.27 & 0.06 & -0.05 & -0.10 \\ 
  Lives in medium city & 0.23 & 0.08 & -0.04 & 0.03 \\ 
  Previous job in secondary sector & -0.10 & 0.22 & -0.06 & -0.07 \\ 
  Mother tongue in canton's language & 0.06 & 0.09 & -0.18 & -0.08 \\ 
  Qualification: skilled without degree & 0.15 & 0.09 & -0.00 & -0.17 \\ 
  Caseworker education: above vocational training & -0.01 & 0.09 & 0.11 & 0.16 \\ 
  Caseworker education: tertiary track & 0.00 & -0.15 & -0.12 & -0.16 \\ 
  Caseworker cooperative & 0.01 & -0.06 & 0.15 & -0.02 \\ 
  Allocation of unemployed to caseworkers: by employability & -0.12 & 0.10 & 0.02 & 0.05 \\ 
  Caseworker education: vocational degree & -0.12 & -0.01 & -0.11 & -0.01 \\ 
  Caseworker tenure & 0.04 & 0.02 & -0.05 & -0.10 \\ 
  Allocation of unemployed to caseworkers: by age & -0.05 & 0.00 & 0.10 & 0.04 \\ 
  Allocation of unemployed to caseworkers: by industry & 0.10 & -0.09 & -0.02 & -0.03 \\ 
  Missing caseworker characteristics & 0.05 & -0.09 & 0.05 & -0.09 \\ 
  Previous job: self-employed & 0.06 & 0.00 & -0.02 & 0.03 \\ 
  Caseworker has own unemployemnt experience & 0.03 & -0.02 & -0.04 & 0.06 \\ 
  Allocation of unemployed to caseworkers: other & -0.04 & 0.03 & 0.01 & 0.00 \\ 
    \bottomrule
    \end{tabular}%
         \begin{tablenotes} \item \textit{Note:} Table shows the differences in means of normalized covariates between the fifth and the first quintile of the respective estimated IATE distribution. Variables are ordered according to the largest absolute difference. \end{tablenotes}  
\end{threeparttable}
\end{table}
\doublespacing

\clearpage

\subsection{Optimal treatment assignment} \label{sec:app-op}

\begin{figure}[h]
    \centering
        \caption{Optimal decision tree of depth three with five covariates}
    \includegraphics[width=0.99\linewidth]{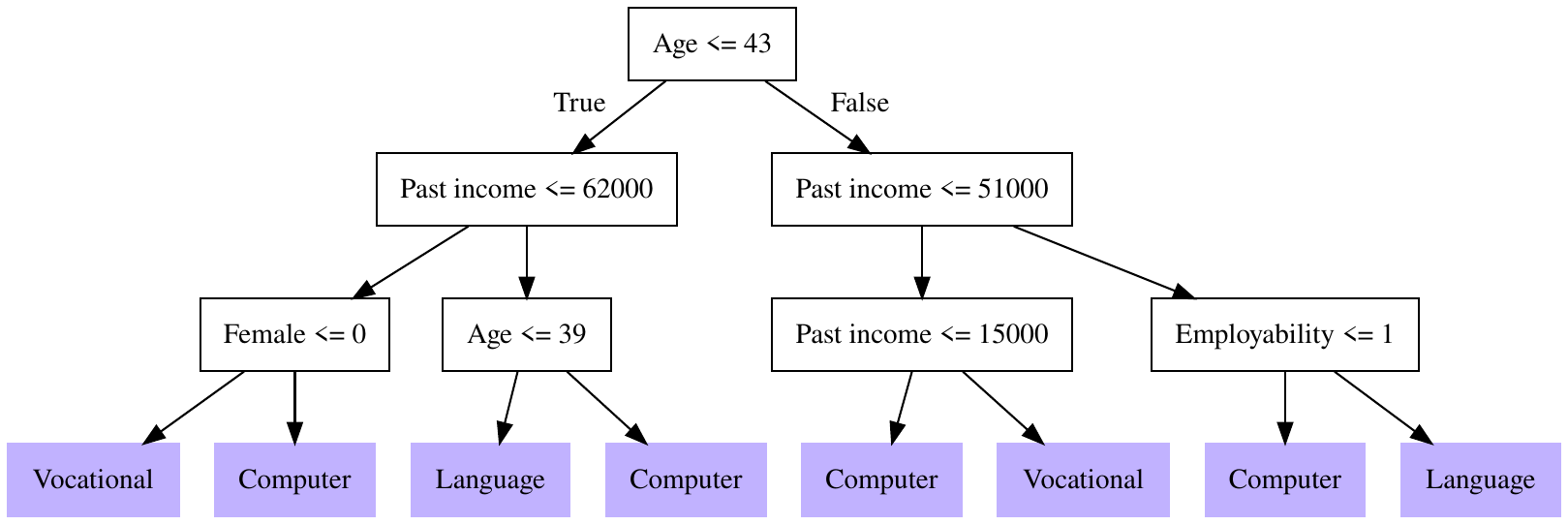}
\caption*{\textit{Notes:} Optimal assignment rules estimated following the procedure defined in Section 3.4.}
\end{figure}

\begin{figure}[h]
    \centering
        \caption{Overlap of cross-validated policy rules with five covariates}
    \includegraphics[width=0.7\linewidth]{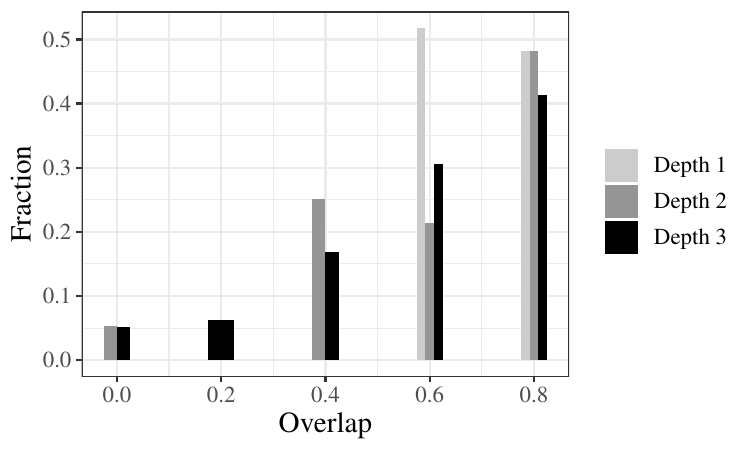}
    \label{fig:op-overlap}
\caption*{\textit{Notes:} Figure shows the fraction of cross-validated policies that agree with the full sample policy.}
\end{figure}

\begin{figure}[h]
    \centering
        \caption{Optimal decision tree of depth three with 16 covariates}
    \includegraphics[width=0.99\linewidth]{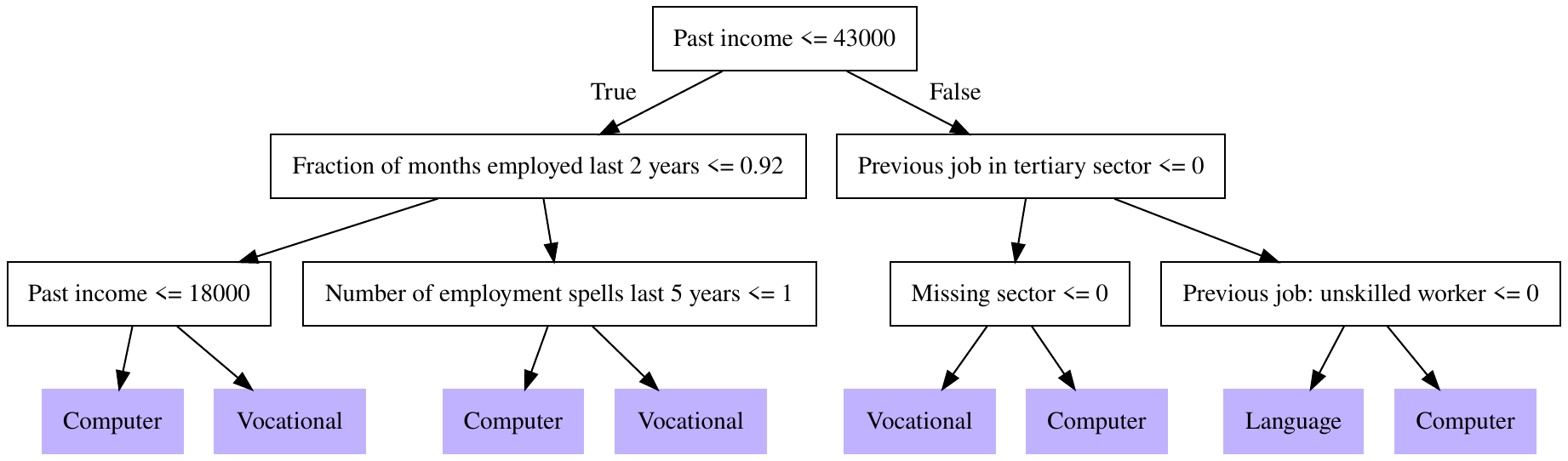}
\caption*{\textit{Notes:} Optimal assignment rules estimated following the procedure defined in Section 3.4.} 
\end{figure}

\begin{figure}[h]
    \centering
        \caption{Overlap of cross-validated policy rules with 16 covariates}
    \includegraphics[width=0.7\linewidth]{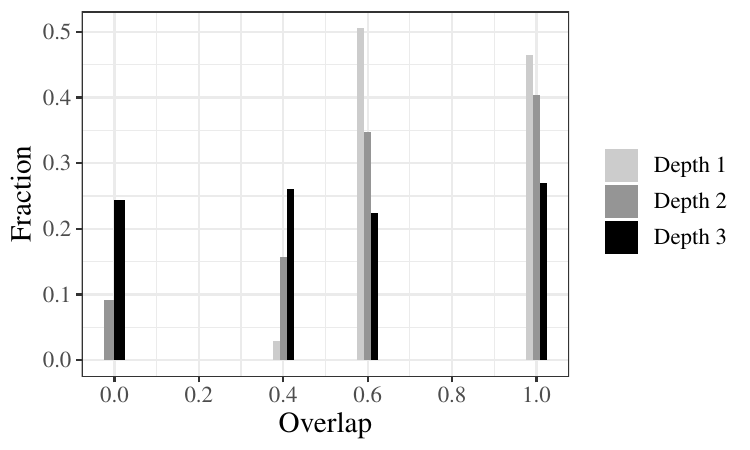}
    \label{fig:op-overlap-high}
\caption*{\textit{Notes:} Figure shows the fraction of cross-validated policies that agree with the full sample policy.}
\end{figure}
\end{appendices}

\end{document}